\documentclass[aps,prx,preprint,onecolumn,citeautoscript,superscriptaddress,nofootinbib,
eqsecnum]{revtex4}
\usepackage{amsmath}
\usepackage{amssymb}
\usepackage{bbm}
\usepackage{bm}
\usepackage{tikz-feynman}
\usepackage{comment}
\usepackage{graphicx}
\usepackage{color}
\usepackage[papersize={8.5in,11in}]{geometry}
\usepackage[colorlinks=true]{hyperref}
\hypersetup{
    bookmarks=true,         
    unicode=false,          
    pdftoolbar=true,        
    pdfmenubar=true,        
    pdffitwindow=false,     
    pdfstartview={FitH},    
    pdftitle={Large N theory of critical Fermi surfaces},    
    pdfauthor={Ilya Esterlis, Haoyu Guo, Aavishkar A. Patel, Subir Sachdev},     
    pdfsubject={},   
    pdfcreator={},   
    pdfproducer={}, 
    pdfkeywords={} {} {}, 
    pdfnewwindow=true,      
    colorlinks=true,       
    linkcolor=magenta, 
    citecolor=blue,        
    filecolor=magenta,      
    urlcolor=blue           
}

\usepackage{amsfonts}
\usepackage{upgreek}
\usepackage{slashed}
\usepackage{latexsym}
\usepackage{dsfont}
\usepackage{subcaption}
\newcommand{\beq}{\begin{equation}}
\newcommand{\eeq}{\end{equation}}
\def\bea{\begin{eqnarray}}
\def\eea{\end{eqnarray}}

\DeclareMathOperator{\Tr}{Tr}
\newcommand{\nn}{\nonumber \\}
\renewcommand{\vec}[1]{\boldsymbol{#1}}

\newcommand{\be}{\begin{equation}}
\newcommand{\ee}{\end{equation}}

\usepackage{braket}

\renewcommand{\approx}{\simeq}
\renewcommand{\Re}{\text{Re}}

\definecolor{wrongultramarine}{rgb}{1,0.5,0}

\newcommand{\rd}{{\rm d}}
\newcommand{\sgn}{{\rm sgn\,}}
\newcommand{\calG}{{\mathcal G}}
\usepackage{tikz-3dplot}
\usetikzlibrary{calc}
\tikzset{
  mid arrow/.style={postaction={decorate,decoration={
        markings,
        mark=at position .575 with {\arrow[#1]{stealth}}
      }}},
  near arrow/.style={postaction={decorate,decoration={
        markings,
        mark=at position .275 with {\arrow[#1]{stealth}}
      }}},
   far arrow/.style={postaction={decorate,decoration={
        markings,
        mark=at position .800 with {\arrow[#1]{stealth}}
      }}},
  sigma snake/.style={decorate,decoration={snake, amplitude=0.5pt, segment length=2pt}
      },
}

\begin{document}

\preprint{\href{https://arxiv.org/abs/2103.08615}{arXiv:2103.08615}}

\title{Large $N$ theory of critical Fermi surfaces}

\author{Ilya Esterlis}
\affiliation{Department of Physics, Harvard University, Cambridge MA 02138, USA}

\author{Haoyu Guo}
\affiliation{Department of Physics, Harvard University, Cambridge MA 02138, USA}

\author{Aavishkar A. Patel}
\affiliation{Department of Physics, University of California Berkeley, Berkeley CA 94720, USA}

\author{Subir Sachdev}
\affiliation{Department of Physics, Harvard University, Cambridge MA 02138, USA}

\date{\today}

\begin{abstract}
We describe the large $N$ saddle point, and the structure of fluctuations about the saddle point, of a theory containing a sharp, critical Fermi surface in two spatial dimensions. The theory describes the onset of Ising order in a Fermi liquid,  and closely related theories apply to other cases with critical Fermi surfaces. We employ random couplings in flavor space between the fermions and the bosonic order parameter, but there is no spatial randomness: consequently, the $G$-$\Sigma$ path integral of the theory is expressed in terms of fields bilocal in spacetime.
The critical exponents of the large $N$ saddle-point are the same as in the well-studied non-random RPA theory; in particular, the entropy density vanishes in the limit of zero temperature. We present a full numerical solution of the large $N$ saddle-point equations, and the results agree with the critical behavior obtained analytically. Following analyses of Sachdev-Ye-Kitaev models, we describe scaling operators which descend from fermion bilinears around the Fermi surface. This leads to a systematic consideration of the role of time reparameterization symmetry, and the scaling of the Cooper pairing and $2k_F$ operators which can determine associated instabilities of the critical Fermi surface. We find no violations of scaling from time reparameterizations. We also consider the same model but with spatially random couplings: this provides a systematic large $N$ theory of a marginal Fermi liquid with Planckian transport.
\end{abstract}
\maketitle
\tableofcontents

\section{Introduction}
\label{sec:intro}

The problem of the critical Fermi surface without quasiparticle excitations has received extensive interest in recent years \cite{Lee_ARCMP}, given its central role in the theory of the half-filled Landau level, quantum critical points (QCPs) in metals, and for Fermi surfaces of fractionalized particles coupled to emergent gauge fields in gapless spin liquids \cite{PALee89,Polchinski:1993ii,HLR,Kim94,Altshuler94,Nayak:1994ng,sungsik1,metlitski1,mross,sungsik3,metlitski5,Hartnoll:2014gba,Patel_viscosity,HolderMetzner1,HolderMetzner2,Raghu1,Raghu2,Torroba1,Torroba2,Hooley15,Patel2018mag,Chowdhury:2018sho,Moon2010,Chubukov1,Chubukov2,Chubukov3,Berg1,Berg2,Berg3,DCBerg,DebanjanAPS,Patel:2018eak,Ilya1,Ilya2,Wang:2020dtj,Altman1,Patel:2016wdy,Patel:2019qce,oganesyan2001}. Much has been understood about the structure of the theory, but a fully systematic analysis of the critical theory, its operator spectrum, and possible low temperature instabilities to symmetry-broken or topological states has remained elusive. Initially, it was assumed that a theory with a large number of flavors, $N$, of fermions could provide such a systematic theory. But, in an influential analysis, Sung-Sik Lee \cite{sungsik1} showed that certain higher loop corrections implied a break down of the large $N$ expansion, and that the theory remains strongly coupled even at large $N$. Various workarounds have been proposed since then, but none are fully satisfactory: they either involve deploying additional expansion parameters, introduce non-analytic terms not present in the original theory, or depend upon choosing $N$-dependent energy scales.

In this paper, we wish to apply insights gained by the study of another class of realizations of compressible quantum matter without quasiparticle excitations: the Sachdev-Ye-Kitaev (SYK) class of models \cite{SY92,kitaev2015talk,SS15,JMDS16}. These model involve random couplings, but after an average over randomness, the resulting theory realizes a compressible non-Fermi liquid which is amenable to a systematic large $N$ expansion via a path integral over bilocal fields \cite{RSY95,GPS,JMDS16} often called the $G$-$\Sigma$ theory (the low energy limit of the $G$-$\Sigma$ theory is a theory of two-dimensional quantum gravity \cite{SS10,kitaev2015talk,JMDS16,nearlyads2,Kitaev:2017awl}). Furthermore, the model is strongly self-averaging in the non-Fermi liquid phase
{\it i.e.\/} for sufficiently large $N$, the properties of a {\it single} sample are indistinguishable from those of the average, and so it is technically far easier, and permissible, to work with the average theory. Recent works \cite{Kim:2019upg,Klebanov:2020kck,Tikhanovskaya:2020elb,Tikhanovskaya:2020zcw} have used this method to obtain operator spectra and instabilities of non-Fermi liquids realized by SYK models. The idea of simplification realized by an average over similar strongly-coupled theories is also playing an important role in recent investigations in quantum gravity,
and averages over random matrices or conformal field theories yield systematic large $N$ holographic realizations of the path integral of simple theories of gravity \cite{Saad:2019lba,Stanford:2019vob,Afkhami-Jeddi:2020ezh,Maloney:2020nni,Perez:2020klz,Cotler:2020ugk,Engelhardt:2020qpv,Datta:2021ftn}. 

In the context of finite-dimensional systems with Fermi surfaces, Aldape {\it et al.}~\cite{Altman1} introduced the idea of coupling the fermions to a boson, and making the associated Yukawa-like fermion boson coupling a random function of flavor indices, but not of space. In such a model, the sharp Fermi surface is maintained even in the presence of randomness, and a systematic large $N$ saddle-point with regimes of non-Fermi liquid behavior is obtained. They used this strategy to describe non-Fermi liquid behavior across a transition from a Fermi liquid to a fractionalized Fermi liquid (FL*). We also note Ref.~\cite{Kim:2020jpz} which applies this idea to critical Dirac fermions in 1+1 dimensions, and obtained evidence for maximal chaos in the large $N$ limit with vanishing entropy density in the zero temperature limit. We emphasize that in all these cases there is no spatial disorder, and the disorder is entirely in coupling-constant space. A given sample with a particular set of couplings will not be identical to another sample, but the difference will vanish in the large $N$ limit. To compute the differences between samples we have to include fluctuations of the replica off-diagonal components of the bilocal Green's functions \cite{RSY95}, but we will not do that here. The replica off-diagonal components vanish in the $N=\infty$ saddle point theory, and it is even possible that the replica off-diagonal fluctuations do not modify the universal critical properties at higher orders in $1/N$, because we do not expect the critical properties to be sensitive to the microscopic values of the couplings.

We follow these recent works \cite{Altman1,Kim:2020jpz}, and examine the non-Fermi liquid with a critical Fermi surface formed at the QCP involving the onset of Ising order a two-dimensional Fermi liquid. We note that the location of the critical Fermi surface in momentum space obeys an extended Luttinger theorem \cite{Huijse:2011hp,ElseSenthil1}, even in the presence of the random couplings. Our large $N$ theory of a critical Fermi surface has the same critical behavior at $N=\infty$ as already anticipated in the early work \cite{Lee_ARCMP,PALee89,Polchinski:1993ii,HLR}. Specifically, there is no extensive entropy in the zero temperature limit in our approach, unlike previous studies of critical Fermi surfaces employing a large $N$ limit with random couplings \cite{Patel2018mag,Chowdhury:2018sho,Patel:2019qce,DCBerg}. And there is anisotropic dynamic scaling on the Fermi surface, with the frequency $\omega \sim q_{\perp}^{3/2}$ for momenta $q_\perp$ normal to the Fermi surface, and $\omega \sim q_\parallel^3$ for momenta parallel to the Fermi surface.

We introduce our model, its averaged effective action, and the setup of the large-$N$ expansion in Section~\ref{sec:rpaffs}. This involves a path integral over a $G$-$\Sigma$ action in which the fields are bilocal in spacetime; this should be compared to previous studies \cite{RSY95,GPS,JMDS16} in which the fields were bilocal only in time.
The large $N$ critical theory is obtained in Section~\ref{sec:rpapch} by taking the low energy limit on separate patches on the Fermi surface. We then present a full numerical solution of the large $N$ equations for the complete Fermi surface in Section~\ref{sec:num}, and find results which are in agreement with the analytic patch analysis in Section~\ref{sec:rpapch}.

We begin our discussion of the fluctuations about the large $N$ saddle point in Section~\ref{sec:fluc} by a consideration of the role of time reparameterization symmetry. The special role of this symmetry, and an associated soft mode, was first noted for the SYK model \cite{kitaev2015talk,JMDS16,Kitaev:2017awl,Bagrets:2016cdf,Bagrets:2017pwq,Altland:2019czw,Kruchkov:2019idx,Kobrin:2020xms} where it leads to a violation of scaling at times of order $N$. The symmetry is also present in the saddle-point action for the critical Fermi surface.
However, we shall find here that there is no corresponding soft mode for the critical Fermi surface, and no associated violation of scaling. This time reparameterization analysis is carried out in a single patch theory, which on its own realizes a `chiral non-Fermi liquid' \cite{sungsik3}. The soft mode analysis involves examination of the eigenmodes of a ladder operator, which determine composite operators in the particle-hole sector. We limit our consideration to momenta orthogonal to the Fermi surface, in which case the eigenmode equations simplify to one-dimensional integral equations. We do not find any non-trivial operators in this sector, apart from the conserved density operator whose correlations were studied by Kim {\it et al.} \cite{Kim94}.

One operator that could have appeared in a single patch theory is the fermion pair operator associated with Amperean pairing \cite{LeeAmperean1,LeeAmperean2}. This requires consideration of physics beyond the scaling limit, and is discussed in Appendix~\ref{app:amperean}.

Section~\ref{sec:diagram} describes the structure of the $G$-$\Sigma$ theory beyond the large $N$ saddle point. We will obtain formal expressions for the fermion self energy at order $1/N$ in terms of the inverse of the ladder operator of the large $N$ theory. These considerations will be quite general, and can be applied equally to the single patch theory of Section~\ref{sec:fluc}, the lattice theory of Section~\ref{sec:rpapch}, or the antipodal patch theory to be considered in Section~\ref{sec:bilinears}, also bears a resemblance to the corresponding analysis of the SYK model in Ref.~\cite{Kitaev:2017awl}.

Section~\ref{sec:bilinears} turns to the examination of the scaling structure for the non-chiral case, with a closed Fermi surface. Then, a number of singular effects arise from antipodal pairs of patches on the Fermi surface:
\begin{itemize}
    \item Ref.~\cite{metlitski1} showed that three-loop diagrams lead to a small correction to the fermion anomalous dimension, $\eta_\psi$. Our $1/N$ expansion will contain a similar correction to the value of $\eta_\psi$, but it will be a systematic contribution at order $1/N$, unlike the previous analysis which was not controlled in their large $N$ limit.
    \item In the particle-particle sector, fermions on antipodal patches can undergo a Cooper-pairing instability. A renormalization group analysis of this was presented in Ref.~\cite{metlitski5} employing an expansion \cite{mross} which combined large $N$ with a bare long-range interaction controlled with small $\epsilon$. In our large $N$ expansion, the $N=\infty$ equations for the scaling dimension of the Cooper pair operator reduce to integral equations which appeared in the $\gamma$-model of Chubukov and collaborators \cite{Moon2010,Chubukov1,Chubukov2,Chubukov3}, and also coincide with equations studied in the SYK model \cite{Klebanov:2016xxf,Klebanov:2018fzb}. We find either a non-trivial scaling dimension for the Cooper pair operator, or an instability to a paired ground state with no critical Fermi surface.
    \item In the particle-hole sector, fermions on antipodal patches yield density fluctuations at the $2k_F$ wavevector. The scaling dimension of the $2 k_F$ operator was computed in the combined large $N$/small $\epsilon$ expansion in Ref.~\cite{mross}. Our large $N$ theory yields integral equations in momentum and frequency for the scaling dimension of particle-hole operators on antipodal patches, and we solve these equations numerically. These equations have not been studied previously.
\end{itemize}

Finally, in Section~\ref{sec:spatial} we consider a model in which the Yukawa coupling is spatially random, in addition to the randomness in the flavor space. This model also provides a systematic large $N$ theory of a sharp Fermi surface, but for a marginal Fermi liquid. The physical properties of this model turn out to be quite similar to a different model considered by Aldape {\it et al.}~\cite{Altman1}, including Planckian transport discussed in Section~\ref{sec:planckian}.

\section{Lattice model and effective action}
\label{sec:rpaffs}

This section will introduce a lattice model for the onset of Ising order in a Fermi liquid on the square lattice, and described the structure of its large $N$ saddle point.

We consider fermions $\psi_{ik}$ with a flavor index $i=1 \ldots N$ and momenta $k$ obeying periodic boundary conditions. These fermions are coupled to soft Ising fields $\phi_{ik}$ representing Ising-nematic order corresponding to a breaking of $C_4$ rotational symmetry. In imaginary time ($\tau$), the lattice action is
\begin{align}
&S = \int d\tau\sum_k\sum_{i=1}^N\psi^\dagger_{ik}(\tau)\left[\partial_\tau-2t(\cos k_x+\cos k_y)-\mu\right]\psi_{ik}(\tau) \nn
&+\frac{1}{2}\int d\tau \sum_q\sum_{i=1}^N \phi_{iq}(\tau)\left[-\partial_\tau^2-2J(\cos q_x+\cos q_y-2)+m_b^2\right]\phi_{i,-q}(\tau) \nn
&+\int d\tau \sum_{k,q}\left(\cos k_x - \cos k_y \right)\sum_{i,j,l=1}^N\left[\frac{g_{ijl}}{N}\psi^\dagger_{i,k+q}(\tau)\psi_{jk}(\tau)\phi_{lq}(\tau)\right]\,.
\label{eq:latticeaction}
\end{align}
Here $t$ is the fermion hopping, $\mu$ is chemical potential, the coupling $J$ determines the dispersion of the boson, and $m_b$ is the bare boson mass.  We will henceforth set the nematic form factor $\cos k_x - \cos k_y$ to $1$ for simplicity, since it does not qualitatively affect any of the physics we will be interested in apart from at a measure zero set of special points in momentum space with $k_x=\pm k_y$. In the absence of this form factor, the $\phi_{ik}$ field no longer has an interpretation as a fluctuating order parameter, as condensing $\phi_{ik}$ does not break any symmetries. However, one may still think of $\phi_{ik}$ as a phonon field, in which case a $q=0$ instability corresponds to phase separation.

The novelty in our approach is that the Yukawa couplings $g_{ijl}$ are independent, translationally-invariant complex Gaussian random variables with zero mean and variance $g^2$, and $g_{jil}=g^\ast_{ijl}$. We will comment where needed on the differences that appear upon taking real $g_{ijl}$. Upon performing an SYK-like disorder average at large $N$, we obtain
\begin{align}
&S = \int d\tau\sum_k\sum_{i=1}^N\psi^\dagger_{ik}(\tau)\left[\partial_\tau-2t(\cos k_x+\cos k_y)-\mu\right]\psi_{ik}(\tau) \nn
&+\frac{1}{2}\int d\tau \sum_q\sum_{i=1}^N \phi_{iq}(\tau)\left[-\partial_\tau^2-2J(\cos q_x+\cos q_y-2)+m_b^2\right]\phi_{i,-q}(\tau) \nn
&+N\frac{g^2}{2}\int d\tau d\tau' \sum_{k,q}G(k,\tau-\tau')G(k+q,\tau'-\tau)D(\tau-\tau',q) \nn
&-N\int d\tau d\tau'\sum_k\Sigma(k,\tau'-\tau)\left[G(k,\tau-\tau')+\frac{1}{N}\sum_{i=1}^N\psi_{ik}(\tau)\psi^\dagger_{ik}(\tau')\right] \nn
&+\frac{N}{2}\int d\tau d\tau'\sum_q\Pi(q,\tau'-\tau)\left[D(q,\tau-\tau')-\frac{1}{N}\sum_{i=1}^N\phi_{iq}(\tau)\phi_{i,-q}(\tau')\right].
\end{align}
Here, we have introduced fermion (boson) Green's functions $G$ ($D$) and self energies $\Sigma$ ($\Pi$) as dynamical degrees of freedom by employing them as Lagrange multipliers. We have already assumed the saddle point structure in which the $G$, $\Sigma$, $D$, $\Pi$ fields are functions only of differences in spacetime positions; but in the full path integral, these fields are bilocal in spacetime.
The action is quadratic in fermions and bosons, and integrating them out yields the $G$-$\Sigma$-$D$-$\Pi$ action of the theory with a prefactor of $N$.

The saddle point equations $\delta S/\delta \Sigma = \delta S/\delta G = 0 = \delta S/\delta D= \delta S/\delta \Pi$, yield the familiar RPA-Dyson equations, exact at large $N$:
\begin{align}
&\Sigma(r,\tau) = g^2 D(r,\tau)G(r,\tau),~~\Pi(r,\tau) = -g^2 G(-r,-\tau)G(r,\tau), \nn
&G(k,i\omega_n) = \frac{1}{i\omega_n+2t(\cos k_x+\cos k_y)+\mu-\Sigma(k,i\omega_n)}, \nn
&D(q,i\Omega_m) = \frac{1}{\Omega_m^2-2J(\cos q_x+\cos q_y -2)+m_b^2-\Pi(q,i\Omega_m)}.
\label{eq:saddle_pt_eqs}
\end{align}
These equations may be solved numerically by using Fourier transforms in space and time, and iterative updates, as will be described in Section~\ref{sec:num}. Care needs to be taken that $D^{-1}(q,i\Omega_m)>0$ throughout the iterative procedure. At criticality, we must first determine $m_b^2$ by requiring that $m_b^2-\Pi(0,0)=0$ at infinite system size and zero termperature, and then use this value of $m_b^2$ in the finite-size and finite-temperature problem. The number of points in the numerical analysis can be reduced somewhat by exploiting the KMS conditions $G(\tau+\beta)=-G(\tau)$, $D(\tau+\beta)=D(\tau)=D(-\tau)$, and the spatial $C_4$ symmetry $G(x,y)=G(\pm x,\pm y)$, $D(x,y)=D(\pm x,\pm y)$.

\subsection{Issues with the boson thermal mass}
\label{sec:thfluc}

In this RPA theory, criticality is achieved when $m_b^2-\Pi(0,0)|_{T=0}^{L=\infty}=0$, where $L$ is the system length. At $T\neq0$ and/or finite $L$, there is a `thermal mass' for the boson given by $M^2(T,L)=m_b^2-\Pi(0,0)=\Pi(0,0)|_{T=0}^{L=\infty}-\Pi(0,0)$. This follows the finite temperature/length correction to the free fermion compressibility, which is very small, and, for the nearest-neighbor square lattice considered here, is negative at low $T$, which causes a first-order transition at small $T\neq0$ at generic fermion fillings. This is undesirable.

To remedy this issue, we include a fixed length constraint $\sum_q\sum_{i=1}^N \phi_{iq}(\tau)\phi_{i,-q}(\tau)=N/\gamma$. This can arise from the $U\rightarrow\infty$ limit on the quartic boson self-interaction $\sum_r U/(2N)(\sum_{i=1}^N\phi_{i,r}(\tau)\phi_{i,r}(\tau)-N/\gamma)^2$, which in turn is generated by integrating out fermions in the full, non-RPA theory. Performing a Hubbard-Stratonovich transformation and sending $U\rightarrow\infty$ gives the action with a Lagrange multiplier $\lambda_r(\tau)$,
\beq
S_1 = S + \frac{1}{2}\int d\tau\sum_{r}i\lambda_r(\tau)\left(\sum_{i=1}^N\phi_{ir}(\tau)\phi_{ir}(\tau)-\frac{N}{\gamma}\right).
\eeq
At the large $N$ saddle point, $i\lambda_r(\tau)=m_b^2$, and we get the model described in Sec.~\ref{sec:rpaffs}, with an additional constraining equation in the set of Dyson equations: $D(r=0,\tau=0)=1/\gamma$. We can now control the phase diagram by tuning $\gamma$; $m_b^2$ is adjusted along with $L$ and $T$ to keep $\gamma$ fixed. At $\gamma=\gamma_c$, $m_b^2|_{T=0}^{L=\infty}-\Pi(0,0)|_{T=0}^{L=\infty}=0$. We therefore tune $\gamma$ so that $M^2(T,L)$ vanishes at the lowest numerically accessible $T$ and largest $L$. Our numerical solution, as well as analysis with the simpilfied continuum RPA boson propagator $D(q,i\Omega_m)=1/(\Omega_m^2+q^2+|\Omega_m|/|q|+M^2(T))$, find $M^2(T,\infty) \sim T\ln(1/T)$ at low $T$ at criticality, which is parametrically smaller than the expected RPA quantum critical scaling $M^2(T,\infty)\sim T^{2/3}$ \cite{millis,Hartnoll:2014gba}. This anomalous thermal mass also leads to an anomalous contribution to the fermion self energy, from the thermal ($\Omega_m=0$) fluctuations of the bosons, that doesn't obey the expected quantum critical scaling; however, this was shown to be in qualitative agreement with recent analysis of quantum Monte Carlo simulation data \cite{Meng2020} from the full, non-RPA theory, which indicates that the RPA model is a reasonable description of the actual physical problem, at least above some temperature scale.

\subsection{Thermodynamics}
\label{sec:thd}

The RPA grand canonical free energy, exact at the large $N$ saddle point, is given by ($\mathcal{Z}=e^{-S_1}$)
\begin{align}
&\mathcal{F}/N = -T\sum_{k,\omega_n}\ln\left[\frac{i\omega_n+2t(\cos k_x+\cos k_y)+\mu-\Sigma(k,i\omega_n)}{i\omega_n+2t(\cos k_x+\cos k_y)+\mu}\right]-T\sum_k\ln\left[1+e^{(2t(\cos k_x+\cos k_y)+\mu)/T}\right] \nn
&+\frac{T}{2}\sum_{q,\Omega_m}\ln\left[\frac{\Omega_m^2-2J(\cos q_x+\cos q_y-2)+m_b^2-\Pi(q,i\Omega_m)}{\Omega_m^2-2J(\cos q_x+\cos q_y-2)+m_b^2}\right]+T\sum_q\ln\left[1-e^{-(m_b^2-2J(\cos q_x+\cos q_y-2))^{1/2}/T}\right] \nn
&-T\sum_{k,\omega_n}\Sigma(k,i\omega_n)G(k,i\omega_n)-L^2\frac{m_b^2}{2\gamma} + I_0.
\label{frenerg}
\end{align}
Here expressions for free fermion and free boson free energies have been added and subtracted to ensure numerical convergence, and $L$ is the system size. Note that, for the boson contribution to the free energy, we are subtracting  $I_T=(T/2)\sum_{q,\Omega_m}\ln[\Omega_m^2-2J(\cos q_x + \cos q_y -2)+m_b^2]$, which is unregulated, but only adding the (regular) difference $I_T-I_0=T\sum_q\ln[1-e^{-(m_b^2-2J(\cos q_x+\cos q_y-2))^{1/2}/T}]$, where $I_0=(1/2)\int_{\Omega_m}\sum_q\ln[\Omega_m^2-2J(\cos q_x + \cos q_y -2)+m_b^2]$. This difference represents the free energy of free boson {\it excitations}, and vanishes as $T\rightarrow0$. We therefore also add back the formally infinite constant $I_0$ on the third line of (\ref{frenerg}), which is physically just the ground state energy of the collection of dispersive free boson harmonic oscillators with mass $m_b^2$:
\beq
I_0 = \frac{1}{2}\sum_q\sqrt{m_b^2-2J(\cos q_x +\cos q_y -2)}.
\eeq
The entropy is then given by $\mathcal{S}/N=-(1/N)(\partial \mathcal{F}/\partial T)|_{\mu,\gamma}$, where the derivative is taken numerically at fixed inverse length $\gamma$ and chemical potential $\mu$.

\section{Large $N$ critical theory}
\label{sec:rpapch}

In this section, we analyze the low-energy version of the quantum critical RPA model around a single patch of the Fermi surface. We take $L=\infty$ to begin with, and will be concerned only about quantum fluctuations here. We hence do not bother about using the fixed length constraint to determine the boson thermal mass, and we just discard the purely thermally fluctuating boson modes of the theory by hand.

\begin{figure}
\begin{center}
\includegraphics[width=0.4\textwidth]{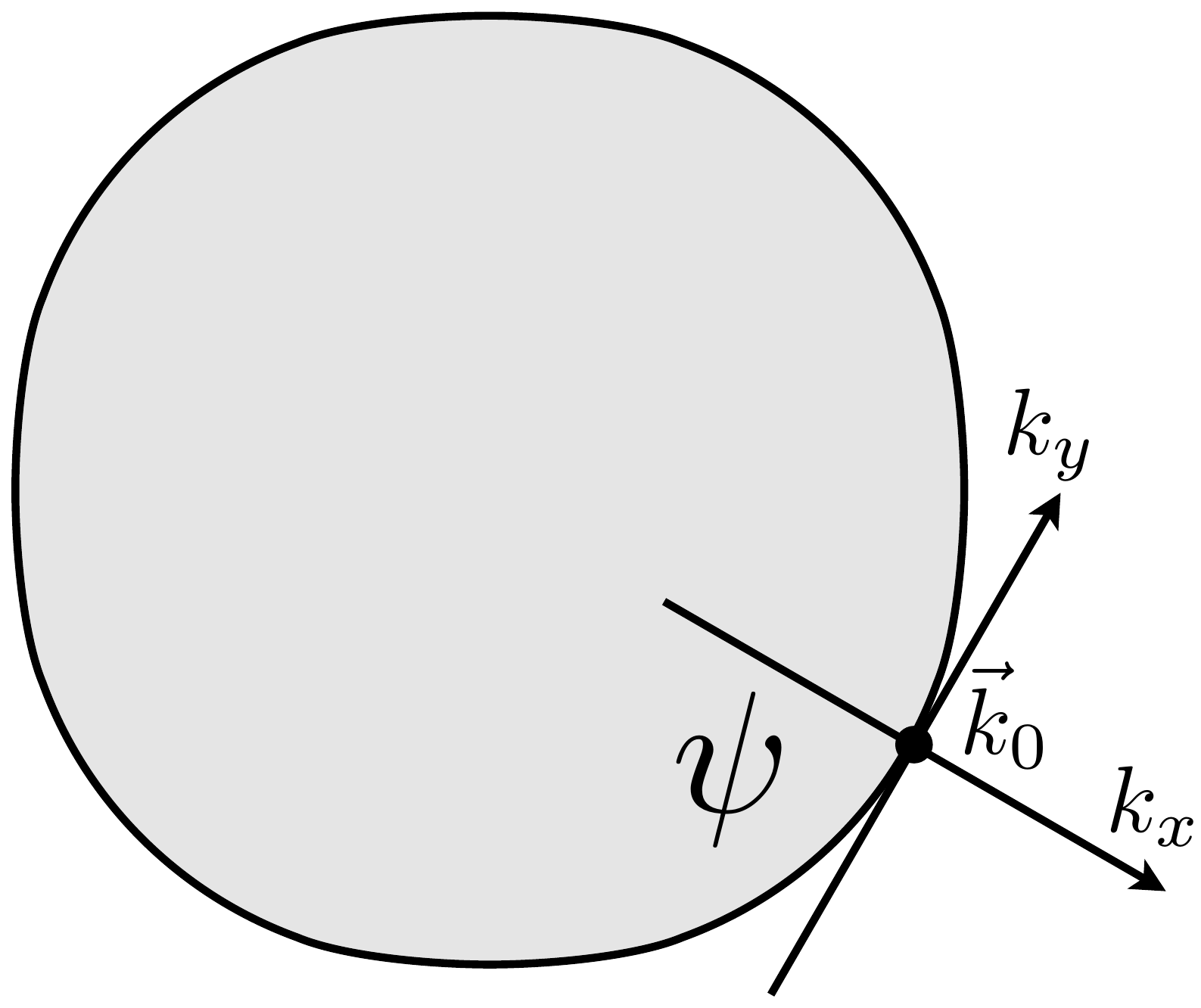}
\caption{Patch theory for fermions in the vicinity of the points $\vec{k}_0$ on the Fermi surface.}
\label{fig:patch}
\end{center}
\end{figure}

The critical singularties of the lattice $N=\infty$ theory of Section~\ref{sec:rpaffs} are described by a continuum theory which focuses on a single patch of the Fermi surface around a chosen wavevector $k_0$ on the Fermi surface \cite{metlitski1,Polchinski:1993ii} -- see Fig~\ref{fig:patch}.
We choose a point along the $x$ axis, and then the fermion dispersion near this point is $k_x + k_y^2$, where we have scaled the $k_x$ and $k_y$ axes so that the co-efficients are unity. In this manner, we obtain the action for single patch theory is ($\int_k \equiv \int dk_x dk_y/(2\pi)^2$)
\begin{align}
&S = \int d\tau\int_k\sum_{i=1}^N\psi^\dagger_{ik}(\tau)\left[\partial_\tau+k_x+k_y^2\right]\psi_{ik}(\tau) +\frac{1}{2}\int d\tau \int_q\sum_{i=1}^N \phi_{iq}(\tau)\left[q_y^2\right]\phi_{i,-q}(\tau) \nn
&+N\frac{g^2}{2}\int d\tau d\tau' \int_{k,q}G(k,\tau-\tau')G(k+q,\tau'-\tau)D(\tau-\tau',q) \nn
&-N\int d\tau d\tau'\int_k\Sigma(k,\tau'-\tau)\left[G(k,\tau-\tau')-\frac{1}{N}\sum_{i=1}^N\psi_{ik}(\tau)\psi^\dagger_{ik}(\tau')\right] \nn
&+\frac{N}{2}\int d\tau d\tau'\int_q\Pi(q,\tau'-\tau)\left[D(q,\tau-\tau')-\frac{1}{N}\sum_{i=1}^N\phi_{iq}(\tau)\phi_{i,-q}(\tau')\right].
\label{pchact}
\end{align}
However, the $\phi_{iq}$ fields are now defined to have $\phi_{iq}(i\Omega_m=0)=0$ as the thermal fluctuations are excluded.

The saddle point equations are
\begin{align}
&\Sigma(r,\tau) = g^2 D(r,\tau)G(r,\tau),~~\Pi(r,\tau) = -g^2 G(-r,-\tau)G(r,\tau), \nn
&G(k,i\omega_n) = \frac{1}{i\omega_n-k_x-k_y^2-\Sigma(k,i\omega_n)}, \nn
&D(q,i\Omega_m\neq 0) = \frac{1}{q_y^2-\Pi(q,i\Omega_m)}.
\label{empch}
\end{align}
We can solve the saddle point equations analytically at $T\neq0$. First we note that $\mathrm{sgn}(\omega_n-\mathrm{Im}[\Sigma(k,i\omega_n)])= \mathrm{sgn}(\omega_n)$. We then assume $\Sigma(k,i\omega_n)=\Sigma(i\omega_n)$ is momentum-independent. Then
\begin{align}
&\Pi(q,i\Omega_m) = -g^2T\sum_{\omega_n}\int_k\frac{1}{i\omega_n-k_x-k_y^2-\Sigma(i\omega_n)}\frac{1}{i\omega_n+i\Omega_m-k_x-q_x-(k_y+q_y)^2-\Sigma(i\omega_n+i\Omega_m)} \nn
&=-ig^2\frac{T}{2}\sum_{\omega_n}\int_{k_y}\frac{\mathrm{sgn}(\omega_n+\Omega_m)-\mathrm{sgn}(\omega_n)}{i\Omega_m-q_x-2k_yq_y-q_y^2+\Sigma(i\omega_n)-\Sigma(i\omega_n+i\Omega_m)} \nn
&=g^2\frac{T}{8|q_y|}\sum_{\omega_n}\mathrm{sgn}(\omega_n)\left(\mathrm{sgn}(\omega_n+\Omega_m)-\mathrm{sgn}(\omega_n)\right) = -\frac{g^2}{8\pi}\frac{|\Omega_m|}{|q_y|}.
\label{patchypie}
\end{align}
Further,
\begin{align}
&\Sigma(k,i\omega_n) = g^2T\sum_{\Omega_m\neq0}\int_q\frac{1}{q_y^2+\frac{g^2}{8\pi}\frac{|\Omega_m|}{|q_y|}}\frac{1}{i\omega_n+i\Omega_m-k_x-q_x-(k_y+q_y)^2-\Sigma(i\omega_n+i\Omega_m)} \nn
&=-ig^2\frac{T}{2}\sum_{\Omega_m\neq0}\int_{q_y}\frac{\mathrm{sgn}(\omega_n+\Omega_m)}{q_y^2+\frac{g^2}{8\pi}\frac{|\Omega_m|}{|q_y|}} \nn
&=-2ig^{4/3}\pi^{1/3}\frac{T}{3\sqrt{3}}\sum_{\Omega_m\neq0}\frac{\mathrm{sgn}(\omega_n+\Omega_m)}{|\Omega_m|^{1/3}} = -i\mathrm{sgn}(\omega_n)2^{5/3}g^{4/3}\frac{T^{2/3}}{3\sqrt{3}}H_{1/3}\left(\frac{|\omega_n|-\pi T}{2\pi T}\right). \nn
&\Sigma(k,i\omega_n,T=0) = -i\mathrm{sgn}(\omega_n)\frac{g^{4/3}}{\pi^{2/3}\sqrt{3}}|\omega_n|^{2/3}.
\label{ftfse}
\end{align}
The function $H_{1/3}(\mathrm{x})$ is \verb|Mathematica|'s \verb|HarmonicNumber[x,1/3]| and is related to generalized Reimann zeta functions \cite{Patel:2016wdy}. We can therefore see that the assumptions we made about $\Sigma(k,i\omega_n)$ are self-consistent. The fermion self-energy vanishes at the first Matsubara frequencies $\omega_n=\pm\pi T$.

\section{Numerical solution of the lattice model}
\label{sec:num}

In this  section we describe the numerical solution of the saddle point equations \eqref{eq:saddle_pt_eqs} for the lattice model, along with the fixed-length constraint
	\begin{equation}
	D(r=0,\tau=0) = 1/\gamma.  \label{eq:fixed_len}
	\end{equation}
We solve these equations on a square lattice with periodic boundary conditions and consider systems of linear dimensions $L=32 - 256$. 
As mentioned previously, the equations are solved efficiently by employing fast Fourier transforms in both space and imaginary time. Furthermore, we find convergence of the iterative procedure is significantly enhanced by solving the equations progressively form high to low temperature, using the solution from the previous higher temperature as a seed for a given temperature. This procedure is typically started at a relatively high temperature, $T \sim t$.

\subsection{Lattice parameters}

From the bare fermion and boson dispersion in \eqref{eq:saddle_pt_eqs} we identify the following energy scales:
	\be
	\{t, ~ \sqrt J, ~ g^2/J \}.
	\ee
Other relevant electronic energy scales are the bandwidth $W=8t$ and the density of states (DOS) at the Fermi energy $N_0 \sim 1/W$.
We define a dimensionless coupling constant as
	\be
	\lambda_0 = \frac{g^2}{JW}.
	\ee
The other important dimensionless parameter is $\sqrt J/t \sim c/v_F$, where $c$ and $v_F$ are the boson and fermion velocities, respectively. In terms of lattice parameters, $v_F \sim t a$ and $c \sim \sqrt J a$, where $a$ is the lattice constant. Here we focus on the regime $c \sim v_F$. For all the data presented below we fix the following parameters: $\lambda_0 = 0.125$, $\sqrt J /t = 2$, and $\mu=-0.5t$. For reference, this chemical potential corresponds to a Fermi energy $\epsilon_F = 3.5t$ and DOS $N_0 \approx 0.19/t$. As explained in Sec.~\ref{sec:thfluc}, the boson mass, $m_b$, is determined self-consistently for a fixed value of $\gamma$.

\subsection{Results}

\subsubsection{Phase diagram}

\begin{figure}[t]
\begin{center}
\includegraphics[width=\textwidth]{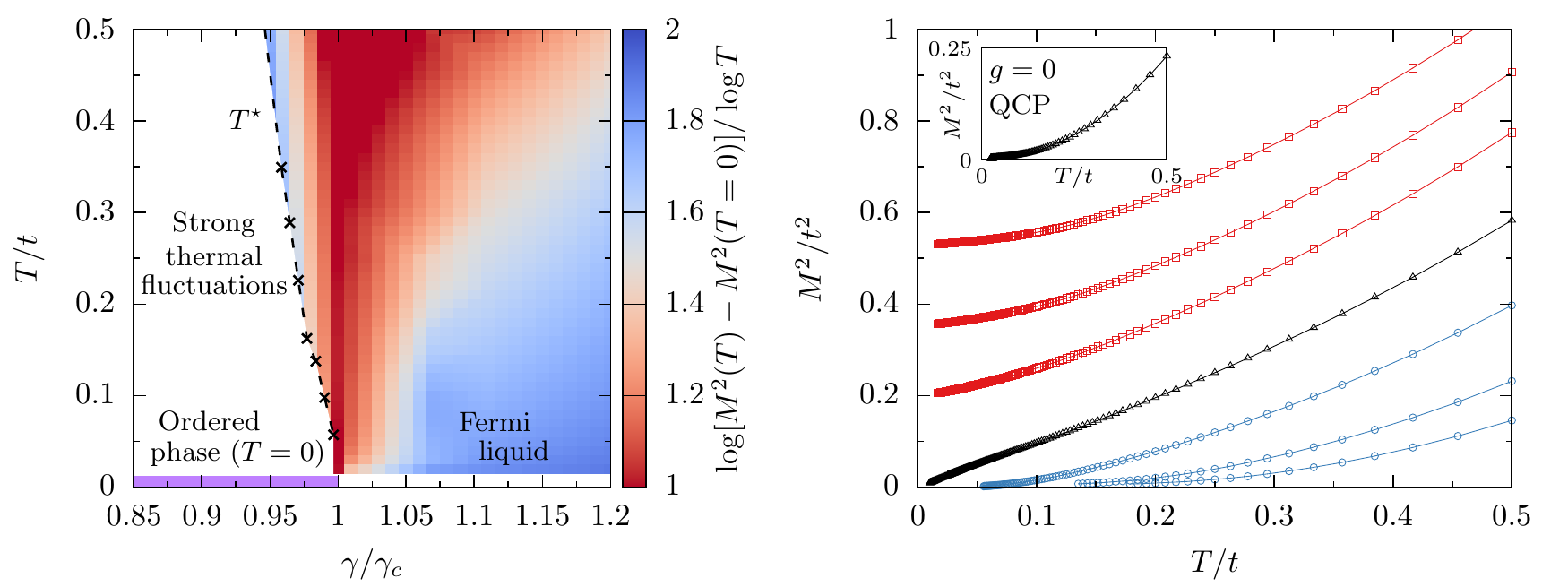}
\caption{Left: Phase diagram as a function of $\gamma$ in units of $\gamma_c \approx 0.15$ and temperature $T$. The color scale shows the exponent, $x$, with which the boson mass approaches its $T=0$ value: $M^2(T) - M^2(T=0) \sim T^x$. Up to logarithmic corrections, we expect $x=1$ at the QCP and $x=2$ in the Fermi liquid regime. In the region $T < T^\star$, $M^2(T) \sim e^{-T^\star/T}$ and the behavior of the system is governed by the soft thermal fluctuations of the boson. For $\gamma < \gamma_c$, the system is ordered only at $T=0$. Right: Boson mass as a function of $T$ for $\gamma > \gamma_c$ (red), $\gamma = \gamma_c$ (black), and $\gamma < \gamma_c$ (blue). The inset shows the behavior of the boson mass at the $g=0$ QCP, where $M^2\sim T^2$.}
\label{fig:phase_diagram+boson_mass}
\end{center}
\end{figure}

\begin{figure}[t]
\begin{center}
\includegraphics[width=\textwidth]{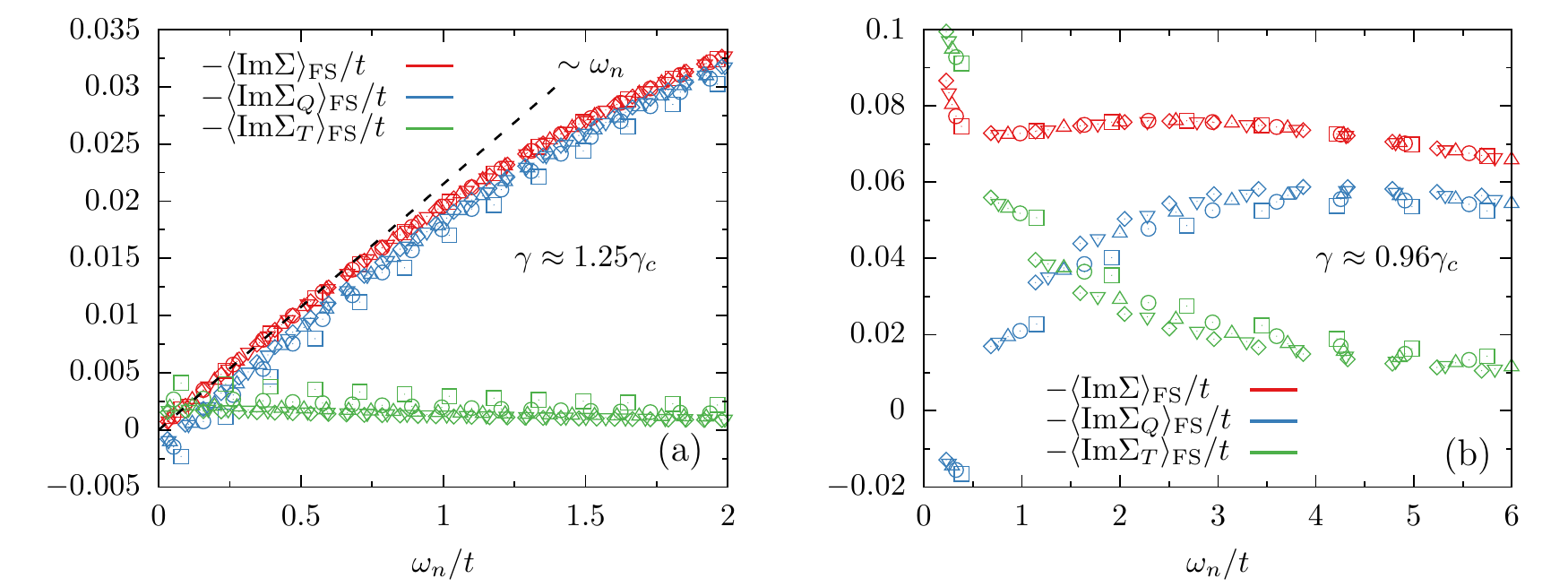}
\caption{(a) Imaginary part of the fermion self-energy averaged over the Fermi surface, as a function of Mastubara frequency, $\omega_n$, on the Fermi liquid side ($\gamma \approx 1.25 \gamma_c$) for $\beta t = 40 - 100$. Red shows the full self-energy, $-\langle\mathrm{Im}\Sigma\rangle_\mathrm{FS}$, while blue and green show the quantum, $-\langle\mathrm{Im}\Sigma_Q\rangle_\mathrm{FS}$, and thermal, $-\langle\mathrm{Im}\Sigma_T\rangle_\mathrm{FS}$, contributions, respectively. Different symbols correspond to different temperatures. For the full $\Sigma$, the points for various temperatures fall essentially on the same curve, indicating $\Sigma$ has converged to its $T=0$ value. The components $\Sigma_Q$ and $\Sigma_T$ show a more significant temperature dependence. We find $-\mathrm{Im}\Sigma \sim \omega_n$ at small frequency, consistent with a Fermi liquid. In this regime, the contribution from $\Sigma_T$ is small. \\
(b) Same as (a) but on the ordered side ($\gamma \approx 0.96 \gamma_c$) for $\beta t = 8 - 14$. In this regime $-\mathrm{Im}\Sigma$ is an increasing function of $\omega_n$ at small frequency, where it is dominated by $\Sigma_T$.}
\label{fig:sigma_away_from_qcp}
\end{center}
\end{figure}

To access the QCP, we first map out the phase diagram of the model as a function of $\gamma$ and temperature $T$. The results are shown in the left panel of Fig.~\ref{fig:phase_diagram+boson_mass} and are based primarily on the behavior of the renormalized boson mass,
    \be
	M^2(T)  = m_b^2 - \Pi(0,0),
	\ee
shown in Fig.~\ref{fig:phase_diagram+boson_mass} (right). We find a QCP at the value $\gamma_c \approx 6.7t$, where we observe the boson mass vanishes approximately linearly, $M^2 \sim T$. Up to logarithmic corrections, which are hard to detect numerically, this is consistent with earlier analytic calculations \cite{millis, Hartnoll:2014gba}. This scaling holds in the quantum-critical fan above the QCP, as is seen from the color scale in Fig.~\ref{fig:phase_diagram+boson_mass}. For $\gamma > \gamma_c$, the system is a Fermi liquid at low temperatures, with the boson mass behaving as $M^2 - M^2(T=0) \sim T^2$. For $\gamma < \gamma_c$, the system is ordered only for $T=0$, with the boson mass vanishing according to $M^2 \sim \exp(-T^\star/T)$ (the $T^\star$ line in the figure is obtained by fitting the $M^2(T)$ to this functional form). The absence of a finite temperature ordering transition is a result of the Hohenberg-Mermin-Wagner theorem \cite{Hohenberg66,Mermin1966}, as such a transition corresponds to spontaneously breaking the $O(N)$ symmetry of the original (disorder averaged) model. The exponentially vanishing mass as $T \to 0$ is known from the behavior of the $O(N)$ model at large-$N$ in two dimensions. We remark that the boson mass behaves in the same way even in the absence of the fixed length constraint, \eqref{eq:fixed_len}. Below the temperature scale $T^\star$, the system crosses over into a regime governed by soft thermal fluctuations of the boson, and, as will be further discussed below, the fermion self-energy has a form distinct from that of a Fermi liquid. For comparison, we also show the behavior of the boson mass at the $g=0$ QCP of the decoupled model in the inset of the right panel of Fig.~\ref{fig:phase_diagram+boson_mass}, in which case $M^2 \sim T^2$ (even for $g=0$, the bosonic sector is self-interacting due to the fixed length constraint, \eqref{eq:fixed_len}).

To better characterize the single-fermion properties of the system across the phase diagram, we decompose the fermion self-energy as
    \be
    \Sigma(k,i\omega_n) = \Sigma_T(k,i\omega_n) + \Sigma_Q(k,i\omega_n),
    \ee
where subscripts $T$ and $Q$ denote the ``thermal'' and ``quantum'' contributions, respectively. The thermal contribution is defined as that from the zero Matsubara frequency transfer term in the self-consistent equation for $\Sigma$, while the quantum contribution comes from non-zero Matsubara frequency transfer: 
    \begin{align}
        \Sigma_T(k,i\omega_n) &= \frac{g^2 T}{L^2} \sum_{k'} D(k-k',\Omega =0) G(k',i\omega_n), \\
        \Sigma_Q(k,i\omega_n) &= \frac{g^2 T}{L^2} \sum_{n'\neq n}\sum_{k'} D(k-k',i\omega_n - i\omega_{n'}) G(k',i\omega_{n'}).
    \end{align}
This decomposition has been used to analyze finite-$T$ corrections to quantum-critical scaling \cite{Metzner2006,Berg2,Meng2020,Torroba1,Torroba2} and here we find it particularly useful in analyzing the behavior both at the QCP, $\gamma = \gamma_c$, and on the ordered side, $\gamma < \gamma_c$.

We first discuss the behavior of the fermion self-energy away from the QCP. In Fig.~\ref{fig:sigma_away_from_qcp}, we show the negative imaginary part of the fermion self-energy, $\langle-\mathrm{Im}\Sigma(\omega_n) \rangle_\mathrm{FS}$, where the brackets denote averaging momentum dependence over the Fermi surface (in general we find $\Sigma$ is weak function of the direction of $k$ along the Fermi surface; see Fig.~\ref{fig:sigma_on_FS_qcp}a. Fig.~\ref{fig:sigma_away_from_qcp}a shows the self-energy for $\gamma > \gamma_c$, where we find a linear frequency dependence, consistent with a Fermi-liquid, corresponding to fermion mass enhancement. Data are shown for a set of temperatures in the range $\beta t = 40 - 100$. The data points for different temperatures fall on the same curve, indicating $\Sigma$ has essentially converged to its $T=0$ value. In the Fermi liquid regime, the contribution from $\Sigma_T$ is small. Fig.~\ref{fig:sigma_away_from_qcp}b shows the self-energy on the ordered side, $\gamma < \gamma_c$. In this case, the low-frequency behavior is significantly affected by $\Sigma_T$. On the ordered side, the exponentially small boson mass makes it challenging to numerically access low temperatures, and the data shown in the figure are for $\beta t = 8 -14$. Even here the full self-energy is essentially temperature-independent, while the separate contributions, $\Sigma_Q$ and $\Sigma_T$, show a stronger temperature dependence. In contrast to the Fermi-liquid regime, on the ordered side and at low frequency, the self-energy is essentially constant over the frequency range $\omega_n\approx (1 - 3)t$ and at lower frequency becomes an increasing function of decreasing frequency. The small boson mass, satisfying $M \ll T$, and large thermal self-energy, $\Sigma_T$, explain why we denote region $T<T^*$ in Fig.~\ref{fig:phase_diagram+boson_mass} as being characterized by ``strong thermal fluctuations''. We note that the data on the ordered side of the transition are only very slightly tuned away from the critical point, $\gamma \approx 0.96 \gamma_c$, yet the behavior of the self-energy is nevertheless drastically different from that at the QCP (to be further discussed in the next section), indicating a rapid crossover in the behavior of the system on the ordered side. Finally, we remark that in both regimes, $\langle-\mathrm{Im}\Sigma_Q(\pi T)\rangle_\mathrm{FS}$ is negative; this curious fact has been explained in \cite{Berg2}.

\subsubsection{Behavior at the QCP and comparison with the patch theory}

\begin{figure}[t]
\begin{center}
\includegraphics[width=0.5\textwidth]{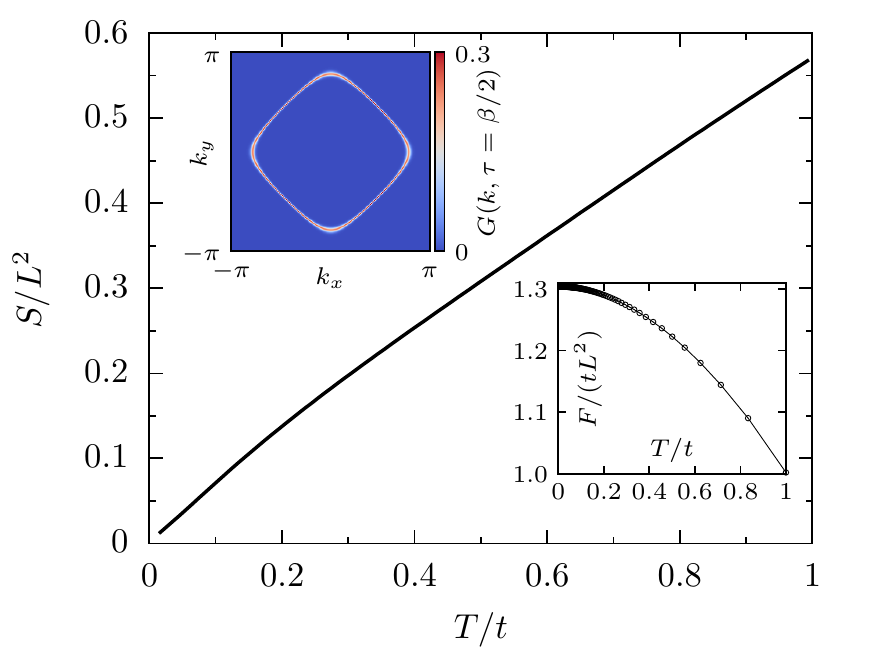}
\caption{Entropy density, $S/L^2$, as a function of $T$ at the QCP. This is expected to vanish as $\sim T^{2/3}$ as $T \rightarrow 0$, but our data is not precise enough to distinguish from a linear $T$ dependence. The entropy is obtained by numerical differentiation of the free energy density, $F/tL^2$, shown in the bottom inset. The top inset is $G(k, \tau=\beta/2)$, which is essentially the fermion spectral function averaged over an energy window of order $T$ about the Fermi energy, for $\beta t = 100$. We see the Fermi surface remains sharply defined at the QCP, in accord with Luttinger's theorem.}
\label{fig:entropy_at_qcp}
\end{center}
\end{figure}

\begin{figure}[t]
\begin{center}
\includegraphics[width=\textwidth]{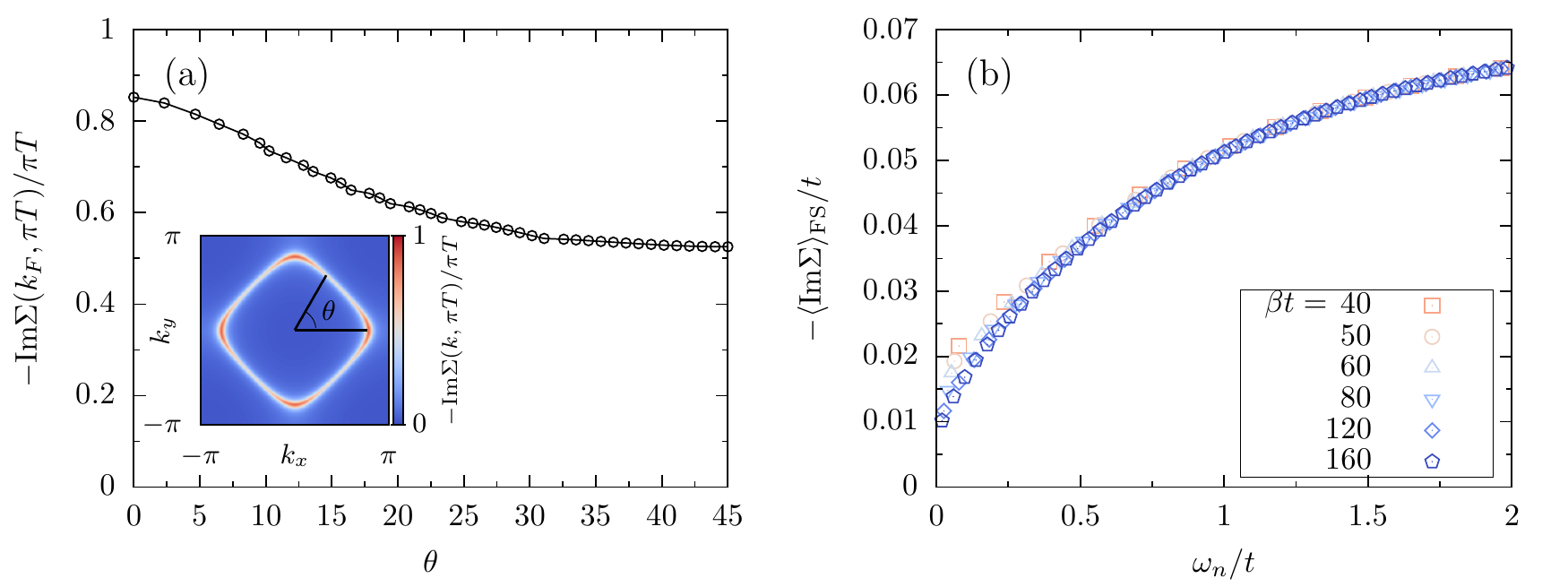}
\caption{(a) Imaginary part of fermion self-energy at the QCP, normalized by the first Matsubara frequency $\omega_0 = \pi T$, as a function of angle, $\theta$, along the Fermi surface for $\beta t = 100$. The variation as a function $\theta$ is small, essentially tracking the density of states of the non-interacting band structure. Inset shows the same quantity through the whole first Brillouin zone, where we see it is sharply peaked at the Fermi surface. (b) Imaginary part of fermion self-energy at the QCP, averaged over the Fermi surface, for the range $\beta t = 40 -160$. For all but the lowest frequencies, $\Sigma$ has essentially converged to its $T=0$ behavior by $\beta t \approx 40$.}
\label{fig:sigma_on_FS_qcp}
\end{center}
\end{figure}

\begin{figure}[t]
\begin{center}
\includegraphics[width=\textwidth]{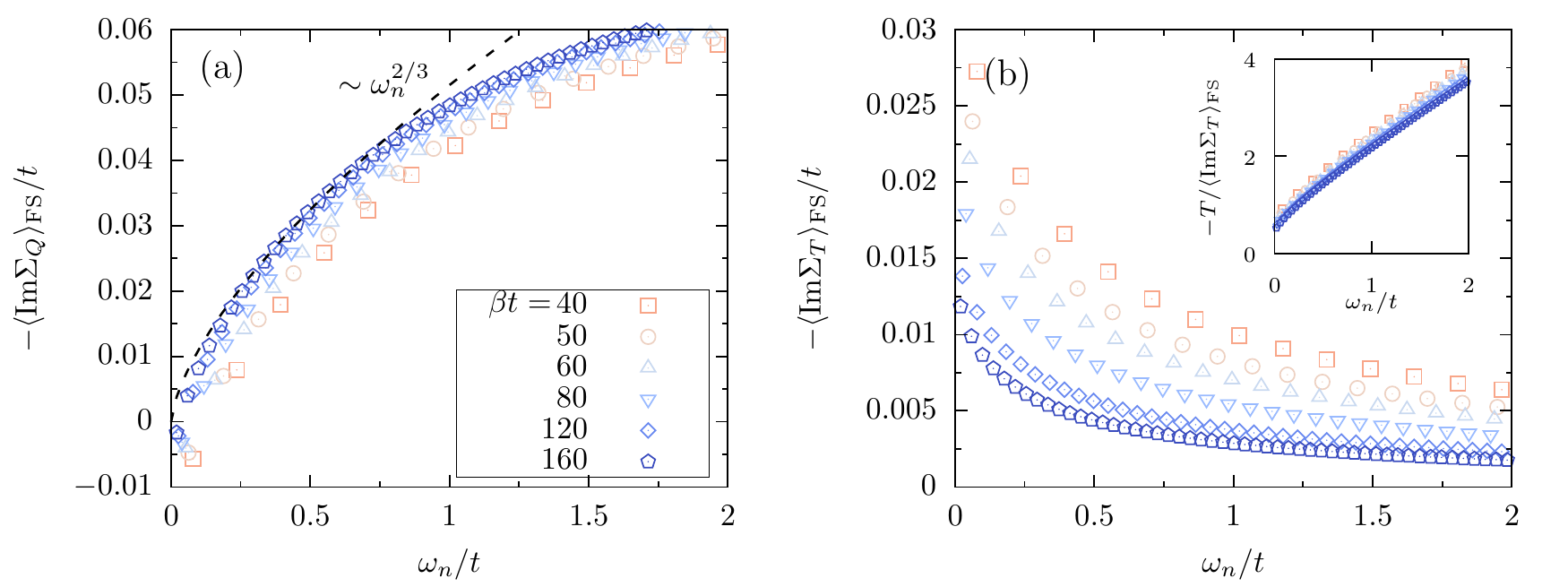}
\caption{Decomposition of imaginary part of fermion self-energy at the QCP, averaged over the Fermi surface, for the range of inverse temperatures $\beta t = 40 -160$. (a) The quantum part, $\mathrm{Im}\Sigma_Q$, shows more significant $T$-dependence than the full $\mathrm{Im}\Sigma$, tending toward the predicted $\omega_n^{2/3}$ behavior as $T\to 0$. (b) Thermal part, $\mathrm{Im}\Sigma_T$, displays a significant temperature dependence, tending slowly to zero as $T \to 0$. The inset shows the thermal part obeys the scaling behavior $-\mathrm{Im}
\Sigma_T \sim T/\omega_n$ \cite{Berg2,Meng2020}.}
\label{fig:sigma_decomp_at_qcp}
\end{center}
\end{figure}

\begin{figure}[t]
\begin{center}
\includegraphics[width=0.5\textwidth]{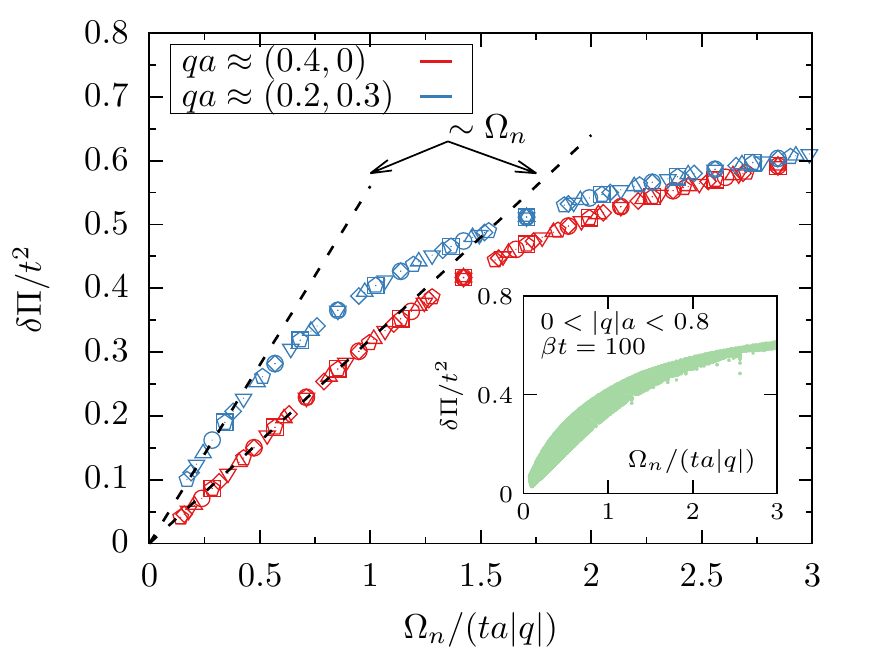}
\caption{Dynamical part of boson self-energy, $\delta \Pi$, as a function of  $\Omega_n/ta|q|$, where $ta \sim v_F$. Data are shown for representative $q$ vectors in the inverse temperature range $\beta t = 50-100$, where $\delta \Pi$ has essentially converged to its $T=0$ limit. The small frequency behavior is linear, as predicted from the low-energy analysis in Sec.~\ref{sec:rpapch}, albeit with a slope that depends on the direction of $q$, as would be expected for an anisotropic Fermi surface. Inset shows $\delta \Pi$ for $q$ vectors in the range $0<|q|a<0.8$ for $\beta t = 100$.}
\label{fig:pi_at_qcp}
\end{center}
\end{figure}

We now describe the behavior at the QCP, $\gamma = \gamma_c$. Fig.~\ref{fig:entropy_at_qcp} shows the entropy density, $S/L^2$, at the QCP, which we find vanishes as $T \to 0$. The entropy is computed by numerical differentiation of the free energy, $F$, which is shown in the bottom inset of Fig.~\ref{fig:entropy_at_qcp} and computed according to (\ref{frenerg}). We also find the Fermi surface remains sharp at the QCP, in accord with Luttinger's theorem. This can be seen in the top inset of Fig.~\ref{fig:entropy_at_qcp}, where we show an imaginary-time proxy for the fermion spectral weight at zero energy:
\be
G(k, \tau=\beta/2) = \int_{-\infty}^\infty d\omega \frac{1}{2\cosh(\beta \omega/2)}A(k, \omega),
\ee
where $A(k,\omega)$ is the fermion spectral function. This quantity is essentially the spectral function averaged over an energy range of order $T$ about the Fermi energy and has been frequently used in numerical studies, as it avoids the need for analytic continuation of Matsubara frequency data \cite{schattner16,gerlach17,Berg4}.

The behavior of the fermion self-energy at the QCP is shown in Figs.~\ref{fig:sigma_on_FS_qcp} and \ref{fig:sigma_decomp_at_qcp}. Fig.~\ref{fig:sigma_on_FS_qcp}a shows the variation of $-\mathrm{Im}\Sigma$ along the Fermi surface. The dependence on angle along the Fermi surface is weak, and essentially tracks the behavior of the non-interacting DOS. The inset of the figure shows $-\mathrm{Im}\Sigma$ across the entire Brillouin zone and we find it is peaked on the Fermi surface. These observations are in line with the analytic predictions of Sec.~\ref{sec:rpapch}. Fig.~\ref{fig:sigma_on_FS_qcp}b shows the Fermi-surface average of $-\mathrm{Im}\Sigma$. The data are shown for a set of temperatures in the range $\beta t = 40-160$. Fig.~\ref{fig:sigma_decomp_at_qcp} shows the decomposition into $\Sigma_Q$ and $\Sigma_T$. We find the temperature dependence of the full $\Sigma$ is weak at all but the lowest frequencies, where the thermal contribution, shown in Fig.~\ref{fig:sigma_decomp_at_qcp}b, is still sufficiently large to obscure the expected $\omega_n^{2/3}$ scaling. The quantity $\mathrm{Im}\Sigma_Q$ precisely removes this thermal contribution and, from Fig.~\ref{fig:sigma_decomp_at_qcp}a, we see that, although $\mathrm{Im}\Sigma_Q$ shows a stronger temperature dependence than $\mathrm{Im}\Sigma$, the low-frequency behavior is indeed compatible with $\omega_n^{2/3}$ scaling as $T \to 0$, as in Eq.~\eqref{ftfse}. In the temperature regime shown in Fig.~\ref{fig:sigma_decomp_at_qcp}, earlier work has predicted that the the thermal self-energy should behave as $-\mathrm{Im}\Sigma_T \sim T/\omega_n$, \cite{Berg2,Meng2020}. We find this scaling is indeed well satisfied, as seen in the inset of Fig.~\ref{fig:sigma_decomp_at_qcp}b.

Finally, Fig.~\ref{fig:pi_at_qcp} shows the ``dynamical part'' of the boson self-energy, defined as $\delta \Pi(q,\Omega_n) = \Pi(q,\Omega_n) - \Pi(q,\Omega_n = 0)$, for two $q$ vectors and range of temperatures $\beta t = 50-100$. We find the expected linear scaling with frequency, $\delta\Pi \sim \Omega_n$, when $\Omega_n < v_F|q|$ (recall $v_F \sim ta$). The scaling with $\Omega_n/v_F|q|$, as in Eq.~\eqref{patchypie}, is not perfectly satisfied due to the anisotropy of the Fermi surface (in the low-energy calculations, a circular Fermi surface is assumed). The dependence of $\delta\Pi$ on the angle of $q$ may be seen in the inset of Fig.~\ref{fig:pi_at_qcp}, where $\delta \Pi$ is shown for a larger set of $q$'s, in the range $0<|q|a<0.8$.

\section{Single patch theory and time reparameterizations}
\label{sec:fluc}

We now turn to a characterization of the fluctuations about the large $N$ saddle point in the spatially uniform model with a critical Fermi surface described in Sections~\ref{sec:rpaffs} and~\ref{sec:rpapch}. Here we will focus on the single patch critical theory in (\ref{pchact}), and will defer consideration of the special role of antipodal patches \cite{metlitski1} to Section~\ref{sec:bilinears}.

In the SYK model, the flucutations are characterized by the structure of the 4-point correlators of the large $N$ saddle point \cite{JMDS16,Kitaev:2017awl}, and this is dominated by the time reparameterization mode. Formally, it appears that the single patch saddle-point of the large $N$ theory in Section~\ref{sec:rpapch} has a time reparameterization symmetry, and so we examine it here for a corresponding soft mode. However, as we show below, the spatial structure of the critical Fermi surface theory does play an important role, and we find there is no special contribution from time reparameterizations. Instead, we find that the 4-point correlators are controlled by response functions of the conserved fermion density, which were explored earlier in Ref.~\cite{Kim94}.

Integrating out the fermions from the single patch continuum action (\ref{pchact}),  leads us to  consider the following $G-\Sigma$ action
\begin{align}\label{GSigmaS}
  \frac{S}{N}= & -\Tr\ln(\partial_\tau\delta^{(3)}-s_f(i\partial_x+\partial_y^2)\delta^{(3)}+\Sigma)+\frac{1}{2}\Tr\ln(-s_b\partial_y^2\delta^{(3)}-\Pi) \nonumber \\
  &~~~~+\frac{g^2}{2}\Tr(G\cdot G D)-\Tr(\Sigma \cdot G)+\frac{1}{2}\Tr(\Pi\cdot D).
\end{align}
The $\Tr(\cdot)$  is defined on the indices of spacetime $(x,y,\tau)$, similar to \cite{Gu:2019jub}:
\begin{equation}\label{}
  \Tr (A\cdot B)\equiv\int \rd^3 x_1 \rd ^3 x_2 A(x_1,x_2)B(x_2,x_1). \nonumber
\end{equation}
 We have also inserted two parameters $s_f,s_b$ in front of momenta for convenience of the analysis below.

The saddle point equations are (we focus at zero temperature and ignore the zero-mode subtraction for bosons)
\begin{eqnarray}
  \Sigma(r,\tau) &=& g^2 G(r,\tau) D(r,\tau) \\
  \Pi(r,\tau) &=& -g^2 G(-r,-\tau) G(r,\tau) \\
  G(k,i\omega_n) &=& \frac{1}{i\omega_n-s_f (k_x+k_y^2)-\Sigma(k,i\omega_n)} \\
  D(q,i\Omega_n) &=& \frac{1}{q_y^2-\Pi(q,i\Omega_n)}
\end{eqnarray}
We recall the solutions of these equations at zero temperature obtained in Section~\ref{sec:rpapch}:
\begin{eqnarray}
  \Sigma(k,i\omega) &=& -i\sgn(\omega)2^{5/3}g^{4/3}\frac{T^{2/3}}{3\sqrt{3}}H_{1/3}\left(\frac{|\omega|-\pi T}{2\pi T}\right) \quad (T\to 0)\,, \\
  \Pi(q,i\Omega) &=& -\frac{g^{2}}{8\pi}\frac{|\Omega|}{|q_y|}\,.
\end{eqnarray}

We describe the structure of fluctuation around the saddle point. We introduce a collective notation for Green's functions $\calG=(D,G)$ and self energies $\Xi=(\Pi,\Sigma)$, and let $\Lambda=\mathrm{diag}(-1/2,1)$ acting on the two component space of $(D,G)$ or $(\Pi,\Sigma)$. Following derivations in \cite{Gu:2019jub, Guo:2020aog}, we can expand the $G-\Sigma$ action around the saddle point to quadratic order as
\begin{equation}\label{eq:deltaS1}
  \delta S=\frac{1}{2}\begin{pmatrix}
                        \delta \Xi^T & \delta \calG^T
                      \end{pmatrix}\Lambda
                      \begin{pmatrix}
                        W_\Sigma & -1 \\
                        -1 & W_G
                      \end{pmatrix}
                      \begin{pmatrix}
                        \delta \Xi \\
                        \delta \calG
                      \end{pmatrix}\,,
\end{equation} where $T$ means transpose on spacetime indices. Here $W_\Sigma$ and $W_G$ are defined as
\begin{equation}\label{}
  W_\Sigma=\frac{\delta \calG_*[\Xi]}{\delta \Xi},\qquad W_G=\frac{\delta \Xi_*[\calG]}{\delta\calG}\,,
\end{equation} where $\calG_*[\Xi]$ is the saddle point expression of $\calG$ viewed as a functional of $\Xi$, and similarly for $\Xi_*[\calG]$.

We can further integrate out $\delta\Xi$ in \eqref{eq:deltaS1}, to obtain
\begin{equation}\label{eq:S=dGdG}
  \delta S=\frac{1}{2}\delta \calG^T \Lambda W_\Sigma^{-1}(\underbrace{W_\Sigma W_G}_{K_G}-1)\delta \calG\,,
\end{equation} where we have defined the kernel $K_G=W_\Sigma W_G$. Therefore, soft fluctuations are related to unit eigenvalue of $K_G$.

\subsection{Time Reparameterization}

We now note the time reparameterization symmetry of (\ref{GSigmaS}).
Consider the following reparameterization
\beq
\tau=f(\sigma), \quad \rd x=(f'(\sigma))^{1/z}\rd \tilde{x}, \quad \rd y=(f'(\sigma))^{1/(2z)}\rd
\tilde{y}.
\eeq
Then ignoring the irrelevant $\partial_\tau$ term, the action is invariant under the change of variables
\begin{eqnarray}
  && G(x,x',y,y',\tau,\tau') = \frac{1}{(f'(\sigma)f'(\sigma'))^a} \label{eq:rep_G} \tilde{G}(\tilde{x},\tilde{x}',\tilde{y},\tilde{y}',\sigma,\sigma')\\
  && \Sigma(x,x',y,y',\tau,\tau') = \frac{1}{(f'(\sigma)f'(\sigma'))^{1+3/(2z)-a}} \tilde{\Sigma}(\tilde{x},\tilde{x}',\tilde{y},\tilde{y}',\sigma,\sigma')\\
  && D(x,x',y,y',\tau,\tau') = \frac{1}{(f'(\sigma)f'(\sigma'))^{1+3/(2z)-2a}} \tilde{D}(\tilde{x},\tilde{x}',\tilde{y},\tilde{y}',\sigma,\sigma')\\
  && \Pi(x,x',y,y',\tau,\tau') = \frac{1}{(f'(\sigma)f'(\sigma'))^{2a}} \tilde{\Pi}(\tilde{x},\tilde{x}',\tilde{y},\tilde{y}',\sigma,\sigma')\\
  && s_f(i\partial_x+\partial_y^2)\delta(x,x')\delta(y,y')\delta(\tau,\tau') = \nonumber \\
  &&~~~~~~~~~~\tilde{s}_f \frac{1}{(f'(\sigma)f'(\sigma'))^{1/2+5/(4z)}}(i\partial_{\tilde{x}}+\partial_{\tilde{y}}^2)\delta(\tilde{x},\tilde{x}')\delta(\tilde{y},\tilde{y}')\delta(\sigma,\sigma') \\
  && s_b \partial_y^2\delta(x,x')\delta(y,y')\delta(\tau,\tau')= \tilde{s}_b  \label{eq:rep_q} \frac{1}{(f'(\sigma)f'(\sigma'))^{1/2+5/(4z)}}\partial_{\tilde{y}}^2\delta(\tilde{x},\tilde{x}')\delta(\tilde{y},\tilde{y}')\delta(\sigma,\sigma')
\end{eqnarray}
The consistency with saddle point equation yields $a=2/3,z=3/2$, and $s_b,s_f$ are marginal couplings.

Let us consider the consequence of reparameterization symmetry on the kernel $K_G$. In the low-energy conformal limit (ignoring $\partial_\tau$ term), the saddle point equations can be written as
\begin{equation}
\begin{split}
  G\cdot(s_f(i\partial_x+\partial_y^2)\delta^{(3)}-\Sigma)&=1,\\
  D\cdot(-s_b\partial_y^2\delta^{(3)}-\Pi)&=1.
\end{split}
\end{equation}

Due to the reparameterization symmetry, it's valid to consider an infinitesimal reparameterization  $\delta_\epsilon:f(\tau)=\tau+\epsilon(\tau)$ on both sides. Because the Schwinger-Dyson equation holds, we can rewrite $\delta_\epsilon\Sigma,\delta_\epsilon \Pi$ in terms of $\delta_\epsilon G,\delta_\epsilon D$, and therefore we obtain
\begin{equation}\label{eq:(1-K)G}
 (1-K_G)\delta_\epsilon \calG=W_\Sigma \delta_\epsilon k.
\end{equation}

This is a two component equation for $(\delta_\epsilon D,\delta_\epsilon G)$. On the RHS, $\delta_\epsilon k$ is the reparameterization of momentum term
\beq
\delta_\epsilon k=(s_f\delta_\epsilon (i\partial_x+\partial_y^2)\delta^{(3)},-s_b\delta_\epsilon \partial_y^2\delta^{(3)}).
\eeq
In the original SYK model, the RHS is absent and the reparameterization mode is an eigenvector of $K_G$ with eigenvalue one \cite{JMDS16}. This eigenvalue one is responsible for the $\beta J$ enhancement in four-point functions.  In the current model, the presence of $\delta_\epsilon k$ term will destroy the dominance of the unit eigenvalue mode in the action for fluctuations, and the reparameterization fluctuation will not have $\beta J$ enhancements. Therefore, the low-energy theory will contain not only the reparameterization but also other fluctuations.

We can also repeat the above discussion for the $U(1)$ gauge symmetry
\begin{equation}\label{}
\begin{split}
  \delta_{\lambda}G(x,x',y,y',\tau,\tau')&=i(\lambda(\tau)-\lambda(\tau'))G(x,x',y,y',\tau,\tau')\,, \\
  \delta_{\lambda}\Sigma(x,x',y,y',\tau,\tau')&=i(\lambda(\tau)-\lambda(\tau'))\Sigma(x,x',y,y',\tau,\tau')\,.
\end{split}
\end{equation}This symmetry is emergent at low-energy given that the $\partial_\tau$ term in the action is irrelevant. Running the above argument for this $U(1)$ symmetry, we obtain an eigenvector of $K_G$ with unit eigenvalue:
\begin{equation}\label{}
  (1-K_G)\delta_{\lambda} \calG=0\,,
\end{equation} and there is no momentum term on the RHS because the symmetry is uniform in space.
In what follows, we will demonstrate that $\delta_\lambda\calG$ is the only unit eigenvector of $K_G$ that obeys sliding symmetry.

\subsection{Sliding symmetry}
\label{sec:sliding}

An important symmetry of the patched Fermi surface problem is the sliding symmetry
\begin{equation}\label{}
\begin{split}
   \phi(x,y) &\to \phi(x,y+\theta x), \\
   \psi(x,y)  & \to e^{-i[(\theta/2)y+(\theta^2/4)x]}\psi(x,y+\theta x)\,.
\end{split}
\end{equation}
In Fourier space, the representation is simpler
\begin{equation}\label{}
  \begin{split}
     \phi(q_x,q_y) & \to \phi(q_x-\theta q_y,q_y),  \\
     \psi(k_x,k_y)  & \to \psi(k_x-\theta k_y-\frac{\theta^2}{4},k_y+\frac{\theta}{2}).
  \end{split}
\end{equation}

We can see there are two different representations of the sliding group. One is the boson class $[0]$, and another is the fermion class $[1]$. Given two momenta $k_1,k_2$ that transforms under representation $[n_1]$ and $[n_2]$ respectively, we can fuse them into a new representation simply by momentum addition
$$
k_3=\alpha k_1+\beta k_2,
$$which transforms under representation $[n_3]$ and $n_3=\alpha n_1+\beta n_2$. Here $\alpha,\beta$ are rational numbers such that $n_3$ is an integer. In our problem, we will only encounter class $[0]$ and $[1]$.

Each representation is associated with some invariants. For example $\vec{q}$ of $[0]$ has invariant $q_y$ and $\vec{k}$ of $[1]$ has invariant $k_x+k_y^2$.

The eigenfunctions of the kernel $K_G$ are bosonic or fermionic two point functions $B(k,p)$ and $F(k,p)$ (see \eqref{eq:Vgraphic}), where $k$ is the relative momentum and $p$ is the CoM momentum. For $B(k,p)$, both $(k,p)$ are class $[0]$. For $F(k,p)$, $k$ is class $[1]$ and $p$ is class $[0]$. Therefore $B$ is in class $[0]\otimes[0]$ and $F$ is in class $[0]\otimes[1]$. They are in tensor product representations.

By trial and error, we find the following invariants of the above tensor product representations up to quadratic order:
\begin{equation}\label{eq:invarinats}
\begin{split}
   &[0]\otimes [0]: k_y,~p_y,~p_x k_y-p_yk_x \\
     & [0]\otimes [1]: p_y,~k_x+k_y^2,~p_x+2p_yk_y
\end{split}
\end{equation}

\subsection{Feynman Diagrams}
In this part we give a diagrammatic prescription to compute $K_G$.
We have the following diagrammatic representations for $\delta_\epsilon G$ and $\delta_\epsilon D$
\begin{equation}\label{}
  \begin{pmatrix}
    \delta_\epsilon D(3;4) \\
    \delta_\epsilon G(3;4)
  \end{pmatrix}=
  \begin{pmatrix}
    \begin{tikzpicture}[baseline={([yshift=-4pt]current bounding box.center)}]
      \draw[thick] (-15pt,-12pt) rectangle (15pt,12pt);
      \draw[thick] (-20pt,12pt)--(-15pt,12pt);
      \draw[thick] (-20pt,-12pt)--(-15pt,-12pt);
      \node at (-25pt, -12pt) {\scriptsize $4$};
      \node at (-25pt,12pt) {\scriptsize $3$};
      \node at (0pt,0pt) {$\delta_\epsilon D$};
    \end{tikzpicture} \\
    \begin{tikzpicture}[baseline={([yshift=-4pt]current bounding box.center)}]
      \draw[thick] (-15pt,-12pt) rectangle (15pt,12pt);
      \draw[thick] (-20pt,12pt)--(-15pt,12pt);
      \draw[thick] (-20pt,-12pt)--(-15pt,-12pt);
      \node at (-25pt, -12pt) {\scriptsize $4$};
      \node at (-25pt,12pt) {\scriptsize $3$};
      \node at (0pt,0pt) {$\delta_\epsilon G$};
    \end{tikzpicture}
  \end{pmatrix},
\end{equation}
where numbers are short-hands for spacetime coordinates.

Using the saddle point equations, we can also write down the Feynman diagrams for $W_\Sigma$ and $W_G$:
\begin{equation}\label{eq:Wsigmagraph}
  W_\Sigma(1,2;3,4)=\begin{pmatrix}
                      \begin{tikzpicture}[baseline={([yshift=-4pt]current bounding box.center)}]
                     \draw[thick, dashed, mid arrow] (40pt,12pt)--(0pt,12pt);
                     \draw[thick, dashed, mid arrow] (0pt,-12pt)--(40pt,-12pt);
                     \node at (-5pt,12pt) {\scriptsize $1$};
                     \node at (-5pt,-12pt) {\scriptsize $2$};
                     \node at (48pt,12pt) {\scriptsize $3$};
                     \node at (48pt,-12pt) {\scriptsize $4$};
                     \end{tikzpicture}
                        & 0 \\
                      0 &
                      \begin{tikzpicture}[baseline={([yshift=-4pt]current bounding box.center)}]
                     \draw[thick, mid arrow] (40pt,12pt)--(0pt,12pt);
                     \draw[thick, mid arrow] (0pt,-12pt)--(40pt,-12pt);
                     \node at (-5pt,12pt) {\scriptsize $1$};
                     \node at (-5pt,-12pt) {\scriptsize $2$};
                     \node at (48pt,12pt) {\scriptsize $3$};
                     \node at (48pt,-12pt) {\scriptsize $4$};
                     \end{tikzpicture}
                    \end{pmatrix},
\end{equation}
\begin{equation}\label{eq:WGgraph}
  W_G(1,2;3,4)=\begin{pmatrix}
                 0 & -g^2\left(\begin{tikzpicture}[baseline={([yshift=-4pt]current bounding box.center)}]
                                  \draw[thick, dashed] (20pt,12pt)--(0pt,12pt);
                                  \draw[thick, dashed] (0pt,-12pt)--(20pt,-12pt);
                                  \draw[thick, mid arrow] (20pt,12pt)--(20pt,-12pt);
                                  \node at (-5pt,12pt) {\scriptsize $1$};
                                  \node at (-5pt,-12pt) {\scriptsize $2$};
                                  \node at (24pt,12pt) {\scriptsize $3$};
                                  \node at (24pt,-12pt) {\scriptsize $4$};
                               \end{tikzpicture}
                               +
                               \begin{tikzpicture}[baseline={([yshift=-4pt]current bounding box.center)}]
                                  \draw[thick, dashed] (20pt,-12pt)--(0pt,12pt);
                                  \draw[thick, dashed] (0pt,-12pt)--(20pt,12pt);
                                  \draw[thick, mid arrow] (20pt,12pt)--(20pt,-12pt);
                                  \node at (-5pt,12pt) {\scriptsize $1$};
                                  \node at (-5pt,-12pt) {\scriptsize $2$};
                                  \node at (24pt,12pt) {\scriptsize $3$};
                                  \node at (24pt,-12pt) {\scriptsize $4$};
                               \end{tikzpicture}
                               \right) \\
                 \frac{g^2}{2}\left(\begin{tikzpicture}[baseline={([yshift=-4pt]current bounding box.center)}]
                                  \draw[thick, dashed] (20pt,12pt)--(0pt,12pt);
                                  \draw[thick, dashed] (0pt,-12pt)--(20pt,-12pt);
                                  \draw[thick, mid arrow] (0pt,-12pt)--(0pt,12pt);
                                  \node at (-5pt,12pt) {\scriptsize $1$};
                                  \node at (-5pt,-12pt) {\scriptsize $2$};
                                  \node at (24pt,12pt) {\scriptsize $3$};
                                  \node at (24pt,-12pt) {\scriptsize $4$};
                               \end{tikzpicture}+
                                \begin{tikzpicture}[baseline={([yshift=-4pt]current bounding box.center)}]
                                  \draw[thick, dashed] (20pt,-12pt)--(0pt,12pt);
                                  \draw[thick, dashed] (0pt,-12pt)--(20pt,12pt);
                                  \draw[thick, mid arrow] (0pt,-12pt)--(0pt,12pt);
                                  \node at (-5pt,12pt) {\scriptsize $1$};
                                  \node at (-5pt,-12pt) {\scriptsize $2$};
                                  \node at (24pt,12pt) {\scriptsize $3$};
                                  \node at (24pt,-12pt) {\scriptsize $4$};
                               \end{tikzpicture}

                 \right) & g^2\begin{tikzpicture}[baseline={([yshift=-4pt]current bounding box.center)}]
                                  \draw[thick, dashed] (20pt,12pt)--(0pt,12pt);
                                  \draw[thick, dashed] (0pt,-12pt)--(20pt,-12pt);
                                  \draw[thick, dashed, mid arrow] (20pt,-12pt)--(20pt,12pt);
                                  \node at (-5pt,12pt) {\scriptsize $1$};
                                  \node at (-5pt,-12pt) {\scriptsize $2$};
                                  \node at (24pt,12pt) {\scriptsize $3$};
                                  \node at (24pt,-12pt) {\scriptsize $4$};
                               \end{tikzpicture}
               \end{pmatrix}\,,
\end{equation}
where a black arrowed line denotes fermion propagator, a dashed arrowed line denotes boson propagator, and an unarrowed dashed line denotes spacetime $\delta$-function. The first entry is boson and the second entry is fermion. Recalling $\Lambda=\mathrm{diag}(-1/2,1)$, we see that $\Lambda W_\Sigma$ and $\Lambda W_G$ are explicitly symmetric as required by quadratic expansion.
Therefore we can obtain the diagram for $K_G=W_\Sigma W_G$ as
\begin{equation}\label{eq:KGgraphic}
  K_G(1,2;3,4)=
  \begin{pmatrix}
                 0 & -g^2\left(\begin{tikzpicture}[baseline={([yshift=-4pt]current bounding box.center)}]
                                  \draw[thick, dashed, mid arrow] (20pt,12pt)--(0pt,12pt);
                                  \draw[thick, dashed, mid arrow] (0pt,-12pt)--(20pt,-12pt);
                                  \draw[thick, mid arrow] (20pt,12pt)--(20pt,-12pt);
                                  \node at (-5pt,12pt) {\scriptsize $1$};
                                  \node at (-5pt,-12pt) {\scriptsize $2$};
                                  \node at (24pt,12pt) {\scriptsize $3$};
                                  \node at (24pt,-12pt) {\scriptsize $4$};
                               \end{tikzpicture}
                               +
                               \begin{tikzpicture}[baseline={([yshift=-4pt]current bounding box.center)}]
                                  \draw[thick, dashed, far arrow] (20pt,-12pt)--(0pt,12pt);
                                  \draw[thick, dashed, near arrow] (0pt,-12pt)--(20pt,12pt);
                                  \draw[thick, mid arrow] (20pt,12pt)--(20pt,-12pt);
                                  \node at (-5pt,12pt) {\scriptsize $1$};
                                  \node at (-5pt,-12pt) {\scriptsize $2$};
                                  \node at (24pt,12pt) {\scriptsize $3$};
                                  \node at (24pt,-12pt) {\scriptsize $4$};
                               \end{tikzpicture}
                               \right) \\
                 \frac{g^2}{2}\left(\begin{tikzpicture}[baseline={([yshift=-4pt]current bounding box.center)}]
                                  \draw[thick, mid arrow] (20pt,12pt)--(0pt,12pt);
                                  \draw[thick, mid arrow] (0pt,-12pt)--(20pt,-12pt);
                                  \draw[thick, mid arrow] (20pt,-12pt)--(20pt,12pt);
                                  \node at (-5pt,12pt) {\scriptsize $1$};
                                  \node at (-5pt,-12pt) {\scriptsize $2$};
                                  \node at (24pt,12pt) {\scriptsize $3$};
                                  \node at (24pt,-12pt) {\scriptsize $4$};
                               \end{tikzpicture}+
                                \begin{tikzpicture}[baseline={([yshift=-4pt]current bounding box.center)}]
                                  \draw[thick, far arrow] (20pt,-12pt)--(0pt,12pt);
                                  \draw[thick, near arrow] (0pt,-12pt)--(20pt,12pt);
                                  \draw[thick, mid arrow] (20pt,12pt)--(20pt,-12pt);
                                  \node at (-5pt,12pt) {\scriptsize $1$};
                                  \node at (-5pt,-12pt) {\scriptsize $2$};
                                  \node at (24pt,12pt) {\scriptsize $3$};
                                  \node at (24pt,-12pt) {\scriptsize $4$};
                               \end{tikzpicture}

                 \right) & g^2\begin{tikzpicture}[baseline={([yshift=-4pt]current bounding box.center)}]
                                  \draw[thick, mid arrow] (20pt,12pt)--(0pt,12pt);
                                  \draw[thick, mid arrow] (0pt,-12pt)--(20pt,-12pt);
                                  \draw[thick, dashed, mid arrow] (20pt,-12pt)--(20pt,12pt);
                                  \node at (-5pt,12pt) {\scriptsize $1$};
                                  \node at (-5pt,-12pt) {\scriptsize $2$};
                                  \node at (24pt,12pt) {\scriptsize $3$};
                                  \node at (24pt,-12pt) {\scriptsize $4$};
                               \end{tikzpicture},
               \end{pmatrix}
\end{equation}
Examination of these diagrams shows that summing the series $(1 - K_G)^{-1}$ is equivalent to summing ladder diagrams of fermions with the so-called `Maki-Thomson' and `Aslamazov-Larkin' corrections to all orders; the first order terms of this type were examined by Kim {\it et al.\/} \cite{Kim94}.

\subsection{Eigenvalues of the Kernel}

Let us investigate the eigenvalues of the kernel \eqref{eq:KGgraphic}. Unit eigenvalues will correspond to composite scaling operators \cite{Gross:2016kjj,Klebanov:2016xxf,Klebanov:2018fzb,Tikhanovskaya:2020elb,Tikhanovskaya:2020zcw} appearing in the operator product expansion of a particle and a hole in the single patch theory, as they lead to singularities in ladder expansion $(1 - K_G)^{-1}$.

Assume the eigenvectors have the following ansatz
\begin{equation}\label{eq:Vgraphic}
 \begin{tikzpicture}[baseline={([yshift=-4pt]current bounding box.center)}]
      \draw[thick] (-15pt,-12pt) rectangle (15pt,12pt);
      \draw[thick, mid arrow]  (-15pt,12pt)--(-30pt,12pt);
      \draw[thick, mid arrow] (-30pt,-12pt)->(-15pt,-12pt);
      \node at (-30pt, -18pt) {\scriptsize $k+p/2$};
      \node at (-30pt,18pt) {\scriptsize $k-p/2$};
      \node at (0pt,0pt) {$V$};
    \end{tikzpicture}
    =V(\omega,\vec{k},\Omega,\vec{p})=
  \begin{pmatrix}
    B(k,p) \\
    F(k,p)
  \end{pmatrix},
\end{equation}where $k=(\omega,\vec{k})$ is the relative momentum (frequency), and $p=(\Omega,\vec{p})$ is the conserved center of mass momentum (frequency).

We now calculate how $K_G$ acts on the above ansatz, and obtain
\begin{equation}\label{}
  \begin{pmatrix}
    \tilde{B} \\
    \tilde{F}
  \end{pmatrix}
  =K_G\begin{pmatrix}
        B \\
        F
      \end{pmatrix},
\end{equation}
where
\begin{equation}\label{eq:tB}
\begin{split}
  &\tilde{B}(k_1,p)=-g^2D(k_1+p/2)D(k_1-p/2)\int\frac{\rd^3 k_2}{(2\pi)^{3}}\left[G(k_2-k_1)F(k_2,p)+G(k_1-k_2)F(-k_2,p)\right],
\end{split}
\end{equation}
\begin{equation}\label{eq:tF}
\begin{split}
   \tilde{F}(k_1,p) & = g^2G(k_1+p/2)G(k_1-p/2)\\
      & \times\int\frac{\rd^3 k_2}{(2\pi)^{3}}\left[\frac{1}{2}G(k_1-k_2)\left(B(k_2,p)+B(-k_2,p)\right)+D(k_1-k_2)F(k_2,p)\right]
\end{split}
\end{equation}

\subsection{Using Sliding Symmetry}

    We can use sliding symmetries to simplify the kernel $K_G$. The generic sliding symmetric eigenfunctions $B,F$ depend on the invariants discussed in \eqref{eq:invarinats} (we assumed that higher order invariants can be factorized into lower order ones):
\begin{eqnarray}
  B(k_2,p) &=& B(\omega_2,\Omega,k_{2y},p_y,k_{2y}p_x-p_yk_{2x}), \\
  F(k_2,p) &=& F(\omega_2,\Omega,k_{2x}+k_{2y}^2,p_y,p_x+2p_y k_{2y}).
\end{eqnarray}

One can verify that the kernel $K_G$ actually preserves the above sliding symmetric ansatz.

\subsection{Further simplification}

Because the CoM momentum $p$ is conserved by $K_G$, we are free to specify its value.
Similarly for the frequency $\Omega$. We will limit ourselves here to the case $p_y=0$ because then, as shown below, the integral equations can be simplified to one over frequency alone. So we will be restricting attention to longitudinal density fluctuations of the fermions on the Fermi surface, in the terminology of Kim {\it et al.} \cite{Kim94}. The case with $p_y \neq 0$ corresponds to the transverse
`diamagnetic' sector \cite{Kim94}, which we do not analyze below.

The kernel further simplifies if we set $p_y=0$, which simplifies one of the arguments in the ansatz:
\begin{eqnarray}
  B(k_2,p) &=& B(\omega_2,k_{2y},\Omega,p_x) \\
  F(k_2,p) &=& F(\omega_2,k_{2x}+k_{2y}^2,\Omega,p_x)
\end{eqnarray}

Therefore, in \eqref{eq:tB}, for the first term we do $k_{2x}\to k_{2x}-k_{2y}^2$ followed by $k_{2y}\to k_{2y}+k_{1y}$, and for the second term we do $k_{2x}\to k_{2x}+k_{2y}^2$, and then integrate over $k_{2y}$:
\begin{equation}\label{eq:tB2}
\begin{split}
  &\tilde{B}(k_1,p)=-g^2D(k_1+p/2)D(k_1-p/2)\int\frac{\rd^3 k_2}{(2\pi)^{3}}\\
  &\times\Big[G(\omega_2-\omega_1,k_{2x}-(k_{1x}+k_{1y}^2)-2k_{1y}k_{2y})F(\omega_2,k_{2x},p)\\
  &+G(\omega_1-\omega_2,(k_{1x}+k_{1y}^2)-k_{2x}-2k_{1y}k_{2y})F(-\omega_2,-k_{2x},p)\Big]\\
  &=-g^2 D(k_1+p/2)D(k_1-p/2)\int \frac{\rd\omega_2\rd k_{2x}}{(2\pi)^2}\\
  &\times \frac{i}{4|k_{1y}|}\sgn(\omega_1-\omega_2)\left[F(\omega_2,k_{2x},p)-F(-\omega_2,k_{2x},p)\right],
\end{split}
\end{equation} and in the second term of last line we also flipped $-k_{2x}\to k_{2x}$.

For the first term of \eqref{eq:tF}, we directly integrate over $k_{2x}$, and for the second term, we shift $k_{2x}\to k_{2x}-k_{2y}^2$ and then integrate over $k_{2y}$:
\begin{equation}\label{eq:tF2}
\begin{split}
   \tilde{F}(k_1,p) &= g^2G(k_1+p/2)G(k_1-p/2)\int \frac{\rd\omega_2}{2\pi}\\
   &\times\Big[\int \frac{\rd k_{2y}}{2\pi}\frac{-i}{4}\sgn(\omega_1-\omega_2)(B(\omega_2,k_{2y})+B(-\omega_2,k_{2y}))\\
   &+\int\frac{\rd k_{2x}}{2\pi}\frac{4\pi^{1/3}}{3\sqrt{3}g^{2/3}|\omega_1-\omega_2|^{1/3}}F(\omega_2,k_{2x},p)\Big].
\end{split}
\end{equation}

Plugging $p=(\Omega,p_x,0)$ into \eqref{eq:tB2} and \eqref{eq:tF2}, we see that the sliding symmetry is preserved. Furthermore, the action of $K_G$ is highly degenerate because it only cares about the integration of $F,B$ over all spatial momenta. We can therefore integrate out all spatial momenta to get a functional only in frequency space.
Let
\begin{eqnarray}
  B_{I}(\omega) &=& \int\frac{\rd k_y}{2\pi} B(\omega,k_y), \\
  F_I(\omega) &=& \int \frac{\rd k_x}{2\pi} F(\omega,k_x).
\end{eqnarray} For simplicity we have suppressed $(\Omega,p_x)$ dependence in the arguments.

The projected action of $K_G$ is
\begin{equation}\label{eq:tBI}
\begin{split}
  \tilde{B}_I(\omega_1)&=-\frac{8i \pi^{4/3}}{3\sqrt{3}g^{2/3}|\omega_1^2-\Omega^2/4|^{1/3}\left(|\omega_1-\Omega/2|^{2/3}+|\omega_1+\Omega/2|^{2/3}+|\omega_1^2-\Omega^2/4|^{1/3}\right)}\\
  &\times\int \frac{\rd \omega_2}{2\pi}\sgn(\omega_1-\omega_2)\left[F_I(\omega_2)-F_I(-\omega_2)\right],
\end{split}
\end{equation} and
\begin{equation}\label{eq:tFI}
\begin{split}
   \tilde{F}_I(\omega_1) & = \frac{ig^2}{2} \frac{\sgn(\omega_1+\Omega/2)-\sgn(\omega_1-\Omega/2)}{i\Omega-p_x-\left[\Sigma(\omega_1+\Omega/2)-\Sigma(\omega_1-\Omega/2)\right]}\int \frac{\rd\omega_2}{2\pi}\\
   &\times\Big[\frac{-i}{4}\sgn(\omega_1-\omega_2)(B_I(\omega_2)+B_I(-\omega_2))+\frac{4\pi^{1/3}}{3\sqrt{3}g^{2/3}|\omega_1-\omega_2|^{1/3}}F_I(\omega_2)\Big].
\end{split}
\end{equation}

In the conformal limit $|\Omega|,|p_x|\ll g$, we can ignore the $i\Omega-p_x$ term in \eqref{eq:tFI}, and after rescaling $b= g^{2/3}B$, we can rewrite the above equations to be independent of $g$:
\begin{equation}\label{eq:tBI2}
\begin{split}
  \tilde{b}_I(\omega_1)&=-\frac{8i \pi^{4/3}}{3\sqrt{3}|\omega_1^2-\Omega^2/4|^{1/3}\left(|\omega_1-\Omega/2|^{2/3}+|\omega_1+\Omega/2|^{2/3}+|\omega_1^2-\Omega^2/4|^{1/3}\right)}\\
  &\times\int \frac{\rd \omega_2}{2\pi}\sgn(\omega_1-\omega_2)\left[F_I(\omega_2)-F_I(-\omega_2)\right],
\end{split}
\end{equation}
\begin{equation}\label{eq:tFI2}
\begin{split}
   \tilde{F}_I(\omega_1) & = -\frac{i}{2} \frac{\sgn(\omega_1+\Omega/2)-\sgn(\omega_1-\Omega/2)}{\left[\bar{\Sigma}(\omega_1+\Omega/2)-\bar{\Sigma}(\omega_1-\Omega/2)\right]}\int \frac{\rd\omega_2}{2\pi}\\
   &\times\Big[\frac{-i}{4}\sgn(\omega_1-\omega_2)(b_I(\omega_2)+b_I(-\omega_2))+\frac{4\pi^{1/3}}{3\sqrt{3}|\omega_1-\omega_2|^{1/3}}F_I(\omega_2)\Big]\,,
\end{split}
\end{equation} where $\bar{\Sigma}=g^{-4/3}\Sigma$.

\subsection{Unit eigenvalue of $K_G$}

  The action of $K_G$ in \eqref{eq:tBI2},\eqref{eq:tFI2} can be classified into two sectors. The first one is $b_I=0$ and $F_I(\omega)$ even. The second sector is $b_I(\omega)$ even and $F_I(\omega)$ odd.

  \subsubsection{$F_I$ even sector}

   In this case the conformal $K_G$ reduces to \eqref{eq:tFI2} with $b_I=0$. By numerical diagonalization, we found only one mode with unit eigenvector. It is generated by the U(1) gauge symmetry
   \begin{equation}\label{}
\begin{split}
  \delta_{\lambda}G(x,x',y,y',\tau,\tau')&=i(\lambda(\tau)-\lambda(\tau'))G(x,x',y,y',\tau,\tau')\,, \\
  \delta_{\lambda}\Sigma(x,x',y,y',\tau,\tau')&=i(\lambda(\tau)-\lambda(\tau'))\Sigma(x,x',y,y',\tau,\tau')\,.
\end{split}
\end{equation} There is no action on $D,\Pi$ or the kinetic term. The Fourier transform of $\delta_{\lambda}G$ is
\begin{equation}\label{}
  \delta_{\lambda} G(\Omega,\omega,\vec{k})=\lambda_{\Omega}\left[i G(\omega-\Omega/2,\vec{k})-i G(\omega+\Omega/2,\vec{k})\right]\,,
\end{equation}where $\Omega$ is the CoM frequency, $\omega$ is the relative frequency and $\vec{k}$ is the relative momentum. $\lambda_{\Omega}=\int \rd\tau e^{i\Omega\tau} \lambda(\tau)$.

Integrating out the spatial momentum, we get
\begin{equation}\label{}
  F_I(\Omega,\omega)=\int \frac{\rd k_x}{2\pi} \delta_\lambda G(\Omega,\omega,k_x)=\frac{1}{2}\left(\sgn(\omega-\Omega/2)-\sgn(\omega+\Omega/2)\right),
\end{equation} and $B_I(\Omega,\omega)=0$. We can analytically verify that this is an exact eigenvector of the conformal $K_G$ with unit eigenvalue. If we retain the $i\Omega-p_x$ term, the correction is of order $\Omega^{1/3}g^{-4/3}$.

  \subsubsection{$F_I$ odd sector}

  To numerically diagonalize the kernel, we first substitute \eqref{eq:tBI2} into \eqref{eq:tFI2} to eliminate $b_I$:
\begin{equation}\label{eq:tFI3}
  \tilde{F}_I(\omega)=\frac{2}{3}\frac{\theta(\Omega^2/4-\omega^2)}{|\omega-\Omega/2|^{2/3}+|\omega+\Omega/2|^{2/3}}\int_{-\Omega/2}^{\Omega/2} \rd \omega' \left[\frac{1}{|\omega-\omega'|^{1/3}}+h(\omega,\omega')\right]F_I(\omega')\,,
\end{equation} where
\begin{equation}\label{}
  h(\omega,\omega')=\int_{-\infty}^\infty \rd \nu \frac{\sgn(\nu-\omega)\sgn(\nu-\omega')}{|\nu^2-\Omega^2/4|^{1/3}\left(|\nu-\Omega/2|^{2/3}+|\nu+\Omega/2|^{2/3}+|\nu^2-\Omega^2/4|^{1/3}\right)}\,.
\end{equation} Here we have used the explicit form of $\Sigma(\omega)$.

  In \eqref{eq:tFI3}, $\Omega$ is the only external scale so we can safely set $\Omega=1$. By numerically diagonalizing $K_G$, we found no eigenvalue close to 1.

\section{Diagrammatics of $G$-$\Sigma$ theory}
\label{sec:diagram}

In this section we discuss the diagrammatics of the $G$-$\Sigma$ theory, with the goal of developing a systematic $1/N$ expansion.
As the large $N$ limit is expressed as the saddle point of a $G$-$\Sigma$ action, and the self energy does not have a prefactor of $1/N$ in the Dyson equation, the difficulties described in Ref.~\cite{sungsik1} do not arise here. The structure of the expansion for the bilocal fields is dictated by the form of the $G$-$\Sigma$ action, and the bilocal field propagator resums an infinite number of terms from the previous approach \cite{sungsik1}.

    Using notations in Sec.~\ref{sec:fluc}, we can expand the $G$-$\Sigma$ action around the saddle point as
\begin{equation}\label{}
  S=N S_0+N S_2[\delta \calG,\delta\Xi]+ N S_3[\delta \calG,\delta \Xi]+N S_4[\delta \calG,\delta \Xi]+\dots
\end{equation} Here $S_n$ means the term which contains $n$-th power of $\delta \calG$ and $\delta \Xi$. In particular, $S_2$ is given by \eqref{eq:deltaS1}. For convenience of power counting, we decide to make propagators of $\delta\calG$ and $\delta\Xi$ independent of $N$ and push $N$ power counting into vertices. This can be done by rescaling $(\delta\calG,\delta\Xi) \to N^{-1/2}(\delta\calG,\delta\Xi)$, and therefore $NS_n[\delta\calG,\delta\Xi]\to N^{1-n/2}S_n[\delta\calG,\delta\Xi]$.

\subsection{Propagator}

  Using \eqref{eq:deltaS1} and \eqref{eq:S=dGdG}, we can write down the propagator of $\delta \calG$ as
\begin{equation}\label{}
  G_{\calG\calG}(1,2;3,4)\equiv \braket{\delta\calG(1,2)\delta\calG(4,3)}=\left[\frac{1}{K_G-1}W_\Sigma\Lambda^{-1}\right](1,2;3,4)\equiv \begin{tikzpicture}[baseline={([yshift=-4pt]current bounding box.center)}]
                                 \coordinate (v1) at (0pt,12pt);
                                 \coordinate (v3) at (20pt,12pt);
                                 \coordinate (v2) at (0pt,-12pt);
                                 \coordinate (v4) at (20pt,-12pt);
                                 \draw[thick, mid arrow](v2)--(v1);
                                 \draw[thick, mid arrow](v3)--(v4);
                                 \draw[fill=gray, opacity=0.3] (v1) rectangle (v4);
                                  \node at (-5pt,12pt) {\scriptsize $1$};
                                  \node at (-5pt,-12pt) {\scriptsize $2$};
                                  \node at (24pt,12pt) {\scriptsize $3$};
                                  \node at (24pt,-12pt) {\scriptsize $4$};
                               \end{tikzpicture}.
\end{equation} Here the numbers in the argument denote space time indices. For example $1$ means $(x_1,y_1,\tau_1)$. On the RHS we also defined its diagrammatic representation.

    Similarly we can derive the propagator for $\delta\Xi$ to be
\begin{equation}\label{}
  G_{\Xi\Xi}(1,2;3,4)\equiv \braket{\delta\Xi(1,2)\delta\Xi(4,3)}=\left[W_G\frac{1}{K_G-1}\Lambda^{-1}\right](1,2;3,4)\equiv
  \begin{tikzpicture}[baseline={([yshift=-4pt]current bounding box.center)}]
                                 \coordinate (v1) at (0pt,12pt);
                                 \coordinate (v3) at (20pt,12pt);
                                 \coordinate (v2) at (0pt,-12pt);
                                 \coordinate (v4) at (20pt,-12pt);

                                 \draw[mid arrow,draw=none](v2)--(v1);
                                 \draw[ mid arrow,draw=none](v3)--(v4);
                                 \draw[sigma snake](v2)--(v1);
                                 \draw[sigma snake](v3)--(v4);
                                 \draw[fill=gray,draw=none,opacity=0.3] (v1) rectangle (v4);
                                  \node at (-5pt,12pt) {\scriptsize $1$};
                                  \node at (-5pt,-12pt) {\scriptsize $2$};
                                  \node at (24pt,12pt) {\scriptsize $3$};
                                  \node at (24pt,-12pt) {\scriptsize $4$};
                               \end{tikzpicture}.
\end{equation} Here we use thick lines to denote Green's functions $\delta\calG$ and wavy lines to denote self energies $\delta\Xi$.

    There are also mixed correlators between $\delta\calG$ and $\delta\Xi$
\begin{equation}\label{}
  G_{\Xi\calG}(1,2;3,4)\equiv \braket{\delta \Xi(1,2)\delta\calG(4,3)}=-\left[\frac{1}{K_\Sigma-1}\Lambda^{-1}\right](1,2;3,4)\equiv\begin{tikzpicture}[baseline={([yshift=-4pt]current bounding box.center)}]
                                 \coordinate (v1) at (0pt,12pt);
                                 \coordinate (v3) at (20pt,12pt);
                                 \coordinate (v2) at (0pt,-12pt);
                                 \coordinate (v4) at (20pt,-12pt);

                                 \draw[mid arrow,draw=none](v2)--(v1);
                                 \draw[ mid arrow,draw=none](v3)--(v4);
                                 \draw[sigma snake](v2)--(v1);
                                 \draw[thick](v3)--(v4);
                                 \draw[fill=gray, opacity=0.3,draw=none] (v1) rectangle (v4);
                                  \node at (-5pt,12pt) {\scriptsize $1$};
                                  \node at (-5pt,-12pt) {\scriptsize $2$};
                                  \node at (24pt,12pt) {\scriptsize $3$};
                                  \node at (24pt,-12pt) {\scriptsize $4$};
                               \end{tikzpicture}\,,
\end{equation}
\begin{equation}\label{}
  G_{\calG\Xi}(1,2;3,4)\equiv \braket{\delta \calG(1,2)\delta\Xi(4,3)}=-\left[\frac{1}{K_G-1}\Lambda^{-1}\right](1,2;3,4)\equiv\begin{tikzpicture}[baseline={([yshift=-4pt]current bounding box.center)}]
                                 \coordinate (v1) at (0pt,12pt);
                                 \coordinate (v3) at (20pt,12pt);
                                 \coordinate (v2) at (0pt,-12pt);
                                 \coordinate (v4) at (20pt,-12pt);

                                 \draw[mid arrow,draw=none](v2)--(v1);
                                 \draw[ mid arrow,draw=none](v3)--(v4);
                                 \draw[thick](v2)--(v1);
                                 \draw[sigma snake](v3)--(v4);
                                 \draw[fill=gray, opacity=0.3,draw=none] (v1) rectangle (v4);
                                  \node at (-5pt,12pt) {\scriptsize $1$};
                                  \node at (-5pt,-12pt) {\scriptsize $2$};
                                  \node at (24pt,12pt) {\scriptsize $3$};
                                  \node at (24pt,-12pt) {\scriptsize $4$};
                               \end{tikzpicture}\,,
\end{equation} where $K_\Sigma=W_GW_\Sigma$ and it shares the same nonzero spectrum with $K_G=W_\Sigma W_G$.


    The rule of concatenation is that only edges of the same type (solid or wavy) can concatenate with the following restriction on arrow direction:
\begin{enumerate}
  \item For fermionic components $G$ and $\Sigma$, the arrows of the two concatenating edges should be paired in opposite direction.
  \item For bosonic components $D$ and $\Pi$, the arrows can be paired in either direction, but both ways of pairing should be regarded as identical.  This is because $D$ and $\Pi$ are even functions.
\end{enumerate}
\subsection{Vertices}

  There are two kinds of vertices in the theory, they come from expanding the determinant terms and the interaction term in $G$-$\Sigma$ action \eqref{GSigmaS} respectively.

  Expanding the two determinants in \eqref{GSigmaS}, we obtain non-Gaussian terms of the form
\begin{equation}\label{eq:Ssheet}
  S\supset \sum_{n=3}^{\infty}\frac{1}{n}N^{1-n/2} \Tr(\Lambda(\calG \delta\Sigma)^n)\,.
\end{equation} These terms give rise to the ``sheet'' vertices (following the terminology in \cite{Kitaev:2017awl}). At cubic order, the diagrammatic representation is
\begin{equation}\label{}
  \begin{tikzpicture}[baseline=-4pt]
    \draw[mid arrow, draw=none] (60:15pt)--(0:15pt);
    \draw[mid arrow, draw=none](180:15pt)--(120:15pt);
    \draw[mid arrow, draw=none](300:15pt)--(240:15pt);
    \draw[sigma snake](0:15pt)--(60:15pt);
    \draw[sigma snake](120:15pt)--(180:15pt);
    \draw[sigma snake](240:15pt)--(300:15pt);
    \draw[thick](300:15pt)--(0:15pt);
    \draw[thick](60:15pt)--(120:15pt);
    \draw[thick](180:15pt)--(240:15pt);
    \node at (-120:20pt) {\scriptsize $1$};
    \node at (-60:20pt) {\scriptsize $2$};
    \node at (0:20pt) {\scriptsize $3$};
    \node at (60:20pt) {\scriptsize $4$};
    \node at (120:20pt) {\scriptsize $5$};
    \node at (180:20pt) {\scriptsize $6$};
    \node at (-90:18pt) {\scriptsize $a$};
    \node at (30:18pt) {\scriptsize $a$};
    \node at (150:18pt) {\scriptsize $a$};
  \end{tikzpicture}=-\frac{1}{\sqrt{N}}\Lambda_a \calG_a(3,2)\calG_a(5,4)\calG_a(6,1).
\end{equation} The vertex can connect to 3 self energy propagators of the same type $a=\Sigma$ or $a=\Pi$. Here $\Lambda_\Sigma=1$ and $\Lambda_\Pi=-1/2$. 

    The second type of vertex comes from the $G(\tau)G(-\tau)D(\tau)$ term:
\begin{equation}\label{}
  S\supset \frac{g^2}{2\sqrt{N}}\Tr(\delta G\cdot \delta G \delta D).
\end{equation}
     It generates the ``seam'' vertex in \cite{Kitaev:2017awl}, which can be diagrammatically represented as
\begin{equation}\label{}
\tdplotsetmaincoords{70}{105}
\begin{tikzpicture}[tdplot_main_coords,baseline=10pt]
  \coordinate (v2) at (0,0,0);
  \coordinate (v6) at ($(v2)+0.5pt*(1,0,0)$);
  \coordinate (v4) at ($(v2)+0.5pt*(-0.5,-0.866,0)$);
  \coordinate (v8) at ($(v2)+0.5pt*(-0.5,0.866,0)$);
  \coordinate (v1) at ($(v2)+0.5pt*(0,0,2)$);
  \coordinate (v3) at ($(v4)+0.5pt*(0,0,2)$);
  \coordinate (v5) at ($(v6)+0.5pt*(0,0,2)$);
  \coordinate (v7) at ($(v8)+0.5pt*(0,0,2)$);
  \draw[dashed] (v1)--(v2);
  \draw[thick, mid arrow] (v5)--(v6);
  \draw[thick, mid arrow] (v4)--(v3);
  \draw[thick] (v7)--(v8);
  \node[above] at (v1) {\scriptsize $1$};
  \node[below] at (v2) {\scriptsize $2$};
  \draw[fill=gray, opacity=0.3,draw=none] (v1)--(v5)--(v6)--(v2);
  \draw[fill=gray, opacity=0.3,draw=none] (v1)--(v7)--(v8)--(v2);
  \draw[fill=gray, opacity=0.3,draw=none] (v1)--(v3)--(v4)--(v2);
\end{tikzpicture}
=-\frac{g^2}{\sqrt{N}}\,,
\end{equation} where the two arrowed edges connect to fermionic components ($\delta G$) and the unarrowed edge connects to bosonic component ($\delta D$).

\subsection{$1/N$ correction to self energy}

\begin{figure}
\centering
\begin{subfigure}[t]{0.4\textwidth}
  \centering
  \includegraphics[width=0.8\textwidth]{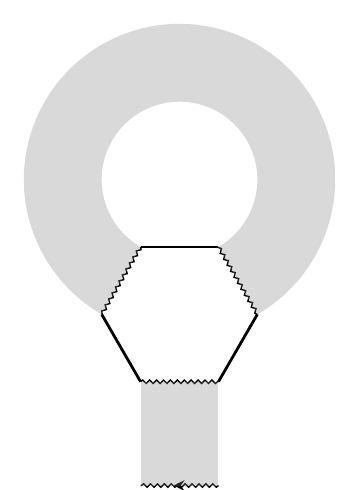}
  \caption{``sheet'' vertex}
\end{subfigure}
\begin{subfigure}[t]{0.4\textwidth}
\centering
\includegraphics[width=0.5\textwidth]{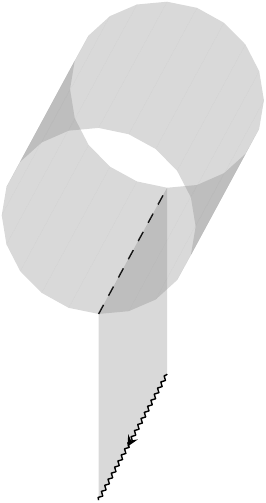}
\caption{``seam'' vertex}
\end{subfigure}
  \caption{Diagrams contributing to self energy at $O(1/N)$. We have suppressed arrows at concatenated edges.}\label{fig:Xidiagrams}
\end{figure}

  Using the above vertices, we can write down the first order $1/N$ corrections to self energies, which is given by a tadpole diagram of $\delta\Xi$:
\begin{equation}\label{}
  (\delta\Xi)_1=\frac{1}{\sqrt{N}}\braket{\delta\Xi}.
\end{equation}
There are two diagrams as shown in  Fig.~\ref{fig:Xidiagrams}, which are due to the ``sheet'' vertex and the ``seam'' vertex respectively.

We focus on the correction of electron self-energy $\Sigma$, and write down the expression based on the diagrams in Fig.~\ref{fig:Xidiagrams}. The ``sheet'' vertex contributes
\begin{equation}\label{}
\begin{split}
  \delta\Sigma_{1a}(1,2)=\frac{1}{N}\int_{3,4,5,6,7,8}&\Big[-G_{\Sigma\Sigma}(1,2;3,4)G_{\Sigma\Sigma}(6,5;2,8)G(3,8)G(5,4)G(7,6)\\
  &+\frac{1}{2}G_{\Sigma\Pi}(1,2;3,4)G_{\Pi\Pi}(6,5;2,8)D(3,8)D(5,4)D(7,6)\Big]\,.
\end{split}
\end{equation}Here numbers in the arguments denote space time coordinates. The notation of the form $G_{\Sigma\Sigma}$ refers to the $\Sigma$-$\Sigma$ or fermion-fermion component of the propagator $G_{\Xi\Xi}$. $G$ and $D$ are the saddle point single particle propagators.

  The ``seam'' vertex contributes
\begin{equation}\label{}
\begin{split}
  \delta\Sigma_{1b}(1,2)=-\frac{g^2}{N}\int_{3,4}&\Big[G_{\Sigma G}(1,2,3,4)G_{GD}(4,3;4,3)\\
  &+\frac{1}{2}G_{\Sigma D}(1,2,3,4)G_{GG}(4,3;4,3)\Big]\,,
\end{split}
\end{equation}where the $1/2$ in the second line is a symmetry factor.

    We note that the above diagrams are the same as in SYK models \cite{Kitaev:2017awl}. It is trivial to obtain corrections for other fields such as $\delta G,\delta D$ and $\delta \Pi$: We merely need to change the first subscript of the $G(1,2;3,4)$ propagator in the above expressions to the corresponding field.

\section{Antipodal patch theory and fermion bilinear operators}
\label{sec:bilinears}

\begin{figure}
\begin{center}
\includegraphics[width=0.4\textwidth]{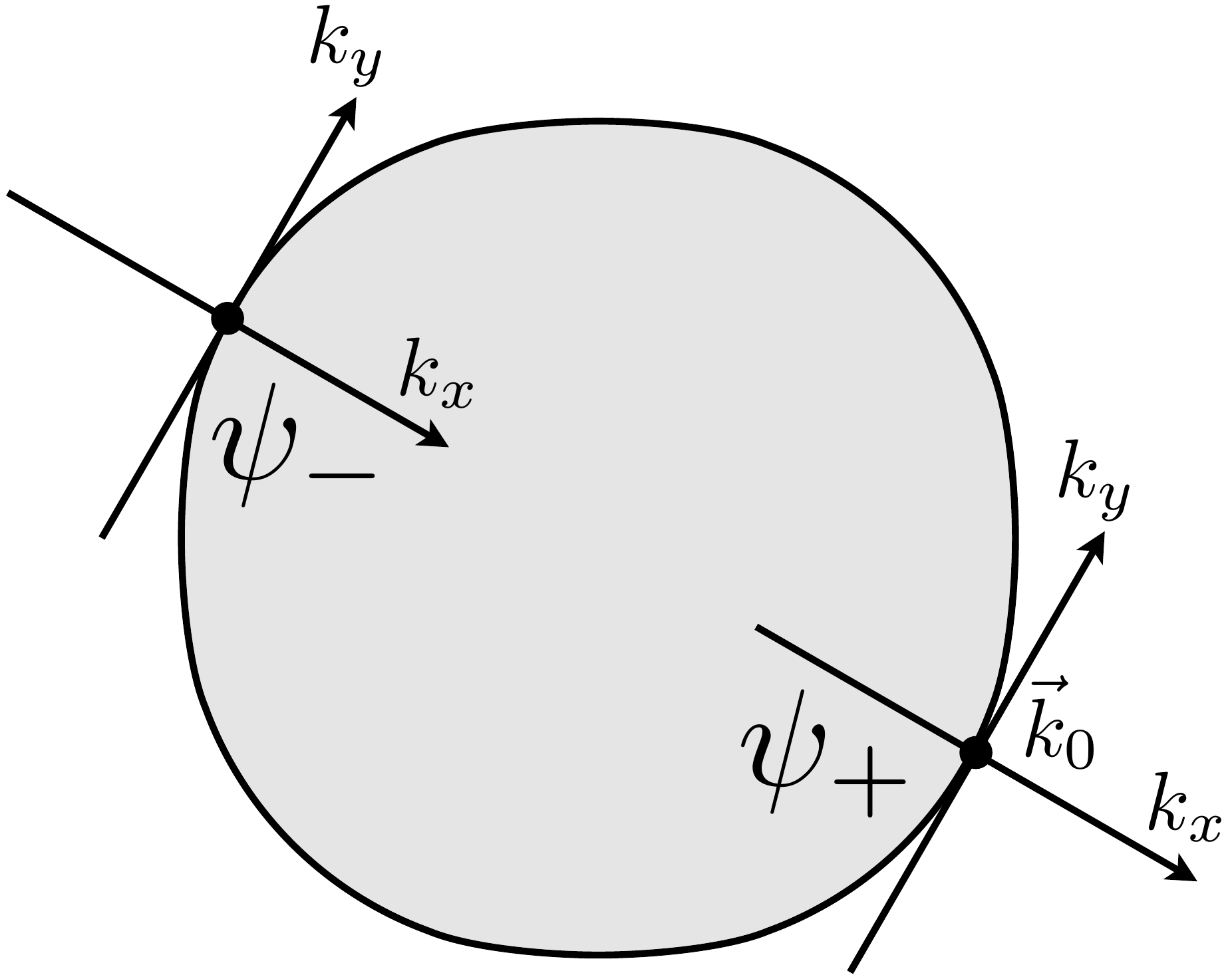}
\caption{Antipodal patches on the Fermi surface with continuum fermions $\psi_s$, $s=\pm 1$.}
\label{fig:2patch}
\end{center}
\end{figure}

This section moves beyond the single patch theory considered so far, and examines the role of antipodal patches around the Fermi surfaces -- see Fig.~\ref{fig:2patch}.

\begin{figure}[htb]
  \centering
  \includegraphics[width=0.4\textwidth]{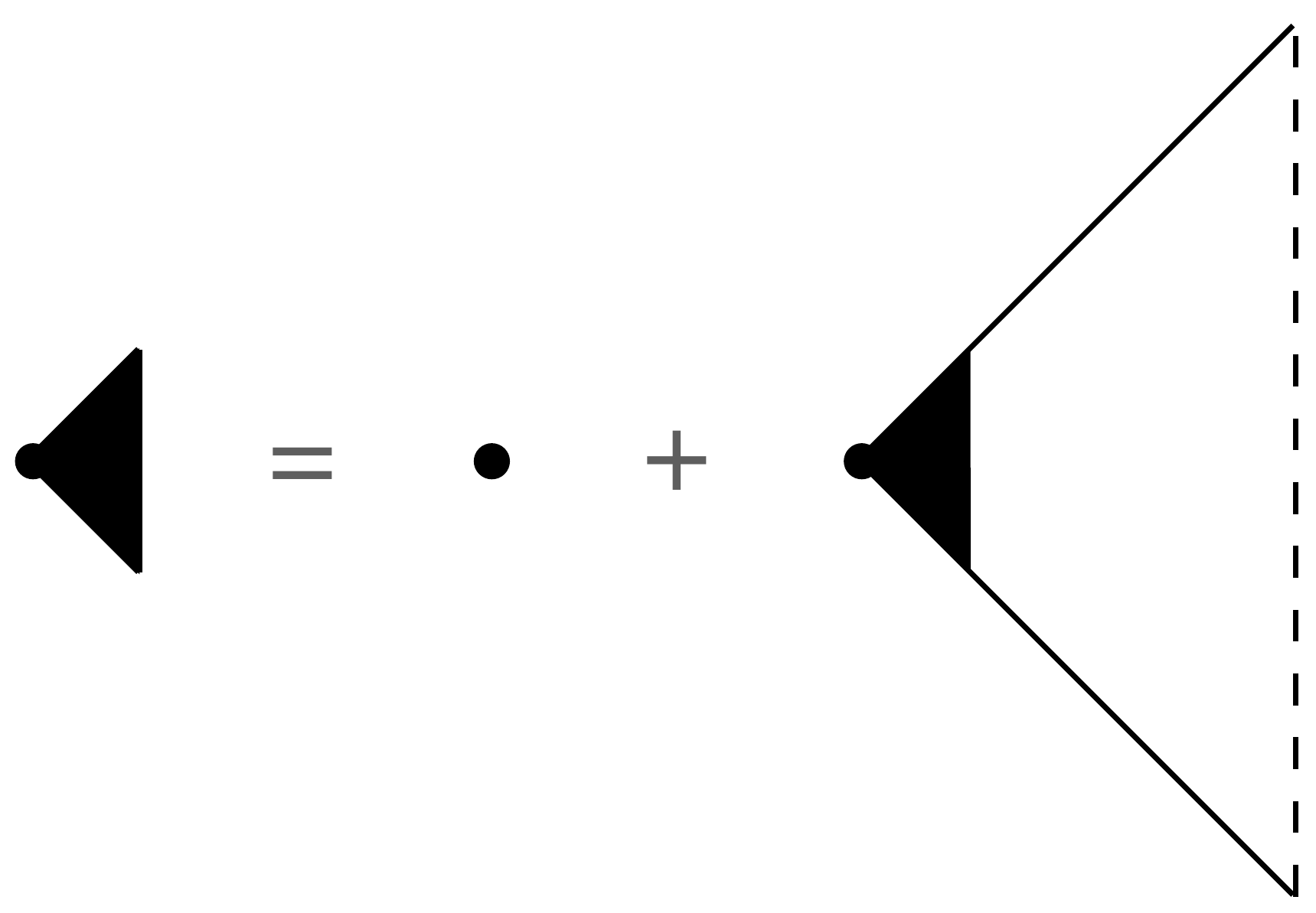}
  \caption{Schematic equation for the two-particle vertex in the large $N$ limit. The full lines are the renormalized fermion Green's functions and the dashed line is the boson Green's function
  }\label{fig:ladder}
\end{figure}

The single patch theory in Section~\ref{sec:fluc} examined the fluctuations about the large $N$ saddle point using a perspective similar to that of the soft mode analysis of Kitaev and Suh \cite{Kitaev:2017awl} for the SYK model. An alternative approach \cite{JMDS16,Klebanov:2016xxf,Klebanov:2018fzb,Klebanov:2020kck,Kim:2019upg,Tikhanovskaya:2020elb,Tikhanovskaya:2020zcw} is to compute all new operators that appear in the operator product expansion of 2 fermions. In the present large $N$ approach, applicable equally to the SYK model and the antipodal patch theory, such a computation is equivalent to computing the eigenmodes of the two-particle Bethe-Salpeter equation for the fermion vertex,  shown schematically in Fig.~\ref{fig:ladder}.

Examination of these diagrams around the Fermi surface shows that the only new operators that appear from this vertex are those corresponding to the Cooper pairing and $2k_F$ operators, as discussed in Refs.~\cite{mross,metlitski5}, and we will discuss these cases in the following subsections.

The case of the pairing operator, discussed in Section~\ref{sec:pairing} turns out to be simpler, because we are able to use the sliding symmetry of Section~\ref{sec:sliding} to simplify the analysis. With this simplification, the resulting integral Bethe-Salpeter equation turns out to act only on frequency space; indeed, it is identical in form to the equations obtained for the SYK model \cite{Polchinski:2016xgd,Gross:2016kjj,JMDS16,Klebanov:2016xxf,Klebanov:2018fzb,Klebanov:2020kck,Kim:2019upg,Tikhanovskaya:2020elb,Tikhanovskaya:2020zcw}.

The case of the $2k_F$ operator, discussed in Section~\ref{sec:2kF}, is more complicated because the reduction to a purely frequency space integral equation is not possible. Instead we have consider an equation involving both the frequency and the tangential momentum on the Fermi surface, whose solution requires a numerical analysis.

For the results of this section, we will consider a more general setting than that of the Ising critical point considered so far. It is known that the Ising boson leads to an attractive interaction between the fermions on antipodal points on the Fermi surface. A very similar theory applies to the problem of a gauge field coupled to a Fermi surface; for a single U(1) gauge field, the interaction between anti-podal Fermi surface points is respulsive. Recent works \cite{YZSS1,LZDC20,YZSS2} have considered problems with multiple gauge fields, and the assignment of gauge charges is such that some gauge fields are repulsive and others are attractive. So we will consider generalization of the theory
(\ref{pchact}) with $N$ flavors of fermions, $M_1$ flavors of bosons which mediate an attractive interaction (in the pairing channel) between antipodal points on the Fermi surface, and $M_2$ flavors of bosons which mediate a repulsive interaction. By rescaling the bosons, we will normalize the mean-square Yukawa coupling for both classes of bosons as in (\ref{pchact}) with the same value $g$; the value of $g$ will drop out in the scaling equations we consider in this section. Having obtained the same Yukawa coupling, we do have to consider the co-efficient of the $(\partial_y \phi)^2$ term in (\ref{pchact}) more carefully \cite{LZDC20,YZSS2}. We take this co-efficient to equal $K_1$ and $K_2$ for the two bosons, and we will see below that the ratio $K_1/K_2$ influences the critical exponents. For the gauge field case, the values of $K_{1,2}$ are equal to the corresponding diamagnetic susceptibility of the system \cite{YZSS2}, and this depends upon the lattice scale properties.

\subsection{Scaling analysis}

\begin{figure}
\begin{center}
\includegraphics[width=0.6\textwidth]{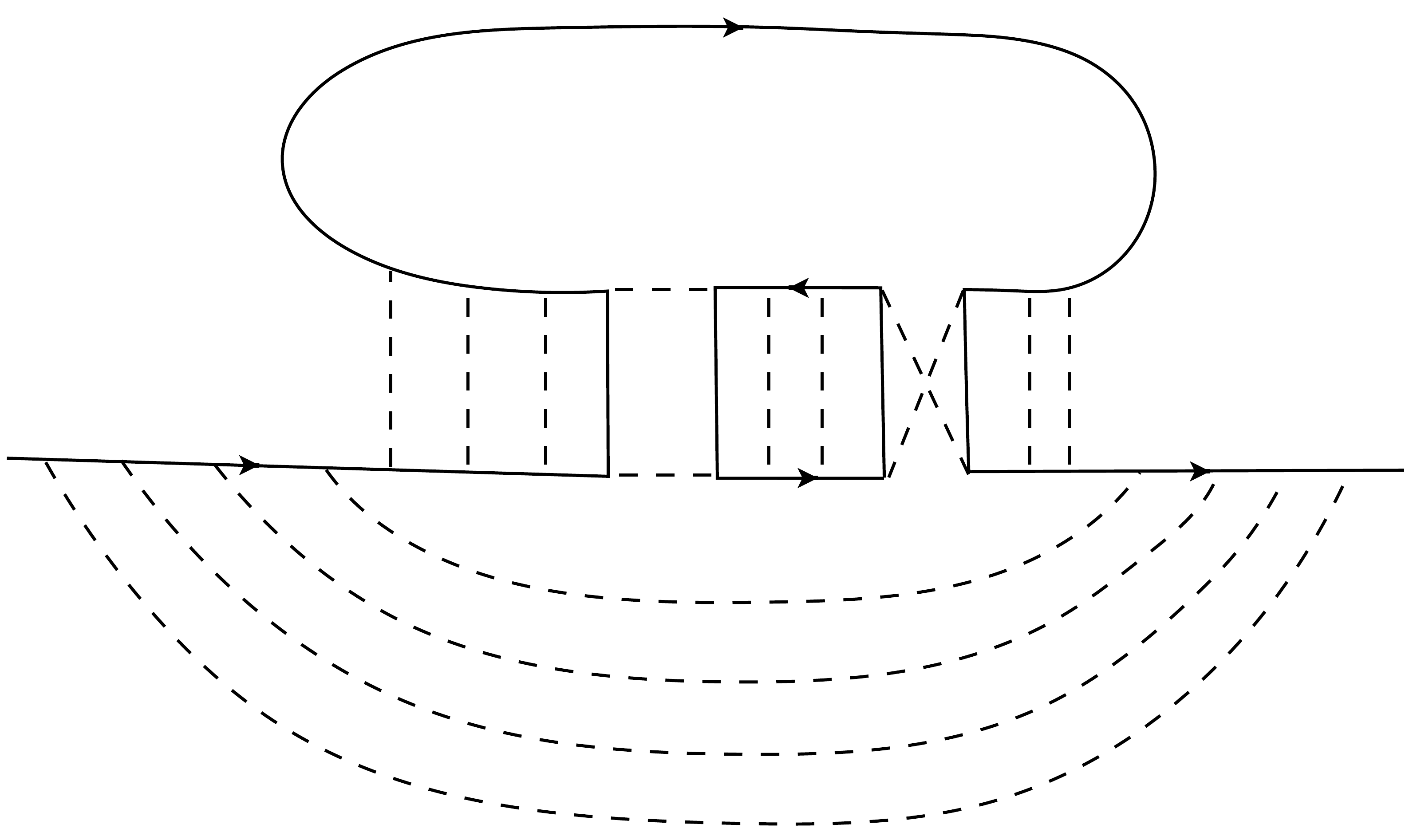}\\
\includegraphics[width=0.5\textwidth]{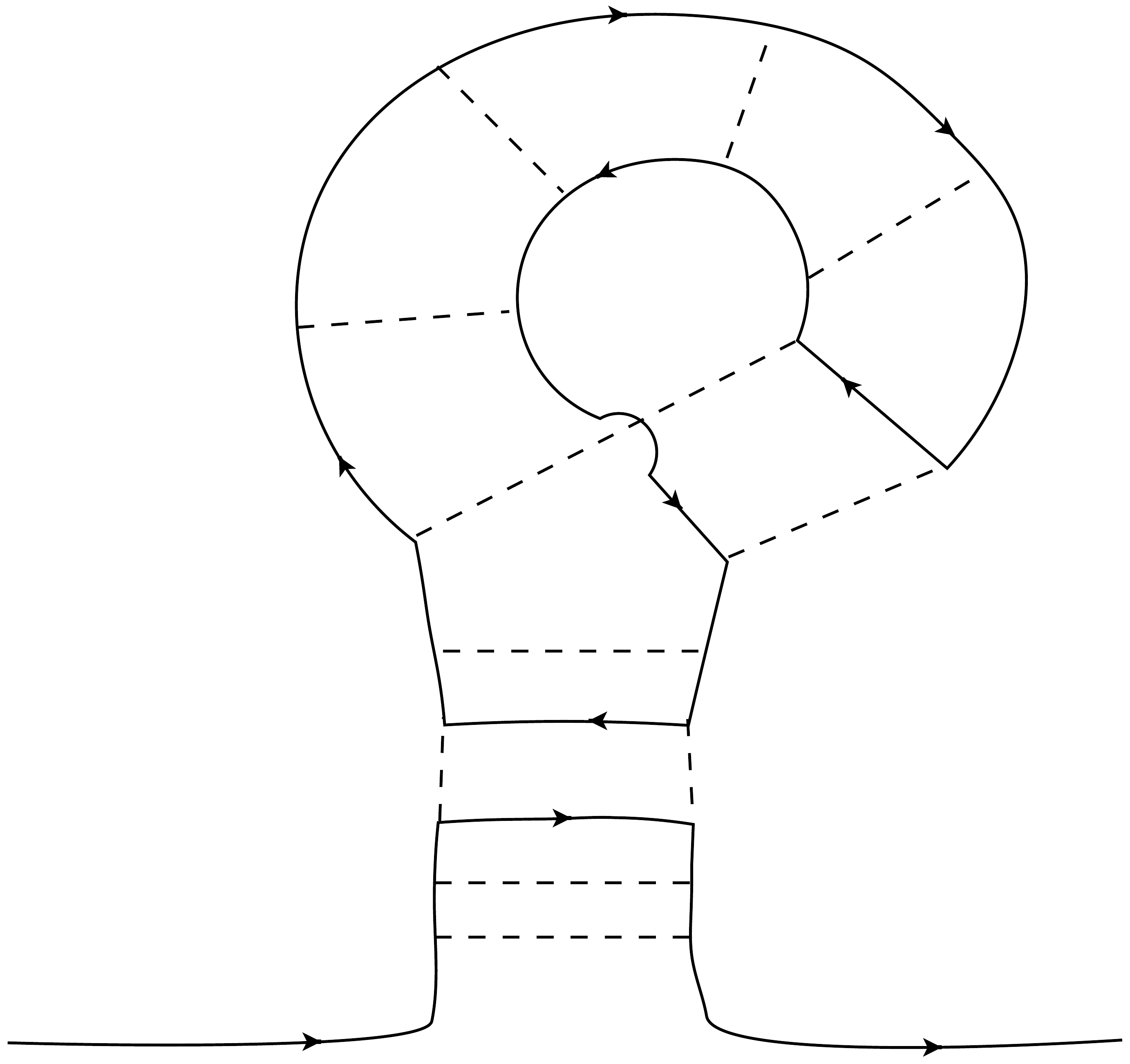}
\end{center}
\caption{Examples of graphs in the sheet and seam contributions in Fig.~\ref{fig:Xidiagrams} to the fermion self energy. As in Fig.~\ref{fig:ladder}, the full line is the fermion, and the dashed line is the boson. For the contribution to $\eta_\psi$, the top fermion line in the first graph should form a closed fermion loop to allow interactions between antipodal patches ({\it i.e.\/} no Aslamazov-Larkin type insertions in the top ladder).}
\label{fig:sheetandseam}

\end{figure}

Let us write down the explicit form of the Lagrangian density of the 2-patch theory, generalizing
the action in (\ref{pchact})
\begin{align}
& \mathcal{L} = \sum_{s=\pm 1} \sum_{i=1}^N \psi^\dagger_{is}\left[\partial_\tau- i s \partial_x - \partial_y^2 \right]\psi_{is} +\sum_{a=1,2} \frac{K_a}{2}\sum_{i=1}^{M_a} \left(\partial_y \phi_{ia} \right)^2 \nn
&~~~~~~~ + \sum_{s=\pm 1}\sum_{a=1}^2 s^{3-a} \sum_{l=1}^{M_a} \sum_{i,j=1}^N \frac{g_{ijl}^a}{N} \psi_{is}^\dagger \psi_{js} \phi_{l a} \,.
\label{pchact2}
\end{align}
Here $s=\pm 1$ is the index of the two anti-podal patches (see Fig.~\ref{fig:2patch}), and $a=1,2$ represents the attractive and repulsive bosons respectively. We now recall the scaling analysis of this theory \cite{metlitski1} under the assignments
\bea
\mbox{dim}[y] &=& -1 \nonumber \\
\mbox{dim}[x] &=& -2 \nonumber \\
\mbox{dim}[\tau] &=& -z \nonumber \\
\mbox{dim}[\psi (r, \tau) ] &=& (1 + z + \eta_\psi)/2
\label{dimxytau}
\eea
This consistently yields
\beq
\mbox{dim}[G(k,\omega)] = -2 +\eta_\psi
\eeq
In the present theory, the large $N$ exponents are $z=3$ and $\eta_\psi = 0$. Earlier work \cite{metlitski1,mross} found a small correction to $\eta_\psi$ at three loop order.
Similar correction will appear in our large $N$ expansion at at first order in $1/N$: an important point is that such a result is fully systematic in our $1/N$ expansion, unlike the result in Ref.~\cite{metlitski1}. The diagrams contributing to the self energy at order $1/N$ were presented in Fig.~\ref{fig:Xidiagrams}, and we show in Fig.~\ref{fig:sheetandseam} examples of the contributions to these diagrams in terms of the fermion and boson Green's functions.
In Ref.~\cite{metlitski1}, the diagrams contributing to $\eta_\psi$ are those in Fig. 10b,c, and these are terms in the infinite series of diagrams in Figs.~\ref{fig:Xidiagrams} and~\ref{fig:sheetandseam};
we also need to include particle-particle ladders in addition to the particle-hole ladders shown, and these appear upon considering the case of real $g_{ijl}$.
As in Ref.~\cite{metlitski1}, we expect that it is important to include antipodal patches in these diagrams to obtain the
contribution to $\eta_\psi$.

Let us now turn to a consideration of the scaling dimensions of the fermion bilinear operators.

\subsubsection{$2k_F$ operator}
\label{sec:scale2kF}

The physical quantity we are interested in is the singular behavior of the $2k_F$ susceptibility, $\chi_{2k_F}$.
However, this is too difficult to compute in our large $N$ theory. So we try an alternative route below by relating its scaling
dimension to that of a vertex function $\Phi_{2 k_F}$. For the $2 k_F$ operator, this is relatively straightforward, as there is no violation of
hyperscaling in the graphs.

Let us define the scaling dimension of the $2 k_F$ operator
\beq
\rho_{2k_F} = \psi_{+}^\dagger \psi_{-}
\eeq
by
\beq
\mbox{dim}[\rho_{2 k_F}(r, \tau)] = \Delta_{2 k_F}
\eeq
The correspondence with the $\Delta$ defined by Mross {\it et al.} \cite{mross} in their (32) is $\Delta = \Delta_{2 k_F}/2$.
At tree level, we have $\Delta_{2 k_F} = 2\, \mbox{dim}[\psi]  = 1 + z + \eta_\psi$.
Then the scaling dimension of the $2k_F$ susceptibility is
\beq
\mbox{dim}[\chi_{2 k_F}(k,\omega)] = 2 \Delta_{2 k_F} - 3 - z \label{dim2kF}
\eeq

We define the vertex function $\Phi_{2 k_F}$ as the 3-point correlator of $\rho_{2k_F}$ with 2 fermion operators, after amputating the external Green's functions. Then
\bea
\mbox{dim}[\Phi_{2 k_F}] &=& \mbox{dim}[\rho_{2 k_F}] -6 -2 z - 2\, \mbox{dim}[G(k,\omega)]  - 2 \, \mbox{dim}[\psi] \nonumber \\
&=& \Delta_{2 k_F} -1 -z - \eta_\psi \label{Phi2kF}
\eea
We can check now that
\bea
\mbox{dim}[\chi_{2 k_F}(k,\omega)] &=& 2 \, \mbox{dim}[\Phi_{2 k_F}] +3 +z + 2\,\mbox{dim}[G(k,\omega)]  \nonumber \\
&=& 2 \, \mbox{dim}[\Phi_{2 k_F}] - 1 +z + 2 \eta_\psi\,. \label{scalechi2kF}
\eea

We expect the solution of the vertex function to scale as
\beq
\Phi_{2 k_F} (\omega) \sim \omega^{\mathrm{dim}[\Phi_{2 k_F}]/z}
\eeq
(and similarly for $\chi_{2k_F}$). We will compute the frequency dependence of $\Phi_{2 k_F} (\omega)$ below in Section~\ref{sec:2kF}, which therefore yields the scaling dimension $\Delta_{2k_F}$ via (\ref{Phi2kF}).

\subsubsection{Cooper pair operator}
\label{sec:scaleCooper}

This is a little more subtle, because the intermediate loop integrals are independent of $k_y$, and so the integral over $k_y$ just yields a factor of $k_F$; this leads to violation of hyperscaling.

Let us define the scaling dimension of the Cooper operator
\beq
\Psi = \psi_+ \psi_-
\eeq
by
\beq
\mbox{dim}[\Psi (r, \tau)] = \Delta_{\Psi}
\eeq
Then the scaling dimension of the Cooper pair susceptibility is
\beq
\mbox{dim}[\chi_{\Psi}(k,\omega)] = 2 \Delta_{\Psi} - 4 - z \label{dimchiPsi}
\eeq
Note that this differs from (\ref{dim2kF}) by an extra -1 on the r.h.s., corresponding to the absence of the $k_y$ integral in evaluating $\chi$.
Without vertex corrrections, evaluation of the Cooperon bubble shows that $\chi_\Psi (k=0, \omega) \sim \omega^{1-2(1- \eta_\psi)/z}$, and so $\mbox{dim}[\chi_{\Psi}(k,\omega)] = z-2 + 2 \eta_\psi$; then (\ref{dimchiPsi}) yields $\Delta_{\Psi} = 1+z + \eta_\psi$, which checks out correctly with $\mbox{dim}[\psi]$ in (\ref{dimxytau}).

For the vertex operator, the relationship remains the same as in (\ref{Phi2kF}) {\it i.e.\/}
\beq
\mbox{dim}[\Phi_{\Psi}] = \Delta_{\Psi} -1 -z - \eta_\psi \label{PhiPsi}
\eeq
Without vertex corrections, we should have $\mbox{dim}[\Phi_{\Psi}] = 0$, and this agrees with the corresponding value of $\Delta_{\Psi}$ quoted above. Another way to think about (\ref{PhiPsi}) is that the 2 fewer $k_y$ integrals in the evaluation of the 3-point correlator cancel with corresponding factors from $\mbox{dim}[G(k, \omega)]$.
We can also check now that
\bea
\mbox{dim}[\chi_{\Psi}(k,\omega)] &=& 2 \, \mbox{dim}[\Phi_{\Psi}] +2 +z + 2\,\mbox{dim}[G(k,\omega)]  \nonumber \\
&=& 2 \, \mbox{dim}[\Phi_{\Psi}] -2 +z + 2 \eta_\psi\,. \label{scalechi}
\eea
We will compute the frequency dependence of
\beq
\Phi_{\Psi} (\omega) \sim \omega^{\mathrm{dim}[\Phi_{\Psi}]/z}
\eeq
next in Section~\ref{sec:pairing}, which determines $\Delta_\Psi$ via (\ref{PhiPsi}).

\subsection{Pairing operator}
\label{sec:pairing}

We consider instabilities towards superconducting pairing. First, we note that with the complex flavor-random Gaussian interaction $g_{ijk}$, no anomalous Green's functions and self energies appear in the large $N$ saddle point, and there is therefore no intrinsic pairing instability, at least at large $N$. To achieve controlled pairing at large $N$ in this approach, we must include an additional attractive $U$ : $-(U/N)\sum_{i,j=1}^N\sum_k\psi^\dagger_{ik}\psi^\dagger_{i,-k}\psi_{j,k}\psi_{j,-k}$. The attractive $U$ is then renormalized exactly by the naive resummation of pairing bubbles, and may also be handled in a saddle-point formalism with static anomalous Green's functions and self-energies \cite{Patel:2018rjp}. In a regular Fermi liquid, this leads to the famous BCS instability even for infinitesimal $U$ as the pairing bubbles diverge as $\sim\ln(1/T)$ in the IR. However, in this non-Fermi liquid, the $\omega^{2/z}$ self-energy dominates in the IR, and the pairing bubble (which is not further dressed by the complex $g_{ijk}$) is no longer IR divergent. Therefore, this model is further resistant to an infinitesimal attractive $U$. The disordered model in Sec. \ref{sec:spatial} with complex random interactions is a marginal Fermi liquid, and this pairing bubble then diverges as $\ln(1/\ln (1/T))$ in the IR, so the infinitesimal attractive $U$ does cause an instability, but it is much weaker than that in a Fermi liquid.

To get intrinsic pairing instabilities at large $N$ without the need for an additional $U$, we consider real Gaussian flavor-random $g_{ijk}$. These now do allow for dynamic anomalous Green's functions and self energies in the large $N$ saddle point itself \cite{Ilya1}, and exact Eliashberg equations can be derived and solved numerically. However, to analyze the pairing instabilities in the metal, we first adapt a simpler approach by assuming the system is a metal, and then looking at the exact renormalization of the pairing vertex at large $N$ \cite{Kim:2019upg}. The pairing vertex may be described by the large $N$ exact self-consistent eigenvalue equation shown in Fig.~\ref{fig:ladder},
\beq
E \Phi_\Psi(q,i\Omega_m) = - \sum_a \frac{M_a \zeta_a g^2}{N}T\sum_{\omega_n\neq\Omega_m}\int_k \Phi_\Psi(k,i\omega_n)G_+(k,i\omega_n)G_-(-k,-i\omega_n)D_a(k-q,i\omega_n-i\Omega_m).
\label{PV}
\eeq
Here $a = 1,2$ sums over the attractive and repulsive bosons and $\zeta_a = 2a-3= -1$ ($+1$) for the attractive (repulsive) interactions. Approaching from high $T$, the transition occurs at $T=T_{sc}$ when the largest eigenvalue $E_\mathrm{max}=1$. Note that this doesn't determine the nature of the transition itself, which instead requires solving the full non-linear Eliashberg equations \cite{Patel:2018rjp} (as detailed in that reference, these non-linearities can sometimes cause some surprises like producing a first-order transition).

We now formulate the theory using two antipodal patches subject to the same real $g_{ijk}$ \cite{metlitski1}. This multiplies (\ref{patchypie}) by 2 and divides (\ref{ftfse}) by $2^{1/3}$. We can then exploit the $\pm k_x+k_y^2$ and $k_y$ dependencies of $G$ and $D$ respectively, i.e. the sliding symmetry, to again see that a self-consistent momentum-independent pairing vertex exists, and its eigenvalue equation is given by
\beq
E \Phi_\Psi(i\Omega_m) = -\sum_a \frac{M_a \zeta_a g^2}{N}\frac{T}{3\sqrt{3}}\sum_{\omega_n\neq\Omega_m}\frac{\Phi_\Psi(i\omega_n)}{|\omega_n+i\Sigma(i\omega_n)|}\frac{(4\pi)^{1/3}}{(gK_a)^{2/3}|\omega_n-\Omega_m|^{1/3}},
\label{PVfreq}
\eeq
At low energies and $T=0$, where we drop the bare $\omega_n$ term in the RHS of (\ref{PVfreq}), because it is irrelevant in the infrared (IR), we obtain a universal equation independent of $g$
\beq
E \Phi_\Psi(i\Omega_m) = \frac{\mathcal{K}}{3}
\int\frac{d\omega_n}{2\pi}\frac{2\pi\Phi_\Psi(i\omega_n)}{|\omega_n|^{2/3}|\omega_n-\Omega_m|^{1/3}}\,,
\label{Deltaclean}
\eeq
where the dimensionless constant
\beq
\mathcal{K} \equiv \frac{M_1 K_2^{2/3} - M_2 K_1^{2/3}}{M_1 K_2^{2/3} + M_2 K_1^{2/3}}\,
\eeq
determines the balance between the attractive and repulsive interactions. Equation (\ref{Deltaclean}) has the same form as that for the $\gamma=1/3$ case of the $\gamma$-model of quantum-critical pairing studied by Chubukov and collaborators \cite{Moon2010,Chubukov1,Chubukov2,Chubukov3}; it also co-incides with equations studied in the SYK model \cite{Polchinski:2016xgd,Gross:2016kjj,JMDS16,Klebanov:2016xxf,Klebanov:2018fzb,Klebanov:2020kck,Kim:2019upg,Tikhanovskaya:2020elb,Tikhanovskaya:2020zcw}.

We now follow \cite{Kim:2019upg}. We assume the eigenvector has the form\footnote{There are also odd parity eigenvectors $\Phi_\Psi(i\Omega_m)=\mathrm{sgn}(\Omega_m)/|\Omega_m|^\alpha$. However, we can see from the diagrams involved in the renormalization of the pairing vertex that the physical eigenvector must be of even parity.}
\beq
\Phi_\Psi(i\Omega_m)=\frac{1}{|\Omega_m|^{\alpha}}\,.
\eeq
In the notation of the scaling analysis of Section~\ref{sec:scaleCooper}, this identifies
\beq
\mathrm{dim}[\Phi_{\Psi}] = -z \alpha\,. \label{scalePhi}
\eeq
We assume $0<\mathrm{Re}~[\alpha]<1/3$ to ensure a convergent integral in (\ref{Deltaclean}), and then we have
\beq
E = \mathcal{K} \frac{\pi ^2 \left(3 \cot \left(\frac{\pi  \alpha }{2}\right)+\sqrt{3}\right) \sec \left(\pi  \left(\alpha +\frac{1}{6}\right)\right)}{9 \Gamma \left(\frac{1}{3}\right) \Gamma (1-\alpha ) \Gamma \left(\alpha +\frac{2}{3}\right)}.
\label{pairingEV}
\eeq
For $\mathcal{K}=1$, Setting $E=1$ indicates a complex scaling dimension $\alpha = 1/6\pm i\times0.53734~...$~, which implies that a pairing instability exists and the ground state is superconducting. As the value of $\mathcal{K}$ is reduced, the magnitude of the imaginary part of $\alpha$ also reduces, going to zero at $\mathcal{K} = \mathcal{K}^\ast = 0.12038~...$, at which point $\alpha=1/6$ exactly. For $\mathcal{K}^\ast>\mathcal{K}>0$~, $E=1$ has two solutions with purely real $\alpha$: $\alpha_1$, with $1/6>\alpha_1>0$ and $\alpha_2=1/3-\alpha_1$, indicating the apparent absence of a superconducting instability arising purely out of the relevant operators in the low energy critical theory, when the repulsive interaction is strong enough \footnote{For $\mathcal{K}<0$, there is no solution for $E=1$ with an even parity eigenvector. Therefore, there is no superconducting instability, and the scaling of the Cooper pair susceptibility is also not renormalized from the pairing bubble value of $\mathrm{dim}[\chi_\Psi(k,\omega)] = 1$}.

\begin{figure}[htb]
  \centering
  \includegraphics[width=0.5\textwidth]{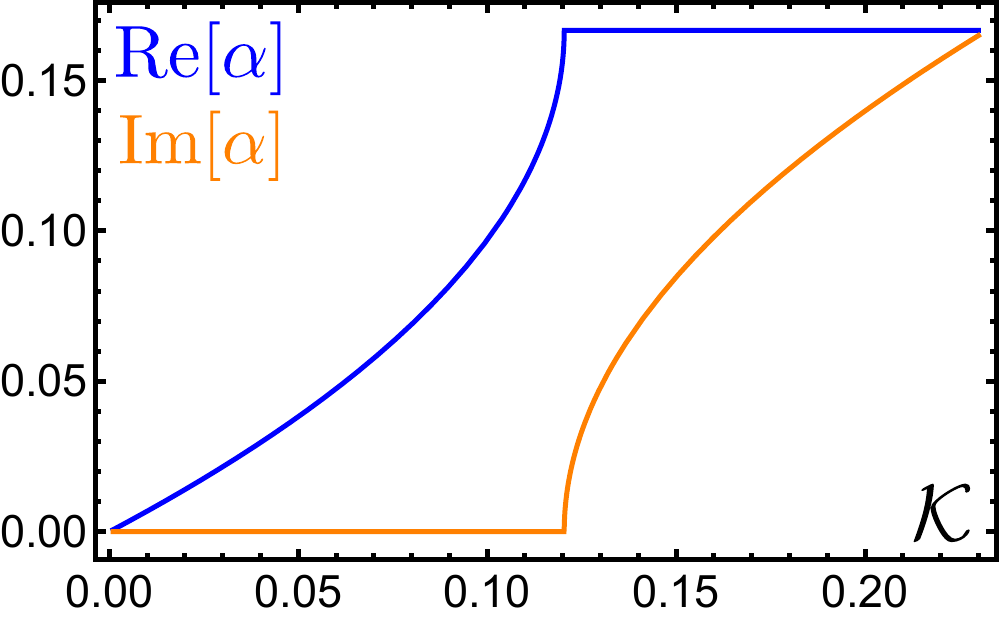}
  \caption{Plot of $\mathrm{Re}[\alpha]$ and $\mathrm{Im}[\alpha]$, for the solutions which have $\mathrm{Re}[\alpha]\leq 1/6$ and $\mathrm{Im}[\alpha]>0$, as a function of $\mathcal{K}$. For $1>\mathcal{K}>\mathcal{K}^\ast$, $\mathrm{Re}[\alpha] = 1/6$ and $\mathrm{Im}[\alpha]\neq 0$.}
  \label{fig:alphaSCplot}
\end{figure}

When $\alpha$ is purely real, both the roots $\alpha_1$ and $\alpha_2$ do not determine the scaling of the Cooper pair susceptbility. In particular, the root $1/3>\alpha_2>1/6$ is not allowed. This may be seen as follows: if we allow for anomalous Green's functions and self energies in the saddle point equations, then the function $\Phi_\Psi(i\Omega_m)$ can be identified as the anomalous component of the self energy. This then leads to a contribution to the saddle point free energy at $\mathcal{O}(\Phi_\Psi^2)$:
\beq
\frac{\mathcal{F}_\Phi}{N} \sim -\int_q \int \frac{d\Omega_m}{2\pi} |\Phi_\Psi(q,i\Omega_m)|^2G(q,i\Omega_m)G(-q,-i\Omega_m) \sim -\int \frac{d\Omega_m}{2\pi} \frac{|\Phi_\Psi(i\Omega_m)|^2}{|\Omega_m+i\Sigma(\Omega_m)|}.
\eeq
The integral diverges in the IR for $\alpha_2$ (but is finite for $\alpha_1$), which makes the free energy of the ground state divergent as $T\rightarrow0$, and therefore unphysical, as the entropy $\mathcal{S}= -\partial F/\partial T$ becomes negative \cite{Chubukov1}. Rejecting $\alpha_2$, and using (\ref{scalechi}) and (\ref{scalePhi}),  we then have the scaling dimension \beq
\mathrm{dim}[\chi_\Psi(k,\omega)] = 1-6\alpha_1,
\eeq
for $M_a,N \rightarrow \infty$. In Fig.~\ref{fig:alphaSCplot}, we show $\alpha_1$ as a function of $\mathcal{K}$.

We may also define a family of models by changing the boson propagator to ($2<z\le3$) \cite{mross}:
\beq
D(q,i\Omega_m) = \frac{1}{\displaystyle |q_y|^{z-1}+\frac{g^2}{4\pi}\frac{|\Omega_m|}{|q_y|}}.
\eeq
When $z=2$, the system is a marginal Fermi liquid, with $\Sigma(i\omega)\sim i\omega \ln(|\omega|)$. We now repeat the above procedure for generic $2<z\le3$. In place of (\ref{pairingEV}) we obtain
\beq
E = -\mathcal{K}\frac{\pi ^2 \csc \left(\frac{\pi  \alpha }{2}\right) \csc \left(\frac{\pi }{z}\right) \sec \left(\pi  \left(\frac{\alpha }{2}+\frac{1}{z}\right)\right)}{4 \Gamma (1-\alpha ) \Gamma \left(-\frac{2}{z}\right) \Gamma \left(\alpha +\frac{2}{z}\right)},
\eeq
with
\beq
\mathcal{K} \equiv \frac{M_1 K_2^{2/z} - M_2 K_1^{2/z}}{M_1 K_2^{2/z} + M_2 K_1^{2/z}}.
\eeq
For $z\rightarrow 2$, the complex scaling dimension is $\alpha = 1/2-1/z \pm i\times \sqrt{z/2-1}$, when $\mathcal{K}=1$. The threshold $\mathcal{K}^\ast=(z-2)/8$ as $z\rightarrow2$. These observations  imply that the superconducting instability still survives in the marginal Fermi liquid limit for $\mathcal{K}>\mathcal{K}^\ast$.

At $T\neq 0$, our approach of neglecting the purely thermal fluctuations of the gauge bosons $\phi$ causes a superconducting instability to occur for all $\mathcal{K}>0$ (but not for $\mathcal{K}\leq 0$), with $T_{sc} \sim g^4/(z-2)$, because the fermion self energy vanishes at the first Matsubara frequencies \cite{Chubukov2}. This is parametrically the same energy scale at which the non-Fermi liquid behavior itself onsets, {\it i.e.} when $|\Sigma(i\omega_n)|$ becomes comparable to $|\omega_n|$. The true physical problem requires a careful consideration of the thermal fluctuations of the massless gauge bosons, and their effects on the fermion Green's function, in two spatial dimensions, along the lines of Ref. \cite{KimLeeWen1995}, as the cancellation of the thermal fluctuations from the equation for the superconducting gap function via Anderson's theorem \cite{Anderson1959} does not occur in the simultaneous presence of attractive and repulsive boson interactions \cite{Ilya2}. We will therefore perform this analysis in future work.

We also briefly comment on the effects of nonzero $T$ in the case where the two $\phi$'s are not gauge bosons, and are therefore allowed to have a thermal mass as in Sec. \ref{sec:rpaffs}, $M^2(T)\sim T\ln(1/T)\ll T^{2/3}$, that arises from operators that are irrelevant in the critical patch theory. This causes both the bosons to induce thermal self energies for the fermions, $\Sigma_{T,a}(i\omega_n)\approx -i\mathrm{sgn}(\omega_n)s_aT^{1/2}\ln^{1/2}(1/T)$, with a non-universal prefactor $s_a$, as in Sec. \ref{sec:num}, that depends on parameters from outside the patch theory. One can then show, following Ref. \cite{Chubukov0}, that the pertinent equation for $\Phi_\Psi$ can be reduced to (we consider $z=3$ here for simplicity, the consequences are similar for other $2<z<3$ as well)
\beq
E \Phi_\Psi(i\Omega_m) \approx -\sum_a \frac{M_a \zeta_a g^2}{N}\frac{T}{3\sqrt{3}}\sum_{\omega_n\neq\Omega_m}\frac{\Phi_\Psi(i\omega_n)}{|\omega_n+i\Sigma_Q(i\omega_n)+2i\Sigma_{T,2}(i\omega_n)|}\frac{(4\pi)^{1/3}}{(gK_a)^{2/3}|\omega_n-\Omega_m|^{1/3}}.
\label{PV2massive}
\eeq
As $T\rightarrow0$, since $i\Sigma_{T,2}(i\omega_{n=\pm 1})\gg \omega_{n=\pm 1}, i\Sigma_Q(i\omega_{n=\pm 1})$, there is no enhancement due to the first Matsubara frequency, and the thermal part of the self energy dominates. The largest eigenvalue therefore scales as $\sim T^{1/6}/\ln^{1/2}(1/T)$, which vanishes as $T\rightarrow0$ instead of diverging. However, while approaching from high $T$, if $s_2$ is small, one still encounters the pairing instability coming from the first Matsubara frequency, but $T_{sc}$ is reduced as $s_2$ is increased, and beyond a certain value of $s_2$, the superconducting transition does not occur.

\subsection{$2k_F$ operator}
\label{sec:2kF}

We can also write down the analog of (\ref{PV}) for a charge density wave instability with wavevector twice the Fermi wavevector, comprising of particle-hole pairs from opposite patches of the Fermi surface\footnote{Since we are considering an operator in the particle-hole channel, non-trivial renormalizations can now occur for both real and complex $g_{ijk}$}:
\bea
&& E \Phi_{2k_F}(q,i\Omega_m) = - \sum_a \frac{M_a\zeta_a g^2}{N}T\sum_{\omega_n\neq \Omega_m}\int_k \frac{1}{i\omega_n-k_x-k_y^2-\Sigma(i\omega_n)}\frac{1}{i\omega_n+k_x-k_y^2-\Sigma(i\omega_n)} \nn
&& ~~~~~~~~~~~~~~~~~~~~~~~\times \frac{1}{\displaystyle K_a(k_y-q_y)^2+\frac{g^2}{4\pi}\frac{|\omega_n-\Omega_m|}{|k_y-q_y|}}\Phi_{2k_F}(k,i\omega_n).
\label{2kFV}
\eea
We can then see that self-consistent eigenvectors $\Phi_{2k_F}(q_y,i\Omega_m)$ exist that do not depend on $q_x$. This simplifies (\ref{2kFV}) at $T=0$ and low energies to
\bea
&& E \Phi_{2k_F}(q_y,i\Omega_m) = - \sum_a \frac{iM_a\zeta_a g^2}{2N}\int_{k_y,\omega_n}\frac{\mathrm{sgn}(\omega_n)}{k_y^2-\frac{ig^{4/3}}{2^{1/3}\pi^{2/3}\sqrt{3}}\mathrm{sgn}(\omega_n)|\omega_n|^{2/3}\left(\frac{M_1}{K_1^{2/3}N}+\frac{M_2}{K_2^{2/3}N}\right)} \nn
&& ~~~~~~~~~~~~~~~~~~~~~~~~\times\frac{|k_y-q_y|}{K_a|k_y-q_y|^3+\frac{g^2}{4\pi}|\omega_n-\Omega_m|}\Phi_{2k_F}(k_y,i\omega_n).
\label{Delta2kFclean}
\eea
We can then rescale $(k_y,q_y)\rightarrow g^{2/3}(k_y,q_y)$ to absorb the coupling $g$, producing a strong-coupling expression analogous to (\ref{Deltaclean}), given by setting $g = 1$ in (\ref{Delta2kFclean}).

We examine a scaling solution for (\ref{Delta2kFclean}) with an eigenvector of the form
\beq
\Phi_{2k_F}(q_y,i\Omega_m) = \frac{1}{|q_y|^\alpha} \Psi \left(\frac{\Omega_m}{|q_y|^3} \right),
\eeq
which will determine an eigenvalue $E(\alpha)$ \footnote{We can again see from the diagrams involved in the renormalization of the $2k_F$ vertex, that the physical eigenvector must be an even function of $q_y$}.
As with the pairing case, we are interested in eigenvalues which solve $E(\alpha) = 1$.
If the solution has $\alpha$ real, then this $\alpha$ will determine the scaling dimension of the $2 k_F$ operator. In the notation of Section~\ref{sec:scale2kF},
\beq
\mathrm{dim}[\Phi_{2 k_F}] = -\alpha \,.
\eeq
A complex $\alpha$ will indicate a charge density wave (CDW) instability.

Upon rescaling $(\omega_n,\Omega_m)\rightarrow(\omega_n|k_y|^3,\Omega_m|q_y|^3)$ and then $k_y\rightarrow k_y q_y$, we transform (\ref{Delta2kFclean}) to a one dimensional integral equation involving only the scaling function $\Psi$:
\bea
&& E \Psi(\Omega_m) = - \sum_a\frac{iM_a\zeta_a}{2N}\int_{-\infty}^{\infty}\frac{d\omega_n}{2\pi}\frac{\Psi(\omega_n)\mathrm{sgn}(\omega_n)}{\displaystyle 1-\frac{i\mathrm{sgn}(\omega_n)|\omega_n|^{2/3}}{2^{1/3}\pi^{2/3}\sqrt{3}}\left(\frac{M_1}{K_1^{2/3}N}+\frac{M_2}{K_2^{2/3}N}\right)}\nn
&&~~~~~~~~~~~~~~~\times\left[\int_{-\infty}^{\infty}\frac{dk_y}{2\pi}\frac{|k_y-1||k_y|^{1-\alpha}}{K_a|k_y-1|^3+\frac{1}{4\pi}|\omega_n|k_y|^3-\Omega_m|}\right],
\label{Delta2kFscaled}
\eea
which we can solve numerically when $0<\mathrm{Re}[\alpha]<2$, which ensures a convergent $k_y$ integral.

If we consider the physically important case of a Fermi surface coupled to a single repulsive gauge field that occurs in some U(1) spin liquids, and thereby set $M_1 = 0$, we can eliminate $K_1$ by rescaling $(\omega_n,\Omega_m)\rightarrow K_1(\omega_n,\Omega_m)$. If we additionally set the number of boson flavors $M_2$ equal to the number of fermion flavors $N$, like we have in most of this paper, then (\ref{Delta2kFscaled}) has a solution for $E = 1$ with $\alpha \approx 1 \pm 0.52i$, indicating an instability to CDW ordering. This instability persists for all $M_2 > N$. As $M_2/N$ is reduced, $\mathrm{Im}[\alpha]$ reduces, and for $M_2/N \lesssim 0.67$ ($M_2/N = 1/2$ for spin-$1/2$ U(1) spin liquids), we once again have two real roots for $E = 1$: $0<\alpha_1<1$, and $\alpha_2 = 2 - \alpha_1$, with $\alpha_1=\alpha_2=1$ at $M_2/N\approx 0.67$, and the instability disappears. An analogous argument about the IR finiteness of the ground state free energy as in Sec. \ref{sec:pairing} requires that $\mathrm{Re}[\alpha]<1$, rejecting the root $\alpha_2$, and using (\ref{scalechi2kF}), the scaling dimension of the $2k_F$ susceptibility is then
\beq
\mathrm{dim}[\chi_{2k_F}(k,\omega)] = 2(1-\alpha_1),
\eeq
in the limit of large $M_a$, $N$. For the spin-$1/2$ U(1) spin liquid, we then have the estimate from our large $N$ strongly coupled theory of $\alpha_1 \approx 0.58$, and $\mathrm{dim}[\chi_{2k_F}(k,\omega)] \approx 0.84$.

\begin{figure}[htb]
  \centering
  \includegraphics[width=0.5\textwidth]{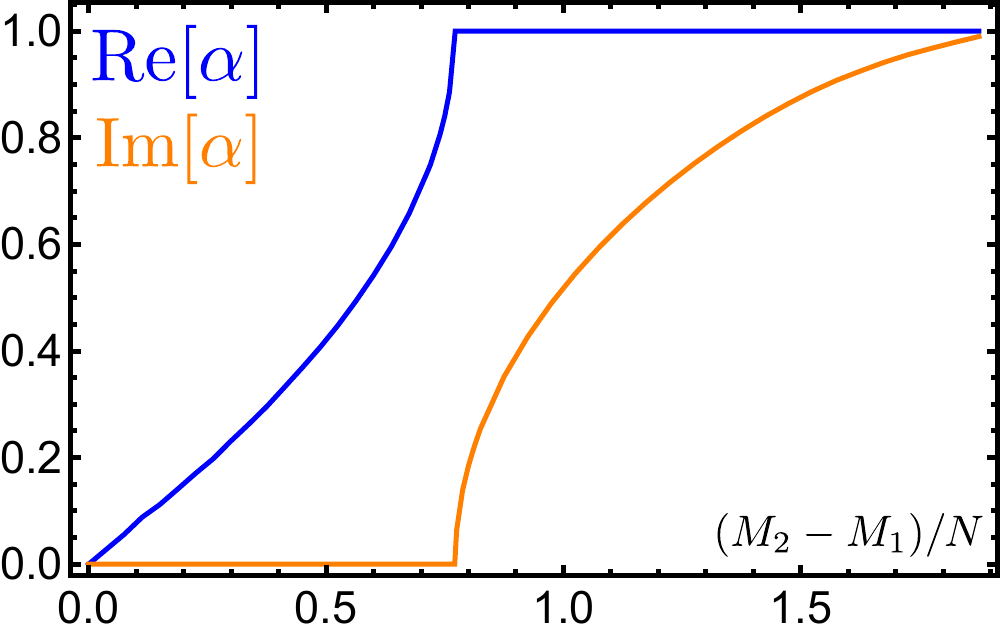}
  \caption{Plot of $\mathrm{Re}[\alpha]$ and $\mathrm{Im}[\alpha]$, for the solutions which have $\mathrm{Re}[\alpha]\leq 1$ and $\mathrm{Im}[\alpha]>0$, as a function of $(M_2-M_1)/N$, with $K_a = 1$ and $(M_1+M_2)/N = 1$. For $0<(M_2-M_1)/N \lesssim 0.77$, there is neither a $2k_F$ instability nor a pairing instability in the scaling theory.}
  \label{fig:alpha2kFplot}
\end{figure}

For a net attractive interaction between the antipodal patches, with $M_1>M_2$, there are no solutions to (\ref{Delta2kFscaled}) with $E=1$, and the scaling dimension, $\mathrm{dim}[\chi_{2k_F}(k,\omega)] = 2$, is thus not renormalized. For other combinations of $M_a,K_a,N$, CDW instabilites can occur, but there are always regimes in which there is no instability even with a net repulsive interaction. In Fig.~\ref{fig:alpha2kFplot}, we show $\alpha_1$ as a function of $(M_2-M_1)/N$ for $K_a=1$ and $(M_1+M_2)/N = 1$, demonstrating this. In particular, reducing the value of $M_2-M_1$ and increasing the value of $N$ both disfavor CDW instabilities, and vice versa.

We can also consider the analog of (\ref{Delta2kFscaled}) for arbitrary $2<z\le3$, as we did in Sec. \ref{sec:pairing}. With a net repulsive interaction, we find that CDW instabilities are disfavored as $z\rightarrow 2$, with $\alpha_1\rightarrow 0$ as $z\rightarrow2$ \cite{mross}, for all values of $M_a,K_a,N$, and favored as $z\rightarrow 3$, although whether or not $\alpha_1$ actually manages to reach $1$ and then move into the complex plane the as $z\rightarrow 3$ depends on the values of $M_a,K_a,N$.

At $T\neq0$, our neglect of the thermal gauge boson fluctuations and the vanishing of the fermion self energy at the first Matsubara frequencies causes the eigenvalue $E$ in (\ref{2kFV}) to diverge as $\sim T^{1/z}$ as $T\rightarrow0$, which causes a $2k_F$ CDW instability for any net repulsive interaction. Therefore, we must carefully consider the effects of the thermal fluctuations of the massless gauge bosons $\phi$. As in the pairing case of Sec. \ref{sec:pairing}, we will consider in detail the effects of the thermally fluctuating gauge boson modes using a gauge invariant formalism in future work.

In the case where the two $\phi$'s are not gauge bosons and are therefore allowed to have a thermal mass $M^2(T)\sim T\ln(1/T) \ll T^{2/z}$, the eigenvalue $E$ does not diverge as $T\rightarrow0$, and one can then have a stable regime for net repulsive interactions even at finite $T$, depending on parameter values.

\section{Spatially Disordered Model}
\label{sec:spatial}

This section will consider a generalization of the model (\ref{eq:latticeaction}) to the case where the couplings $g_{ijl}$ are also random functions of position. This results in a theory in which the non-Fermi liquid effects are weaker, and we obtain a large $N$ expansion of a marginal Fermi liquid. The properties of this marginal Fermi liquid are similar to those studied recently in Ref.~\cite{Altman1} for a different model.

Taking the (complex) Yukawa coupling $g_{ijl}$ in \eqref{eq:latticeaction} to be a Gaussian random in space as well, which satisfies  $g_{ijl}(x)=g_{jil}(x)^*$ and
\begin{equation}\label{}
\left\langle g_{ijk}(x) g_{i'j'k'}(x')^*\right\rangle=g^2\delta(x-x')\delta_{ii'}\delta_{jj'}\delta_{kk'}.
\end{equation}
After performing a disorder average, and inserting self-energies as Lagrange multipliers, we obtain the action
\begin{equation}\label{}
  \begin{split}
     S & =\int \rd\tau\rd^2 x \int \rd \tau'\rd^2 x' \\
       & \sum_i \psi_i^\dagger(\tau,x)(\partial_\tau+\varepsilon_k-\mu)\delta(\tau-\tau')\delta(x-x')\psi_i(\tau',x') \\
       & +\frac{1}{2}\sum_i \phi_i(\tau,x)(-\partial_\tau^2+\omega_q^2+i\lambda(x,\tau))\delta(\tau-\tau')\delta(x-x')\phi_i(\tau',x') \\
       & -\Sigma(\tau',x';\tau,x)\left(N G(\tau,x;\tau',x')+\sum_i \psi_i(\tau,x)\psi_i^\dagger(\tau',x')\right)\\
        &-\frac{1}{2}\Pi(\tau',x';\tau,x)\left(-N D(\tau,x;\tau',x')+\sum_i \phi_i(\tau,x)\phi_i(\tau',x')\right) \\
        & +\frac{g^2 N}{2}G(\tau',x';\tau,x)G(\tau,x;\tau',x')D(\tau,x,\tau',x')\delta(x-x')\\
&- \frac{iN}{2\gamma}\lambda(\tau,x)\delta(x-x')\delta(\tau-\tau')
  \end{split}
\end{equation}
The difference from the translationally invariant model is the extra $\delta$-function in the $g^2$ term; consequently the $G$, $\Sigma$, $D$, $\Pi$ fields are now only bilocal in time, and not bilocal in space. The kinetic terms in the first two lines are differential operators that act on $(\tau,x)$. 
Integrating out $\psi$ and $\phi$, we obtain the $G$-$\Sigma$ action
\begin{equation}\label{}
  \begin{split}
     \frac{S}{N} & = -\ln\det(\partial_\tau+\varepsilon_k-\mu+\Sigma)+\frac{1}{2}\ln\det(-\partial_\tau^2+\omega_q^2+i\lambda-\Pi) \\
       & -\int \rd\tau\rd^2 x \int \rd \tau'\rd^2 x'\left(\Sigma(\tau',x';\tau,x)G(\tau,x;\tau',x')-\frac{1}{2}\Pi(\tau',x';\tau,x)D(\tau,x;\tau',x')\right) \\
       & +\int \rd\tau\rd^2 x \int \rd \tau'\rd^2 x' \frac{g^2}{2}G(\tau,x;\tau',x')G(\tau',x';\tau,x)D(\tau,x;\tau',x')\delta(x-x') \\
       & -\int \rd \tau \rd^2 x \frac{i \lambda(\tau,x)}{2\gamma}.
  \end{split}
\end{equation}
The saddle point equations are (assuming $\lambda$ is constant)
\begin{eqnarray}
  G(\tau,x;\tau',x') &=& \left(\frac{1}{-\partial_\tau+\mu-\varepsilon_k-\Sigma}\right)_{\tau,x;\tau',x'}, \\
  D(\tau,x;\tau',x') &=& \left(\frac{1}{-\partial_\tau^2+\omega_q^2+i\lambda-\Pi}\right)_{\tau,x;\tau',x'},  \\
  \Sigma(\tau,x;\tau',x') &=& \frac{g^2}{2} G(\tau,x;\tau',x')\left[D(\tau,x;\tau',x')+D(\tau',x';\tau,x)\right]\delta(x-x'), \\
  \Pi(\tau,x;\tau',x') &=& -g^2 G(\tau,x;\tau',x')G(\tau',x';\tau,x)\delta(x-x'),\\
  D(0,0;0,0) &=& \frac{1}{\gamma}.
\end{eqnarray}
Unlike the translationally invariant case, the self-energies are momentum independent, and we obtain the following reduced set of equations ($m_b^2=i\lambda$)
\begin{eqnarray}
  \bar{G}(i\omega) &=& \int\frac{\rd^2 k}{(2\pi)^2}\frac{1}{i\omega+\mu-\varepsilon_k-\Sigma(i\omega)}, \label{eq:saddle_kspace1}\\
  \bar{D}(i\nu) &=& \int\frac{\rd^2 q}{(2\pi)^2}\frac{1}{\nu^2+\omega_q^2+m_b^2-\Pi(i\nu)},\label{eq:saddle_kspace2} \\
  \Sigma(\tau) &=& \frac{g^2}{2}\bar{G}(\tau)\left[\bar{D}(\tau)+\bar{D}(-\tau)\right], \\
  \Pi(\tau) &=& -g^2 \bar{G}(\tau)\bar{G}(-\tau),\label{eq:saddle_kspace4}\\
  \frac{1}{\gamma}&=&T\sum_{\nu} \int\frac{\rd^2 q}{(2\pi)^2}\frac{1}{\nu^2+\omega_q^2+m_b^2-\Pi(i\nu)} \label{eq:saddle_constraint}
\end{eqnarray}
We will introduce momentum space cutoffs $\Lambda_k$ for the fermions and $\Lambda_q$ for bosons. They have dimension of energy. $g$ has dimension $[\text{energy}]^{-1/2}$, hence $g^2\Lambda_k$ is dimensionless. $\gamma$ has dimension $[\text{energy}]^{-1}$.

We investigate Eqs.~\eqref{eq:saddle_kspace1}-\eqref{eq:saddle_kspace4} in the patch theory, {\it i.e.\/} setting $\varepsilon_k-\mu=k_x+k_y^2$ and $\omega_q^2=q_x^2+q_y^2$. In this model the two components of the boson momentum scale the same way so we have to retain both. Assuming that the bandwidth is the largest energy scale, we can perform the integrals in \eqref{eq:saddle_kspace1} and \eqref{eq:saddle_kspace2}, which lead to
\begin{eqnarray}
  \bar{G}(i\omega) &=& \Lambda_k\frac{-i\sgn \omega}{2}, \label{eq:barG} \\
  \bar{D}(i\nu) &=& \frac{1}{4\pi}\ln\left(\frac{\nu^2-\bar{\Pi}(i\nu)+\Lambda_q^2}{\nu^2-\bar{\Pi}(i\nu)+\Delta(T)^2}\right),
\end{eqnarray}
where $\Lambda_k=\int{\rd k_y}/({2\pi})$ and the boson propagator is evaluated with Pauli-Villars regulator with cutoff $\Lambda_q$. Here we have subtracted off the zeroth Matsubara frequency from $\Pi$ by defining $\bar{\Pi}(i\nu)=\Pi(i\nu)-\Pi(0)$, and we have also rewritten $m_b$ using the thermal mass $\Delta(T)^2=m_b^2-\Pi(0)$.

At zero temperature, Eq.\eqref{eq:barG} yields
\begin{equation}\label{}
  \bar{G}=-\frac{\Lambda_k}{2\pi\tau},
\end{equation}
and it follows from saddle point equations that
\begin{equation}\label{}
  \Pi(\tau)=\left(\frac{g\Lambda_k}{2\pi\tau}\right)^2.
\end{equation}
To compute $\Pi$ in frequency space, we use the frequency space version of \eqref{eq:saddle_kspace4}:
\begin{equation}\label{}
  \Pi(i\nu)=-g^2 T\sum_{\omega_n}G(i\omega_n)G(i\omega_n+i\nu).
\end{equation}
We subtract off the zeroth Matsubara frequency
\begin{equation}\label{}
  \bar{\Pi}(i\nu)\equiv\Pi(i\nu)-\Pi(0)=\left(\frac{g\Lambda_k}{2}\right)^2 T\sum_{\omega_n}(\sgn(\omega_n)\sgn(\omega_n+\nu)-1)=-\pi|\nu|\left(\frac{g\Lambda_k}{2\pi}\right)^2.
\end{equation}

\subsection{Thermal Mass}

\begin{figure}[htb]
  \centering
  \includegraphics[width=0.5\textwidth]{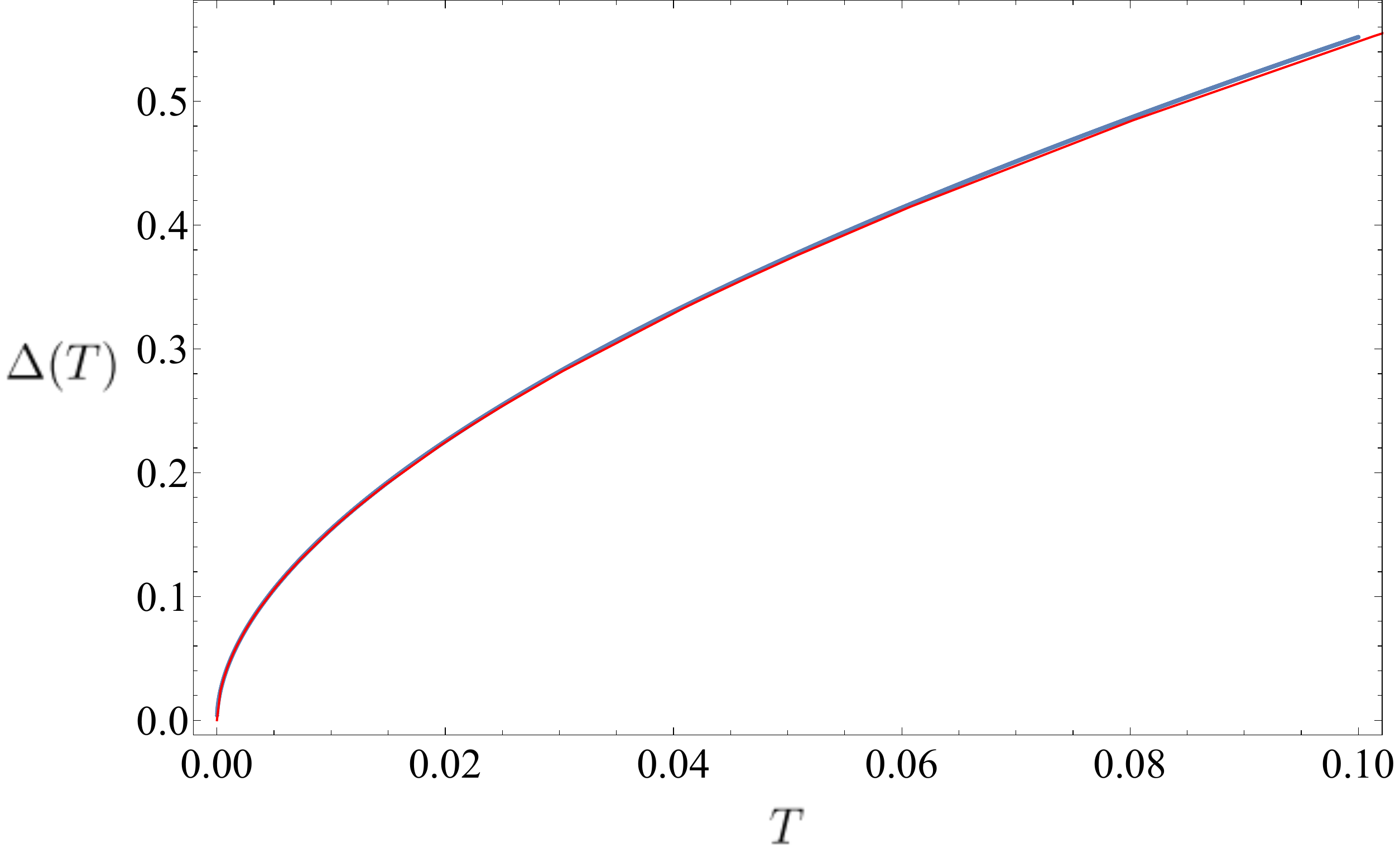}
  \caption{Plot of $\Delta(T)$. $\Delta(0)=0$, $a_0=5$, $\Lambda_q=300$. Blue line is the numerical solution of the thermal mass from  \eqref{eq:DetermineDeltaT1}. Red line is the low temperature asymptotics Eq.~\eqref{eq:Deltaasym}.}\label{fig:thermalmass}
\end{figure}

Using \eqref{eq:saddle_constraint}, we can determine the low-temperature asymptotics of the thermal mass at criticality to be (for details, see Appendix.~\ref{app:spatial})
\begin{equation}\label{eq:Deltaasym}
  \Delta(T)^2=\frac{\displaystyle -\pi a_0 T W_0\left(-\frac{1}{\pi}\ln\left(\frac{2\pi T }{a_0 e^{\gamma}}\right)\right)}{\ln\left( \displaystyle \frac{2\pi T }{a_0 e^{\gamma}}\right)},
\end{equation} where $W_0$ is the principle Lambert W-function and $\gamma$ is Euler's constant. Here $a_0=\pi\left({g \Lambda_k}/({2\pi})\right)^2$. The asymptotic behavior of $W_0(x)$ is
$$
W_0(x\to \infty)\sim \ln x-\ln\ln x\,,
$$
This result indicates that $\Delta(T)\to 0$ as $T\to 0$ slightly faster than $\sqrt{T}$ by some log corrections. A plot of $\Delta(T)$ is given in Fig.~\ref{fig:thermalmass}.

\subsection{Fermion Self Energy}

The electron self-energy is
\begin{equation}\label{eq:Sigma=GD}
  \begin{split}
      \Sigma(i\omega)&=g^2 T\sum_{\mu}\bar{G}(\omega+\mu)\bar{D}(\mu)\\
      &=\frac{-i g^2 \Lambda_k}{4\pi} \sgn(\omega)T\left[\frac{1}{2}\ln\left(\frac{\Lambda_q^2}{\Delta(T)^2}\right)+\sum_{0<\mu<|\omega|}\ln\left(\frac{\mu^2+|\mu|a_0+\Lambda_q^2}{\mu^2+|\mu|a_0+\Delta(T)^2}\right)\right],
  \end{split}
\end{equation}where the sum cancels in pair when $|\mu|>|\omega|$.
The sum can be performed exactly

\begin{equation}\label{}
  \Sigma(i\omega)=\frac{-ig^2 T\Lambda_k}{4\pi}\sgn(\omega)\ln\left[\frac{\Lambda_q P\left(1+\frac{a_0+i\sqrt{4\Lambda_q^2-a_0^2}}{4\pi T},\frac{|\omega|-\pi T}{2\pi T}\right)P\left(1+\frac{a_0-i\sqrt{4\Lambda_q^2-a_0^2}}{4\pi T},\frac{|\omega|-\pi T}{2\pi T}\right)}{\Delta(T)P\left(1+\frac{a_0+\sqrt{a_0^2-4\Delta(T)^2}}{4\pi T},\frac{|\omega|-\pi T}{2\pi T}\right)P\left(1+\frac{a_0-\sqrt{a_0^2-4\Delta(T)^2}}{4\pi T},\frac{|\omega|-\pi T}{2\pi T}\right)}\right]
\end{equation}
where $P(a,b)=\Gamma(a+b)/\Gamma(a)$ is the Pochhammer function. Some limiting cases of $\Sigma(i\omega)$, showing marginal Fermi liquid behavior, are
\begin{equation}\label{}
\begin{split}
  \Sigma(|\omega|\ll T, a_0)&=\frac{-i g^2 T \Lambda_k}{4\pi}\sgn(\omega)\ln \left[\frac{2\pi T}{\Delta(T)}P\left(\frac{1}{2}+\frac{a_0+\sqrt{a_0^2-4\Delta(T)^2}}{4\pi T},\frac{1}{2}\right)\right.\\
  &\times \left.P\left(\frac{1}{2}+\frac{a_0-\sqrt{a_0^2-4\Delta(T)^2}}{4\pi T},\frac{1}{2}\right)\right]\,,
\end{split}
\end{equation}
\begin{equation}
    \Sigma(T\ll |\omega|\ll a_0\ll \Lambda_q)=\frac{-i g^2 T \Lambda_k}{4\pi}\sgn(\omega)\left[\frac{|\omega|}{2\pi T}\left(1+\ln\frac{\Lambda_q^2}{|\omega|a_0}\right)+\frac{1}{2}\ln\frac{a_0 T}{\Delta(T)^2}\right]\,,
\end{equation}
\begin{equation}\label{}
\begin{split}
  \Sigma(T,a_0\ll |\omega|\ll\Lambda_q)=\frac{-i g^2 T \Lambda_k}{4\pi}\sgn(\omega)\left[\frac{|\omega|}{\pi T}\left(1+\ln\frac{\Lambda_q}{|\omega|}\right)-\frac{a_0}{2\pi T}\ln\frac{|\omega|}{2\pi T}+\sigma_0(\Delta(T),T)\right]\,.
\end{split}
\end{equation}
Here $\sigma_0$ is retained to ensure a finite $T\to 0$ limit
\begin{equation}
    \sigma_0(\Delta(T),T)=\ln\frac{T}{\Delta(T)}+\ln\Gamma\left(1+\frac{a_0-\sqrt{a_0^2-4\Delta(T)^2}}{4\pi T}\right)+\ln\Gamma\left(1+\frac{a_0+\sqrt{a_0^2-4\Delta(T)^2}}{4\pi T}\right)\,.
\end{equation}
Notice that in the $\omega\to\infty$ limit, the sum in \eqref{eq:Sigma=GD} is nothing but the constraint \eqref{eq:DeltaTfinite},therefore we have
\begin{equation}\label{}
  \Sigma(|\omega|\gg \Lambda_q)=\frac{-i \Lambda_k}{2}\sgn(\omega)\frac{g^2}{\gamma}.
\end{equation}

\subsection{Free Energy}

The free energy $F$ is given by the value of saddle point action
\begin{equation}\label{}
  \begin{split}
     \frac{\beta F}{N} & = -\ln\det(\partial_\tau+\varepsilon_k-\mu+\Sigma)+\frac{1}{2}\ln\det(-\partial_\tau^2+\omega_q^2+i\lambda-\Pi) \\
       & -\int \rd\tau\rd^2 x \int \rd \tau'\rd^2 x'\left(\Sigma(\tau',x';\tau,x)G(\tau,x;\tau',x')-\frac{1}{2}\Pi(\tau',x';\tau,x)D(\tau,x;\tau',x')\right) \\
       & +\int \rd\tau\rd^2 x \int \rd \tau'\rd^2 x' \frac{g^2}{2}G(\tau,x;\tau',x')G(\tau',x';\tau,x)D(\tau,x;\tau',x')\delta(x-x')\\
&-\int \rd \tau \rd^2 x \frac{i \lambda(\tau,x)}{2\gamma},
  \end{split}
\end{equation} where the variables should be substituted by their saddle point values. The complete evaluation of the above free energy is given in Appendix.~\ref{app:spatial} and we summarize the result here.

\begin{figure}[htb]
  \hspace*{-1.8cm}
  \includegraphics[width=0.6\textwidth]{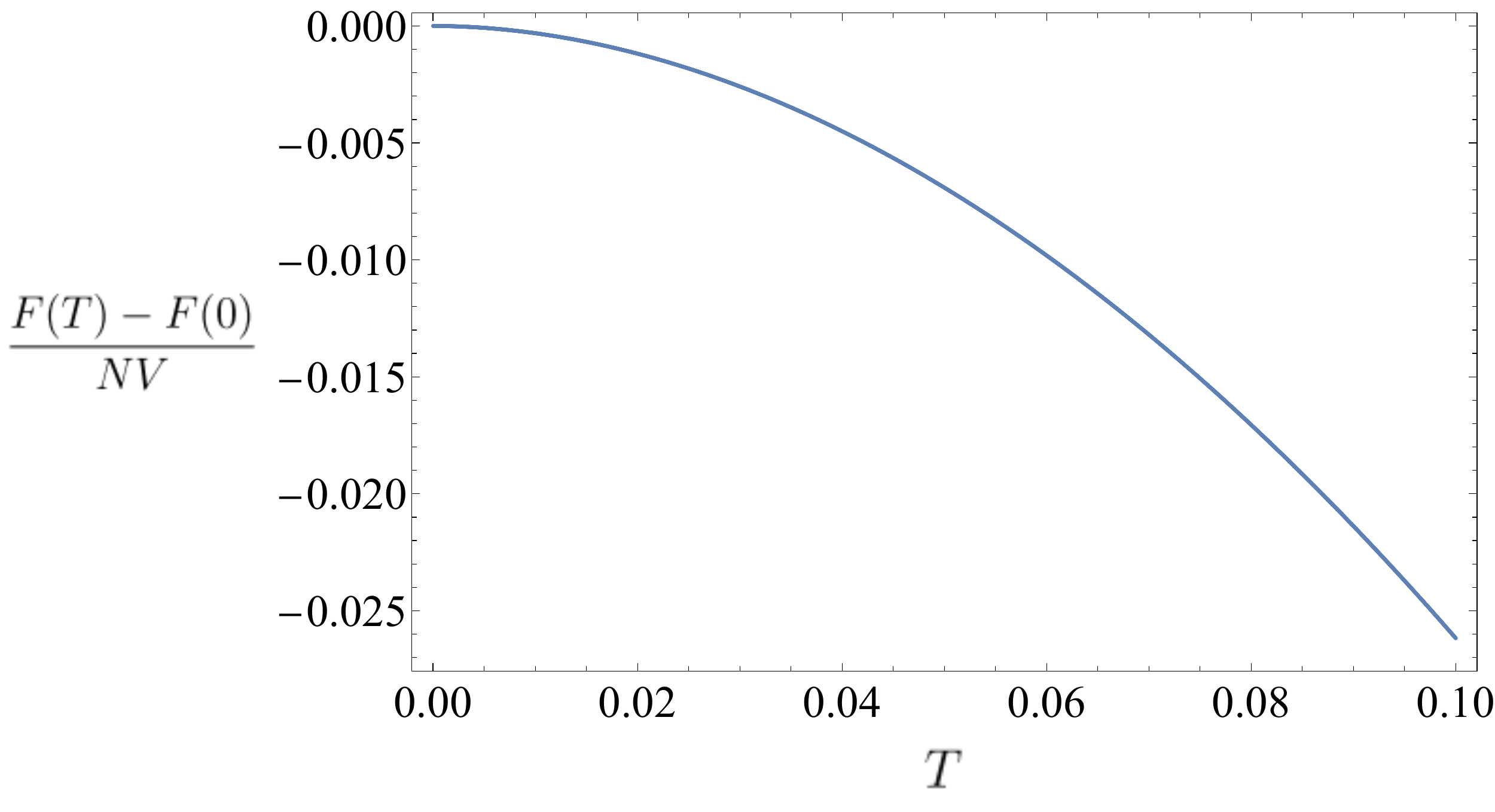}
  \caption{Plot of bosonic contribution to free energy $(F/NV)_{2}$ as a function of $T$. The zero temperature part is subtracted. $\Delta(0)=0$, $a_0=5$, $\Lambda_q=300$. The total free energy also contains a free fermion contribution Eq.\eqref{eq:Ffermi} which is not shown here.}\label{fig:fenergy}
\end{figure}

The free energy contains two contributions $F=F_1+F_2$. The first part is the free energy of free fermion
\begin{equation}\label{eq:Ffermi}
  \frac{F_1}{NV}=-T \Lambda_k \int \frac{\rd k_x}{2\pi}\ln(1+e^{-\beta k_x})\,,
\end{equation} where $V$ is the spatial volume of the system and $\beta=1/T$. The second part $F_2$ is the contribution due to interacting bosons. It has a lengthy analytic expression in Appendix.~\ref{app:spatial}, and the numerical plot is given in Fig.~\ref{fig:fenergy}

\begin{figure}[htb]
  \centering
  \includegraphics[width=0.5\textwidth]{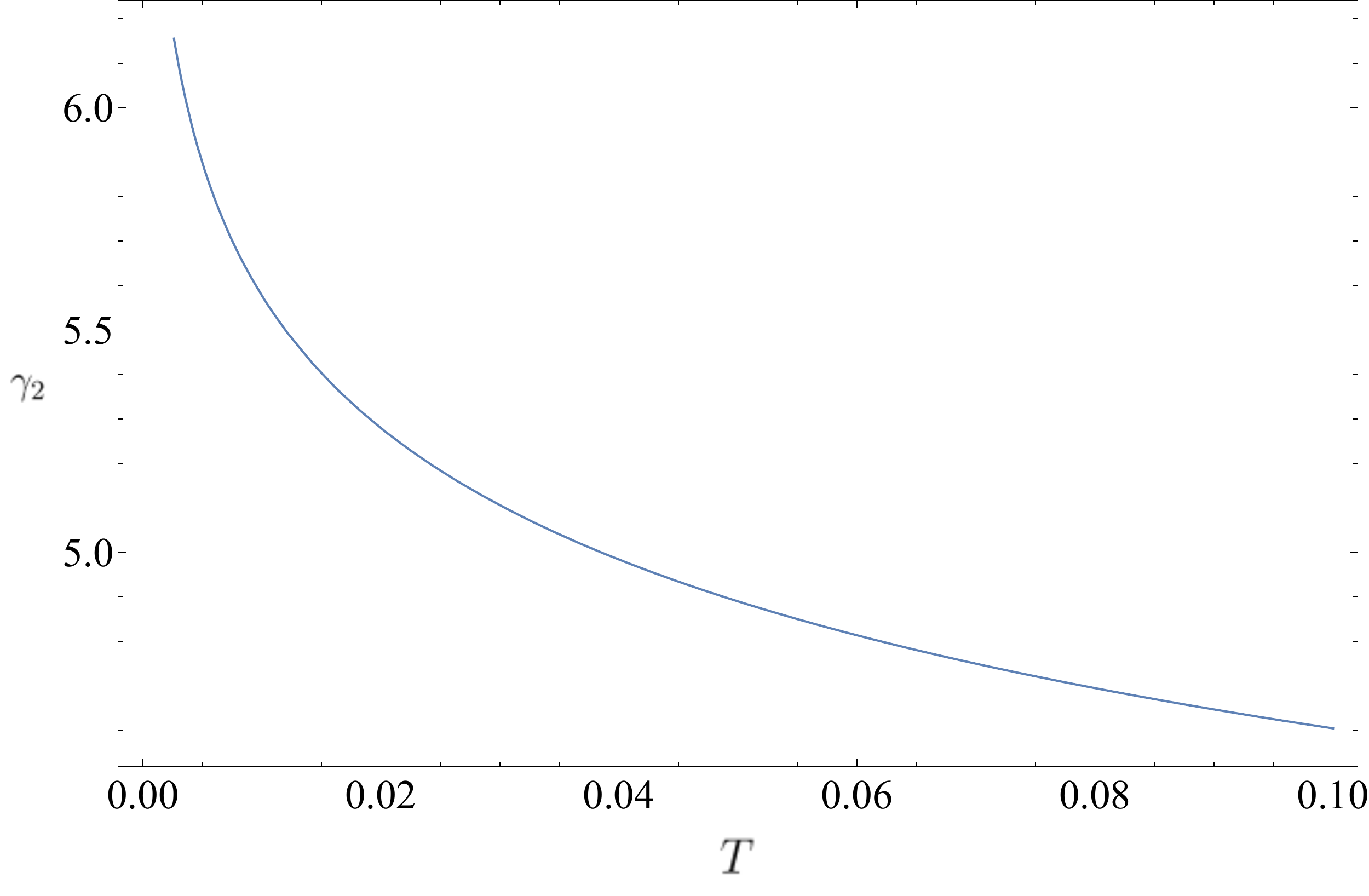}
  \caption{Plot of bosonic contribution to heat capacity $C_2/(NV)=-T/(NV)(\partial^2 F_2/\partial T^2)=T\gamma_2(T)$. $\Delta(0)=0$, $a_0=5$, $\Lambda_q=300$. The total heat capacity is the sum of two contributions $C=NVT(\gamma_1+\gamma_2)$, where $\gamma_1=(\pi/6)\Lambda_k$ takes the free fermion value.}\label{fig:Cv}
\end{figure}

Consequently the heat capacity can be written as $C=NV(\gamma_1+\gamma_2)T$, which corresponds to contributions from $F_1$ and $F_2$ respectively. Here $\gamma_1=(\pi/6)\Lambda_k$, and $\gamma_2$ is plotted in Fig.~\ref{fig:Cv}.

\subsection{Conductivity}
\label{sec:planckian}

The computation of the conductivity in the spatially disordered model is very similar to that in Ref. \cite{Altman1}. In particular, the conductivity $\sigma_\mathrm{DC}$ is governed by the scattering rate set by the imaginary part of the retarded fermion self energy, as vertex corrections vanish due to the isotropic momentum-independent scattering of the fermions off the bosons. The quantity of interest is\footnote{The transport scattering rate is actually set by averaging the lifetime over a frequency range $\sim T$ \cite{Patel2018mag,Altman1}, but, in this case, that only makes a small differnce in its numerical value vs. just using the zero frequency lifetime.}
\beq
\frac{1}{\tau_\mathrm{tr}} \approx -\mathrm{Im}[\Sigma^R(\omega = 0, T\neq 0)] = \frac{g^2\Lambda_k}{8\pi} T L_1,
\label{taubare}
\eeq
where $L_1 \sim \ln (2\ln (a_0e^\gamma/(2\pi T))/\pi)$ is very slowly varying and we therefore just treat it as an $\mathcal{O}(1)$ constant. We have set Planck's and Boltzmann's constants $\hbar = k_B = 1$ so far, but will restore them below.

There have been several experimental claims of ``Planckian dissipation'' in the recent literature \cite{Legros18,Paglione19,Cao2020} occuring at putative QCPs with linear-in-$T$ resistivity in correlated electron materials. The meaning of this statement is that the transport scattering rate defined with respect to an interaction-renormalized effective mass $m^\ast$ is $1/\tau^\ast_\mathrm{tr} \approx k_B T/\hbar$. Therefore
\beq
\frac{1}{\tau^\ast_\mathrm{tr}} = \frac{n e^2}{\sigma_{\mathrm{DC}}m^\ast} = \frac{n e^2}{\sigma_{\mathrm{DC}}m}\frac{m}{m^\ast} = \frac{1}{\tau_\mathrm{tr}}\frac{m}{m^\ast} \approx \frac{k_B T}{\hbar},
\eeq
where $m$ is the bare electron mass, $e$ is the electron charge and $n$ is the density of electrons.

There is also a theoretical argument for the consideration of $\tau^\ast_{\mathrm{tr}}$ as the appropriate time: assuming a momentum independent self energy, $1/\tau_{\mathrm{tr}}$
is related to the imaginary part of the electron self-energy at zero frequency, $\Sigma (0)$, and so its scaling behavior is tied up with the scaling dimension of the electron operator. Only by computing the ratio $\Sigma(0)/(\partial \Sigma/\partial \omega)$ do we obtain a quantity which scales with frequency alone, and this yields $1/\tau^\ast_{\mathrm{tr}}$.

The effective mass $m^\ast$ is determined somewhat away from the QCP in a Fermi liquid regime, where the electron quasiparticle is well defined at low energies, using quantum oscillation measurements, specific heat measurements or measurements of the fermion dispersion near the Fermi surface via angle-resolved photoemission spectroscopy (ARPES). In our model we have, from the fermion propagator in the Fermi liquid phase that arises for $\gamma>\gamma_c$
\beq
\frac{m^\ast}{m} \equiv 1 + i\frac{\partial\Sigma(i\omega)}{\partial\omega}\Bigg|_{\omega\rightarrow0, T\rightarrow0} = 1 + \frac{g^2\Lambda_k}{4\pi^2}L_2,~~L_2 \sim \ln\left(\frac{\Lambda_q}{\sqrt{a_0}}\sqrt{\frac{\gamma\gamma_c}{\gamma-\gamma_c}}\right) \sim \mathcal{O}(1).
\eeq
We consider a regime of strong coupling, where $g^2\Lambda_k\gg 1$. Our results in prior subsections about the various quantities at the QCP then remain valid as long as we restrict ourselves to energy scales $\ll 1/g^2$, where we consider $\Lambda_k\sim \Lambda_q$, as the boson and fermion self energies then remain smaller than their respective bandwidths. Then $m^\ast/m\approx g^2\Lambda_k/(4\pi^2)$. Putting everything together, we then have
\beq
\frac{1}{\tau^\ast_\mathrm{tr}} \approx \frac{\pi}{2} \frac{k_B T}{\hbar} \frac{L_1}{L_2},
\eeq
which is $O(k_B T/\hbar)$, {\it i.e.} ``Planckian''. In fact, some of the $\mathcal{O}(1)$ variations in the measured prefactor in the experimental results may be attributable to where $m^\ast$ is measured relative to the QCP, as that will introduce $\mathcal{O}(1)$ variations in $L_2$, and at what temperatures the measurements are carried out, as that will introduce $\mathcal{O}(1)$ variations in $L_1$.

\subsection{Instabilities}
\label{sec:disinstab}

For complex random $g_{ijk}$, this model does not have any pairing instability at large $N$, as is also the case for the translationally invariant model. However, for real random $g_{ijk}$, there is a pairing instability at low energies, and following the methods of Sec. \ref{sec:pairing}, we estimate the superconducting transition temperature $T_{sc}\sim (\Lambda_q^2/(g^2\Lambda_k^2))e^{-1/(g^2\Lambda_k)}$. This can be appreciably large in the strong coupling regime $g^2\Lambda_k\gg 1$, in which case we also expect superconductivity with real random $g_{ijk}$ to set in at parametrically around the same scale as the Planckian behavior sets in with complex random $g_{ijk}$, and the Planckian behavior may therefore be completely obscured by superconductivity.

Following the analysis in Sec. V D of Ref. \cite{Chowdhury:2018sho}, we can see that the scaling dimension of the $2k_F$ vertex is not renormalized by the momentum independent scattering of fermions. Therefore, there is no $2k_F$ CDW instability at large $N$.

\section{Conclusions}

We have shown that a model with random Yukawa couplings provides a large $N$ theory of a critical Fermi surface. Many existing results are unified in a
systematic perspective, and a formalism is now available to determine $1/N$ corrections.

The primary critical field characterizing the critical Fermi surface is a fermion $\psi$ with anomalous dimension $\eta_\psi$. Its correlations on a single patch of the Fermi surface are characterized by a dynamic scaling exponent $z$, and anisotropic scaling along the spatial directions perpendicular ($x$) and parallel ($y$) to the Fermi surface (see Fig.~\ref{fig:patch}). We define scaling dimensions with the choice $\mathrm{dim}[q_y] = 1$, and then a sliding symmetry of the Fermi surface \cite{metlitski1} implies $\mathrm{dim}[q_x] = 2$. The large $N$ theory has $z=3$ and $\eta_\psi = 0$. A three-loop analysis \cite{metlitski1} found no correction to $z=3$, and a small non-zero value for $\eta_\psi$. In our approach the expansion for $\eta_\psi$ is systematic in $1/N$, and contained in the infinite set of graphs in Figs.~\ref{fig:Xidiagrams} and~\ref{fig:sheetandseam}, which include the graphs in
Ref.~\cite{metlitski1}.
Further loop corrections to the RPA theory have been studied in Refs.~\cite{HolderMetzner1,HolderMetzner2}, and it would be interesting to examine the consequences of bilocal field propagators required by our $1/N$ expansion: it is possible that the scaling described in our analysis will prevail.

Starting from the primary field $\psi$,
we can now build composite operators, as in the SYK model. In the single patch theory of Fig.~\ref{fig:patch}, in the particle-hole sector, we found only the conserved density operators that have been studied in Ref.~\cite{Kim94}. The saddle-point action does have time reparameterization symmetry in the scaling limit, but we showed that, unlike the SYK model, this did not translate into a singular time reparameterization mode because of the non-trivial action of the time reparameterization on the spatial co-ordinates. So there is no corresponding expected violation of scaling here at frequencies of order $1/N$, in contrast to the SYK model \cite{JMDS16,Kitaev:2017awl,Bagrets:2016cdf,Bagrets:2017pwq,Altland:2019czw,Kruchkov:2019idx,Kobrin:2020xms}. The particle-particle sector of the single patch theory is also where the Amperean pairing operator \cite{LeeAmperean1,LeeAmperean2} resides, and we discuss it in Appendix~\ref{app:amperean}.

In a non-chiral system, we have to also consider the role of antipodal patches on the Fermi surface, as in Fig.~\ref{fig:2patch}. In this case, interesting composite operators do arise from fermions on opposite patches, in both the particle-particle and particle-hole sectors.
In the particle-particle sector, we have the Cooper pair operator, and its analysis in our $N=\infty$ theory reduces to that of the $\gamma$ model of Chubukov and collaborators \cite{Moon2010,Chubukov1,Chubukov2,Chubukov3}. Section~\ref{sec:pairing} obtained results for the scaling dimension of the Cooper pair operator for the case where there are multiple scalars coupled to the fermions with both attractive and repulsive interactions, as is needed for the models of Refs.~\cite{YZSS1,LZDC20,YZSS2}.

In the particle-hole sector of the antipodal patch theory, we have the operator associated with charge density waves at the $2k_F$ wavevector. This has been studied earlier by Mross {\it et al.\/} \cite{mross}. Our large $N$ theory leads to integral equations in frequency and momentum, which we numerically solved in the scaling limit in Section~\ref{sec:2kF}. These solutions led to a rich set of possibilities for the scaling dimension of the $2k_F$ density wave operator.

In Section~\ref{sec:num} we presented numerical solutions of the large $N$ saddle point equations while keeping the full Fermi surface in a convenient lattice regularization. At the QCP, we found good agreement with predictions of the low-energy patch theory for the scaling behavior of the fermion and boson Green's functions. Even away from the QCP, the large-$N$ phase diagram is interesting in its own right, where we found the ordered side is characterized by a rapid onset of strong thermal fluctuations, in which $M \ll T$ and the boson is essentially static, behaving in a manner similar to quenched disorder for the fermions. We also note that we have not found a superconducting transition down to the lowest accessible temperatures. However, the superconducting $T_c$ should be finite (it can likely be accessed in numerical calculations by increasing the coupling strength) and it will be interesting to study the nature of superconductivity across the phase diagram presented here.

Finally, in Section~\ref{sec:spatial} we presented a large $N$ theory for a marginal Fermi liquid, obtained by considering a Yukawa coupling which was random in both flavor and position space. The results here are similar to Aldape {\it et al.}~\cite{Altman1} for a different model: there is a nearly linear-in-$T$ contribution to the imaginary self part of the energy of the fermion, and Planckian transport, as described in Section~\ref{sec:planckian}.

We close with some general remarks about `Eliashberg theory', a framework used to solve a variety of problems in condensed matter physics involving the coupling of electrons to a boson with a Yukawa-type coupling \cite{Ilya1,Ilya2,Wang:2020dtj,Marsiglio20,DCBerg,DebanjanAPS}. Two long-standing questions with this framework have been: is there a general systematic expansion whose saddle-point is the Eliashberg theory, and what are the systematic corrections to Eliashberg theory?
We stress the importance of a systematic framework, because only then can we ensure proper treatments of symmetries {\it and} anomalies required for Luttinger-like theorems \cite{ElseSenthil1}.
For problems without spatial randomness, the answer from recent works \cite{Altman1,Kim:2020jpz} and the present paper is that Eliashberg theory is the large $N$ saddle point of a theory in which the Yukawa coupling is a random function of indices in flavor space.
Corrections to this saddle-point are obtained from a $G$-$\Sigma$ theory which is, in general, bilocal in spacetime. The propagators of the bilocal fields resum infinite sets of diagrams in the underlying theory, such as those in Fig.~\ref{fig:Xidiagrams} and~\ref{fig:sheetandseam}.
All of this analysis has close connections to random models in the SYK class \cite{Ilya1,Ilya2,Wang:2020dtj}, which realize the simpler case with $G$-$\Sigma$ fields bilocal only in time. Numerical studies of the models in the SYK class (see {\it e.g.\/} Refs.~\cite{ZiYangMeng20,Kobrin:2020xms,ZiYangMeng21}) have tested the predictions of such large $N$ theories and shown that they are quite accurate at finite $N$.

The structure of our saddle-point equations also have similarities to those of extended dynamical mean field theory \cite{Sengupta95,Smith2000,Haule02,Haule03,Haule07}, which become exact in the limit of large dimensions. Note that their self energies are momentum independent, similar to those in Section~\ref{sec:spatial} in the model with spatial disorder. It would be interesting to extend our methods to obtain systematic corrections to dynamical mean field theories without introducing spatial disorder.

\subsection*{Acknowledgements}

We thank Andrey Chubukov, Sung-Sik Lee, Alex Maloney, and Grigory Tarnopolsky for useful discussions. This research was supported by the National Science Foundation under Grant No.~DMR-2002850. A.A.P. was supported by the Miller Institute for Basic Research in Science. I.~E. acknowledges support from the Harvard Quantum Initiative Postdoctoral Fellowship in Science and Engineering. This work was also supported by the Simons Collaboration on Ultra-Quantum Matter, which is a grant from the Simons Foundation (651440, S.S.).

\appendix

\section{Amperean pairing}
\label{app:amperean}

In this appendix, we consider the fate of the ``Amperean pairing'' instability within a single patch of the critical scaling theory for real random $g_{ijk}$. This instability was proposed by Ref. \cite{LeeAmperean1}, and involves finite momentum pairing of fermions near the same point of the Fermi surface, given by the operator $\Psi_A=\sum_{k,\omega_n}\psi_+(k,i\omega_n)\psi_+(-k,-i\omega_n)$, where both fermions belong to the same patch. Ref. \cite{LeeAmperean1} argued that since the computation of the correlation function $\chi_A = \langle \Psi_A \Psi_A \rangle$ for free fermions (with the frequency integration done first, as performing the $k_x$ integral first incorrectly returns zero) involves momenta with $|k_x|<k_y^2$, the computation in the interacting case must be similar, and they imposed a hard cutoff of $k_y^2$ on the $k_x$ integrals in their analysis of the Amperean pairing instability. However, due to the destruction of the quasiparticle by the $\omega^{2/z}$ self energy, the cutoff on $k_x$ does not turn out to be strictly $k_y^2$, and integration regions with $|k_x|>k_y^2$ also contribute substantially to $\chi_A$. We will therefore implement a cutoff of $\Lambda_A k_y^2$, on $|k_x|$, where $\Lambda_A$ is a dimensionless parameter on which the scaling dimension of $\chi_A$ will depend. We note that $\Lambda_A$ must be taken to infinity at the end of the computations as the range of $k_x$ integration is actually unrestricted, and we will therefore study the behavior of the scaling dimension in this limit.

Considering a single patch and a single type of boson $\phi$, we have the equation for the renormalization of the Amperean pairing vertex $\Phi_A$ at $T=0$~\footnote{Our results for $2<z<3$ are qualitiatively similar to those for $z=3$ below.}:
\bea
E\Phi_A(q_x,q_y,i\Omega_m) &=& \frac{Mg^2}{N}\int_{k_y,\omega}\int_{-\Lambda_A k_y^2}^{\Lambda_A k_y^2} \frac{dk_x}{2\pi}G_+(k,i\omega_n)G_+(-k,-i\omega_n) \nonumber \\
&~&~~~~~~~\times \frac{|k_y-q_y|}{|k_y-q_y|^3+{g^2}|\omega_n-\Omega_m|/(8 \pi)}\Phi_A(k_x,k_y,i\omega_n). \label{amp1}
\eea
We can see that $\Phi_A$ does not depend upon $q_x$, and therefore we can drop the $k_x$ dependence of $\Phi_A$ on the RHS. At low energies, the equation then simplifies to
\bea
E\Phi_A(q_y,i\Omega_m) &=& \frac{Mg^2}{2N\pi}\int_{k_y,\omega}\tanh^{-1}\left(\frac{2\Lambda_A}{1+\Lambda_A^2+{|\Sigma(i\omega_n)|^2}/{k_y^4}}\right) \nonumber \\
&~&~~~~~~~~ \times \frac{|k_y-q_y|}{|k_y-q_y|^3+{g^2}|\omega_n-\Omega_m|/(8 \pi)}\frac{\Phi_A(k_y,i\omega_n)}{k_y^2}.
\eea
Note that the r.h.s. vanishes in the limit $\Lambda_A \rightarrow \infty$: this reflects the fact that, in (\ref{amp1}), the poles in $G_+(k,i\omega_n)$ and $G_+(-k,-i\omega_n)$ as a function of $k_x$ are in the same half-plane, in contrast to the situation with antipodal pairing in Section~\ref{sec:pairing}.
Using the same kind of scaling function and manipulations as in Sec. \ref{sec:2kF}, we obtain a one-dimensional integral equation
\bea
E \Psi_A(i\Omega_m) &=& \frac{M}{2N\pi}\int_{-\infty}^{\infty}\frac{d\omega_n}{2\pi}\Psi_A(i\omega_n)\tanh^{-1}\left(\frac{2\Lambda_A}{1+\Lambda_A^2+\frac{M^2|\omega_n|^{4/3}}{3N^2\pi^{4/3}}}\right) \nn
~~~&\times& \int_{-\infty}^{\infty}\frac{dk_y}{2\pi}\frac{|k_y-1||k_y|^{1-\alpha}}{|k_y-1|^3+|\omega_n|k_y|^3-\Omega_m|/(8\pi)}.
\eea
\begin{figure}[htb]
  \centering
  \includegraphics[width=0.5\textwidth]{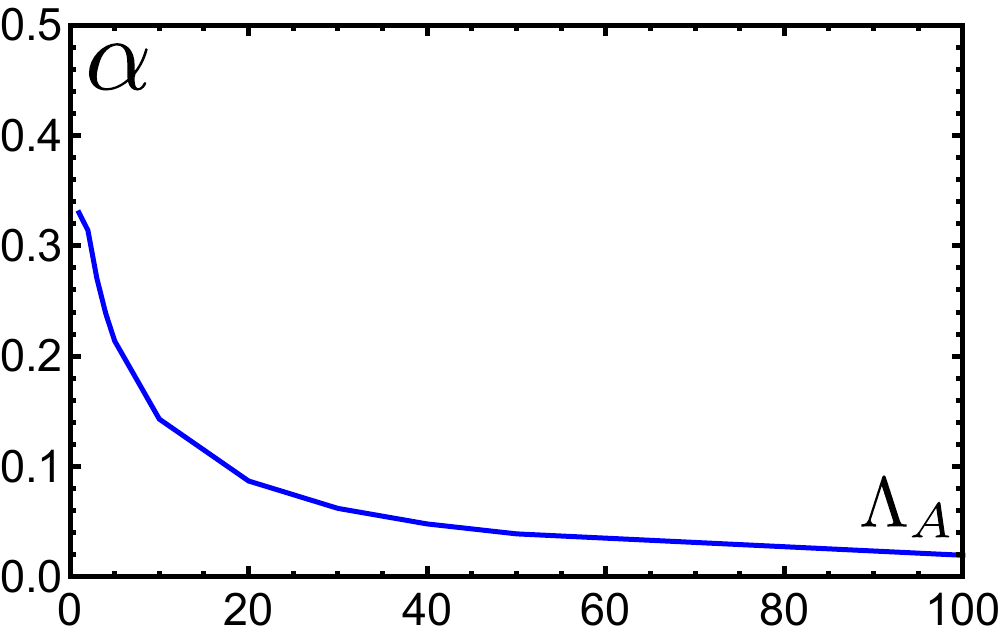}
  \caption{Plot of the scaling dimension $\alpha$ of the Amperean pairing vertex in the single patch critical theory with $M/N=1/2$, vs. the cutoff parameter $\Lambda_A$.}\label{fig:alphaAPplot}
\end{figure}
We are again interested in values of $\alpha$ with $0<\mathrm{Re}[\alpha]\le 1$, for which $E=1$, as in Sec. \ref{sec:2kF}. For the case of the spin-$1/2$ U(1) spin liquid, with $M/N=1/2$, we obtain $\alpha\approx 0.33$ for $\Lambda_A=1$. As the value of $\Lambda_A\rightarrow\infty$, we find that $\alpha\rightarrow0$ (Fig. \ref{fig:alphaAPplot}), independent of $M/N$, implying that there is no non-trivial renormalization of the Amperean pairing operator or Amperean pairing instability in the physical limit, as for $\alpha=0$, $\mathrm{dim}[\chi_A(k,\omega)]=2(1-\alpha)=2$. This continues to be the case at $T\neq0$: any transition temperature is suppressed to zero as $\Lambda_A\rightarrow\infty$. Any occurrence of Amperean pairing therefore requires consideration of physics beyond the low-energy theory discussed in our work.

\section{Large $N$ theory with spatially random couplings}
\label{app:spatial}

This appendix fills in some technical details of Sec.~\ref{sec:spatial}.

\subsection{Thermal Mass}

We determine $\Delta(T)^2=m_b^2-\Pi(0)$ as a function of temperature. In fact $\Pi(0)$ is temperature independent because it diverges as $\sum_{\omega_n} 1$, and sum over a constant is the same as integration over the same constant. Using the constraint \eqref{eq:saddle_constraint}, assuming at zero temperature $\Delta(T=0)=\Delta_0$:
\begin{equation}\label{eq:Delta1}
  \int\frac{\rd \nu}{2\pi}\int\frac{\rd^2 q}{(2\pi)^2}\frac{1}{\nu^2+\omega_q^2+\Delta_0^2-\bar{\Pi}(i\nu)}=T\sum_{\nu}\int\frac{\rd^2 q}{(2\pi)^2}\frac{1}{\nu^2+\omega_q^2+\Delta(T)^2-\bar{\Pi}(i\nu)}=\frac{1}{\gamma}.
\end{equation}
We first evaluate the $q$-integral using Pauli-Villars regularization:
\begin{equation}\label{}
\begin{split}
  \int \frac{\rd ^2 q}{(2\pi)^2}\frac{1}{\nu^2+\omega_q^2+\Delta^2-\bar{\Pi}(i\nu)} &\to \int \frac{\rd ^2 q}{(2\pi)^2}\left(\frac{1}{\nu^2+\omega_q^2+\Delta^2-\bar{\Pi}(i\nu)}-\frac{1}{\nu^2+\Lambda_q^2+\omega_q^2-\bar{\Pi}(i\nu)}\right)\\
  &= \frac{1}{4\pi}\ln\left(\frac{\nu^2-\bar{\Pi}(i\nu)+\Lambda_q^2}{\nu^2-\bar{\Pi}(i\nu)+\Delta^2}\right).
\end{split}
\end{equation}
Eq. \eqref{eq:Delta1} becomes
\begin{equation}\label{eq:DeltaT=0}
  \int \frac{\rd \nu}{2\pi}\ln\left(\frac{\nu^2-\bar{\Pi}(i\nu)+\Lambda_q^2}{\nu^2-\bar{\Pi}(i\nu)+\Delta_0^2}\right)=\frac{4\pi}{\gamma},
\end{equation}and
\begin{equation}\label{eq:DeltaTfinite}
  T\sum_{\nu}\ln\left(\frac{\nu^2-\bar{\Pi}(i\nu)+\Lambda_q^2}{\nu^2-\bar{\Pi}(i\nu)+\Delta(T)^2}\right)=\frac{4\pi}{\gamma}.
\end{equation}
For notational simplicity we substitute $\bar{\Pi}(i\nu)=-a_0|\nu|$, $a_0=\pi(g \Lambda_k/(2\pi))^2$, and we can evaluate the LHS of \eqref{eq:DeltaT=0}
\begin{equation}\label{eq:DeltaT=02}
  \begin{split}
      \int \frac{\rd \nu}{2\pi}\ln\left(\frac{\nu^2-\bar{\Pi}(i\nu)+\Lambda_q^2}{\nu^2-\bar{\Pi}(i\nu)+\Delta_0^2}\right) &= \frac{1}{2\pi}\Bigg[ 2\left(\sqrt{a_0^2-4\Delta_0^2}- a_0\right) \ln \left(\frac{\Lambda_q }{\Delta _0}\right)\\
  &\left.- 2\sqrt{a_0^2-4 \Delta _0^2} \ln
   \left(\frac{2 \Lambda_q}{a_0+\sqrt{a_0^2-4\Delta_0^2}}\right)\right.\\
&+\pi  \sqrt{4 \Lambda_q^2-a_0^2}-2 \sqrt{4 \Lambda_q^2-a_0^2} \tan^{-1}\left(\frac{a_0}{\sqrt{4 \Lambda_q^2-a_0^2}}\right)\Bigg]
  \end{split},
\end{equation} where, to satisfy the constraint \eqref{eq:DeltaT=0}, $a_0$ must be the same order as $\Lambda_k$ and $\Lambda_q$, and we assume $2\Lambda_q>a_0>2\Delta_0$. By properly choosing $\Lambda_k/\Lambda_q$ to be an order one number, there is a solution for $a_0$ around $\Delta_0=0$.

Next we evaluate \eqref{eq:DeltaTfinite}
\begin{equation}\label{eq:DeltaTfinite2}
  \begin{split}
     T\sum_{\nu}&\ln\left(\frac{\nu^2-\bar{\Pi}(i\nu)+\Lambda_q^2}{\nu^2-\bar{\Pi}(i\nu)+\Delta(T)^2}\right) = -2 T \text{log$\Gamma $}\left(\frac{a_0-i \sqrt{4 \Lambda _q^2-a_0^2}}{4 \pi  T}+1\right)-2 T \text{log$\Gamma $}\left(\frac{a_0+i \sqrt{4 \Lambda _q^2-a_0^2}}{4 \pi  T}+1\right)\\
&+2 T \text{log$\Gamma $}\left(\frac{a_0-\sqrt{a_0^2-4 \Delta (T)^2}}{4 \pi  T}+1\right)+2 T \text{log$\Gamma $}\left(\frac{a_0+\sqrt{a_0^2-4 \Delta (T)^2}}{4 \pi
   T}+1\right)+T \log \left(\frac{\Lambda _q^2}{\Delta (T)^2}\right) .
  \end{split}
\end{equation}
We can calculate $\Delta(T)$ by equating \eqref{eq:DeltaT=02} and \eqref{eq:DeltaTfinite2}.
\begin{equation}\label{eq:DetermineDeltaT1}
  \begin{split}
     0= & -2 T \text{log$\Gamma $}\left(\frac{a_0-i \sqrt{4 \Lambda _q^2-a_0^2}}{4 \pi  T}+1\right)-2 T \text{log$\Gamma $}\left(\frac{a_0+i \sqrt{4 \Lambda _q^2-a_0^2}}{4 \pi  T}+1\right) +T\ln\left(\frac{\Lambda_q^2}{a_0^2}\right)\\
       & -\frac{1}{2}  \sqrt{4 \Lambda_q^2-a_0^2}+\frac{1}{\pi} \sqrt{4 \Lambda_q^2-a_0^2} \tan
   ^{-1}\left(\frac{a_0}{\sqrt{4 \Lambda_q^2-a_0^2}}\right)+\frac{a_0}{2\pi}\ln\left(\frac{\Lambda_q^2}{a_0^2}\right)\\
&+2 T \text{log$\Gamma $}\left(\frac{a_0-\sqrt{a_0^2-4 \Delta (T)^2}}{4 \pi  T}+1\right)+2 T \text{log$\Gamma $}\left(\frac{a_0+\sqrt{a_0^2-4 \Delta (T)^2}}{4 \pi
   T}+1\right)+T \ln \left(\frac{a_0^2}{\Delta (T)^2}\right)\\
&+\frac{1}{2\pi}\left[  \left(a_0-\sqrt{a_0^2-4\Delta_0^2}\right) \ln \left(\frac{a_0 ^2}{\Delta _0^2}\right)+ 2\sqrt{a_0^2-4 \Delta _0^2} \ln
   \left(\frac{2 a_0}{a_0+\sqrt{a_0^2-4\Delta_0^2}}\right)\right].
  \end{split}
\end{equation}
We can remove the cutoff by expanding in large $\Lambda_q$ and obtain
\begin{equation}\label{eq:DeltaTnoLambda}
\begin{split}
  &\frac{a_0}{2\pi T}\left(\ln\left(\frac{\Delta(0)}{2\pi T}\right)-1\right)+\frac{\sqrt{a^2-4\Delta(0)^2}}{2\pi T}\ln\left(\frac{a+\sqrt{a^2-4\Delta(0)^2}}{2\Delta(0)}\right)\\
  &+\ln\left(\frac{\Delta(T)}{T}\right)-\ln\Gamma\left(1+\frac{a-\sqrt{a^2-\Delta(T)^2}}{4\pi T}\right)-\ln\Gamma\left(1+\frac{a+\sqrt{a^2-\Delta(T)^2}}{4\pi T}\right)=0.
\end{split}
\end{equation}

At criticality $\Delta_0=0$, and the critical low temperature solution is given by \cite{Patel2014DCR}
\begin{equation}
  \Delta(T)^2=\frac{-\pi a T W_0\left(-\frac{1}{\pi}\ln\left(\frac{2\pi T }{a_0 e^{\gamma}}\right)\right)}{\ln\left(\frac{2\pi T }{a_0 e^{\gamma}}\right)},
\end{equation} where $W_0$ is the principle Lambert W-function and $\gamma$ is Euler's constant. The above result can be obtained by writing $\Delta(T)^2=a_0 T g(T)$ and then expand \eqref{eq:DeltaTnoLambda} in small $T/a_0$, and solve $g(T)$ using the leading order constraint.

\subsection{Free Energy}

We compute the free energy of the theory. The free energy $F$ is given by the value of saddle point action
\begin{equation}\label{}
  \begin{split}
     \frac{\beta F}{N} & = -\ln\det(\partial_\tau+\varepsilon_k-\mu+\Sigma)+\frac{1}{2}\ln\det(-\partial_\tau^2+\omega_q^2+i\lambda-\Pi) \\
       & -\int \rd\tau\rd^2 x \int \rd \tau'\rd^2 x'\left(\Sigma(\tau',x';\tau,x)G(\tau,x;\tau',x')-\frac{1}{2}\Pi(\tau',x';\tau,x)D(\tau,x;\tau',x')\right) \\
       & +\int \rd\tau\rd^2 x \int \rd \tau'\rd^2 x' \frac{g^2}{2}G(\tau,x;\tau',x')G(\tau',x';\tau,x)D(\tau,x;\tau',x')\delta(x-x')\\
&-\int \rd \tau \rd^2 x \frac{i \lambda(\tau,x)}{2\gamma},
  \end{split}
\end{equation} where the variables should be substituted by their saddle point values.

We integrating out CoM coordinates and expand the determinant in momentum space, and also split $\Pi(i\omega)=\bar{\Pi}(i\omega)+\Pi(0)$, $i\lambda=\Delta(T)^2+\Pi(0)$, and use the constraint \eqref{eq:saddle_constraint}, to obtain
\begin{equation}\label{eq:freeenergy3}
  \begin{split}
     \frac{F}{N V} & =-T\sum_{\omega_n}\int\frac{\rd^2 \vec{k}}{(2\pi)^2}\ln\left(\varepsilon_{\vec{k}}-\mu-i\omega_n+\Sigma(i\omega_n)\right) \\
       & +\frac{1}{2} T\sum_{\nu_n}\int\frac{\rd^2\vec{q}}{(2\pi)^2}\ln\left(\nu_n^2+\omega_{\vec{q}}^2+\Delta(T)^2-\bar{\Pi}(i\nu_n)\right) \\
       & -T\sum_{\omega_n}\bar{G}(i\omega_n)\Sigma(i\omega_n)+\frac{T}{2}\sum_{\nu_n}\bar{D}(i\nu_n)\bar{\Pi}(i\nu_n)\\
       &+\frac{g^2}{2}\int\rd\tau \bar{G}(\tau)\bar{G}(-\tau)\bar{D}(\tau)\\
       &-\frac{\Delta(T)^2}{2\gamma}
  \end{split}
\end{equation}
The interaction term cancels the $\bar{D}\bar{\Pi}$ term, leaving a $\Pi(0)$ term which is assumed to be temperature independent. We try to evaluate the remaining terms.

\subsubsection{Fermion determinant}
We regulate the fermion determinant by the free fermion counterpart:
\begin{equation}\label{}
\begin{split}
  \left(\frac{F}{NV}\right)_1=&-T\sum_{\omega_n}\int\frac{\rd^2 \vec{k}}{(2\pi)^2}\ln\left(\frac{\varepsilon_{\vec{k}}-\mu-i\omega_n+\Sigma(i\omega_n)}{\varepsilon_{\vec{k}}-\mu-i\omega_n}\right)\\
  &-T\sum_{\omega_n}\int\frac{\rd^2 \vec{k}}{(2\pi)^2}\ln\left(\varepsilon_{\vec{k}}-\mu-i\omega_n\right)\\
  &-T\sum_{\omega_n}\bar{G}(i\omega_n)\Sigma(i\omega_n)
\end{split}
\end{equation}
The second line is the standard free fermion result
\begin{equation}\label{}
  -T\int\frac{\rd^2\vec{k}}{(2\pi)^2}\ln\left(1+e^{-\beta(\varepsilon_{\vec{k}}-\mu)}\right)=-T\Lambda_k\int\frac{\rd k_x}{2\pi}\ln(1+e^{-\beta k_x}).
\end{equation}
For the first line, we use the sliding symmetry to set
$$
\varepsilon_{\vec{k}}-\mu\to k_x,\qquad \int\frac{\rd^2 \vec{k}}{(2\pi)^2} \to \Lambda_k\int\frac{\rd k_x}{2\pi},
$$
and then we perform the $k_x$-integral in principal value, which yields
\begin{equation}\label{eq:F1}
\begin{split}
   \left(\frac{F}{NV}\right)_1=&-\frac{i}{2}\Lambda_k T\sum_{\omega_n} \sgn(\omega_n)\Sigma(i\omega_n)\\
  &-T\Lambda_k\int\frac{\rd k_x}{2\pi}\ln(1+e^{-\beta k_x})\\
  &-T\sum_{\omega_n}\bar{G}(i\omega_n)\Sigma(i\omega_n)\\
  &=-T\Lambda_k\int\frac{\rd k_x}{2\pi}\ln(1+e^{-\beta k_x})\\
\end{split}
\end{equation}
Here the first line and the third line of \eqref{eq:F1} cancelled, and we are left with a free fermion result. This part yields a specific heat
\begin{equation}\label{eq:gamma1}
    C_1=NV\gamma_1 T,\qquad \gamma_1=\frac{\pi}{6}\Lambda_k\,.
\end{equation}

\subsubsection{Boson Determinant}
We consider the boson determinant term
\begin{equation}\label{}
  \left(\frac{F}{NV}\right)_2 = \frac{1}{2}T \sum_{\nu_n}\int\frac{\rd^2 \vec{q}}{(2\pi)^2}\ln\left(\nu_n^2+\vec{q}^2-\bar{\Pi}(i\nu_n)+\Delta(T)^2\right)-\frac{\Delta(T)^2}{2\gamma}
\end{equation}
The difference from zero temperature can be written as
\begin{equation}\label{}
   \left(\frac{F(T)}{NV}\right)_2- \left(\frac{F(0)}{NV}\right)_2=I_1+I_2,
\end{equation}where
\begin{equation}\label{}
  I_1=\frac{1}{2}\int\frac{\rd \nu}{2\pi}\int\frac{\rd^2\vec{q}}{(2\pi)^2}\left[\ln\left(\frac{\nu^2+\vec{q}^2-\bar{\Pi}(i\nu)+\Delta(T)^2}{\nu^2+\vec{q}^2-\bar{\Pi}(i\nu)+\Delta(0)^2}\right)-\frac{\Delta(T)^2-\Delta(0)^2}{\nu^2-\bar{\Pi}(i\nu)+\vec{q}^2+\Delta(0)^2}\right],
\end{equation}
\begin{equation}\label{}
  I_2=\frac{1}{2}\int \frac{\rd ^2 \vec{q}}{(2\pi)^2}\left[T\sum_{\nu_n}\ln(\nu_n^2+\vec{q}^2-\bar{\Pi}(i\nu_n)+\Delta(T)^2)-\int\frac{\rd \nu}{2\pi}\ln(\nu^2+\vec{q}^2-\bar{\Pi}(i\nu)+\Delta(T)^2)\right]
\end{equation}
Since $I_1$ is sufficiently convergent, we expect we can exchange the order of integrals.
We perform the momentum integral for $I_1$ first (we remind that $a_0=\pi\left(\frac{g \Lambda_k}{2\pi}\right)^2$):
\begin{equation}\label{}
\begin{split}
  I_1&=\frac{1}{8\pi^2}\int_0^\infty \rd \nu\left[(\Delta(T)^2-\Delta(0)^2)-(\Delta(T)^2+\nu^2+a_0\nu)\ln\frac{\nu^2+a_0 \nu+\Delta(T)^2}{\nu^2+a_0\nu+\Delta(0)^2}\right]\\
  &=\frac{1}{96\pi^2}\left[-2a_0(\Delta(T)^2-\Delta(0)^2)+2a_0(6\Delta(T)^2-a_0^2)\ln\left(\frac{\Delta(T)}{\Delta(0)}\right)\right.\\
   &+(a_0^2-4\Delta(T)^2)^{3/2}\ln\left(\frac{a_0-\sqrt{a_0^2-4\Delta(T)^2}}{a_0+\sqrt{a_0^2+4\Delta(T)^2}}\right)
  \\
  &\left.+\sqrt{a_0^2-4\Delta(0)^2}(6\Delta(T)^2-a_0^2-2\Delta(0)^2)\ln\left(\frac{a_0-\sqrt{a_0^2-4\Delta(0)^2}}{a_0+\sqrt{a_0^2-4\Delta(0)^2}}\right)\right].
\end{split}
\end{equation}This expression agrees with numerics.
The $I_2$ can be brought into an integral form
\begin{equation}\label{}
  I_2=-\int\frac{\rd^2\vec{q}}{(2\pi)^2}\int_{0}^{\infty}\frac{\rd z}{\pi}n_B(z)\tan^{-1}\left(\frac{a_0 z}{q^2+\Delta(T)^2-z^2}\right),
\end{equation}and the $\tan^{-1}$ function should vary continuously from $0$ to $\pi$.
We rewrite the $I_2$ integral as
\begin{equation}\label{}
  I_2=\int\frac{\rd ^2\vec{q}}{(2\pi)^2} \int_{\infty}^{\sqrt{\vec{q}^2+\Delta(T)^2}}\eta\rd \eta \int_0^\infty \frac{\rd z}{\pi i}\frac{n_B(z)}{\sqrt{4\eta^2-a_0^2}}\left(\frac{2z}{z^2+u_+^2}-\frac{2z}{z^2+u_{-}^2}\right),
\end{equation}where
\begin{equation}\label{}
  u_{\pm}=\frac{a_0\mp i\sqrt{4\eta^2-a_0^2}}{2}.
\end{equation}
Using the formula
\begin{equation}\label{}
  h(a_0)\equiv \int_0^\infty \rd z n_B(z)\frac{2z}{z^2+a_0^2}=\ln\frac{a_0}{2\pi T}-\frac{\pi T}{a_0}-\psi\left(\frac{a_0}{2\pi T}\right),\quad \Re a_0>0,
\end{equation} we can evaluate the $z$ integral as
\begin{equation}\label{}
  I_2=\int \frac{\rd^2 \vec{q}}{(2\pi)^2} g(\sqrt{\vec{q}^2+\Delta(T)^2}),
\end{equation}and
\begin{equation}\label{}
  \begin{split}
     g(\eta) & =-\frac{1}{4\pi}\left[i \sqrt{4 \eta ^2-a_0^2} \left(\log \left(a_0-i \sqrt{4 \eta ^2-a_0^2}\right)-\log \left(a_0+i \sqrt{4 \eta ^2-a_0^2}\right)\right)\right.\\
     &\left.+4 \pi  T \left(\text{log$\Gamma $}\left(\frac{a_0-i \sqrt{4 \eta ^2-a_0^2}}{4 \pi  T}\right)+\text{log$\Gamma $}\left(\frac{a_0+i \sqrt{4 \eta ^2-a_0^2}}{4 \pi  T}\right)\right)+2 a_0 \log \left(\frac{\pi  T}{\eta }\right)\right.\\
     &\left.+a_0
   (2+\log (4))-4 \pi  T \log \left(\frac{T}{\eta }\right)-8 \pi  T \log (2 \pi )\right]
  \end{split}
\end{equation}
The remaining momentum integral is log-divergent because
\begin{equation}\label{}
  g(\eta)=-\frac{a_0\pi T^2}{6\eta^2}+\mathcal{O}(1/\eta^3),
\end{equation}which yields a term $-\frac{a_0 T^2}{12}\ln\frac{\Lambda_q}{\Delta(T)}$ in the self-energy.
The total contribution of $I_2$ is
\begin{equation}\label{}
\begin{split}
  I_2&=-\frac{a_0 T^2}{12}\ln\frac{\Lambda_q}{\Delta(T)}+\int_{\Delta(T)}^{\infty}\frac{\eta \rd \eta}{2\pi}\left(g(\eta)+\frac{a_0\pi T^2}{6\eta^2}\right)\\
  &=-\frac{a_0 T^2}{12}\ln\frac{\Lambda_q}{\Delta(T)}+\frac{a_0^2 \sqrt{a_0^2-4 \Delta(T) ^2} \coth ^{-1}\left(\frac{a_0}{\sqrt{a_0^2-4 \Delta(T) ^2}}\right)}{48 \pi ^2}\\
  &-\frac{\Delta(T) ^2 \sqrt{a_0^2-4 \Delta(T) ^2} \coth ^{-1}\left(\frac{a_0}{\sqrt{a_0^2-4 \Delta(T) ^2}}\right)}{12 \pi ^2}\\
  &+2 \pi  T^3 \psi ^{(-3)}\left(\frac{a_0-\sqrt{a_0^2-4 \Delta(T) ^2}}{4 \pi  T}\right)+2 \pi  T^3 \psi ^{(-3)}\left(\frac{a_0+\sqrt{a_0^2-4 \Delta(T)
   ^2}}{4 \pi  T}\right)\\
   &+\frac{1}{2} T^2 \sqrt{a_0^2-4 \Delta(T) ^2} \psi ^{(-2)}\left(\frac{a_0-\sqrt{a_0^2-4 \Delta(T) ^2}}{4 \pi  T}\right)\\
   &-\frac{1}{2} T^2 \sqrt{a_0^2-4 \Delta(T) ^2} \psi ^{(-2)}\left(\frac{a_0+\sqrt{a_0^2-4 \Delta(T) ^2}}{4 \pi  T}\right)\\
   &-\frac{a_0^3 \log \left(\frac{2 \pi  T}{\Delta(T) }\right)}{48 \pi ^2}-\frac{a_0^2 T}{16 \pi }-\frac{5 a_0^3}{288 \pi
   ^2}-a_0 T^2 \log (A)+\frac{7 a_0 \Delta(T) ^2}{48 \pi ^2}+\frac{1}{12} a_0 T^2 \log \left(\frac{2 \pi  T}{\Delta(T) }\right)\\
   &+\frac{a_0 T^2}{12}+\frac{a_0 \Delta(T) ^2 \log \left(\frac{2 \pi  T}{\Delta(T) }\right)}{8 \pi ^2}-\frac{T^3 \zeta (3)}{2 \pi }-\frac{\Delta(T) ^2 T}{8 \pi }-\frac{\Delta(T) ^2 T \log \left(\frac{4 \pi ^2 T}{\Delta(T) }\right)}{4 \pi },
\end{split}
\end{equation}where $A$ is Glaisher's constant $\log A=1/12-\zeta'(-1)$.
If we assume $\Delta(0)\neq 0$, the low temperature asymptotics is
\begin{equation}\label{}
  I_2=-\frac{a_0 T^2}{12}\ln\frac{\Lambda_q}{\Delta(T)}+\frac{a_0 \pi^2(a_0^2-6\Delta(T)^2)T^4}{360 \Delta(T)^4}\,,
\end{equation}
and for the critical case $\Delta(0)=0$ there is no simplification.
The boson free energy is the sum of $I_1$ and $I_2$.

\bibliography{fermi}

\end{document}